\documentclass[reprint, superscriptaddress, amsmath, amssymb, aps]{revtex4-2}
\usepackage{graphicx,dcolumn,bm,xcolor,mathptmx,etoolbox,blindtext,upgreek,float,nicefrac}
\usepackage[normalem]{ulem}
\usepackage[utf8]{inputenc}
\usepackage[T1]{fontenc}
\definecolor{prlblue}{RGB}{46, 48, 146}
\usepackage[colorlinks=true,citecolor=prlblue,urlcolor=prlblue,linkcolor=prlblue]{hyperref}
\usepackage{silence}
\newcommand{\microohmcm}{$\mu\Omega$-cm}

\newcommand{\STO}{SrTiO$_3$}
\newcommand{\NNO}{NdNiO$_2$}
\newcommand{\LNO}{LaNiO$_2$}
\newcommand{\PNO}{PrNiO$_2$}

\newcommand{\ILN}{infinite-layer nickelates}
\newcommand{\Tcon}{$T_c^{\mathrm{on}}$}
\newcommand{\rhores}{$\rho_{\mathrm{res}}$}

\begin{document}
\title{Superconductivity in the Parent Infinite-Layer Nickelate NdNiO$_2$}
\author{C. T. Parzyck}
  \affiliation{Laboratory of Atomic and Solid State Physics, Department of Physics, Cornell University, Ithaca, NY 14853, USA}
\author{Y. Wu}
  \affiliation{Laboratory of Atomic and Solid State Physics, Department of Physics, Cornell University, Ithaca, NY 14853, USA}
\author{L. Bhatt}
  \affiliation{School of Applied and Engineering Physics, Cornell University, Ithaca, NY 14853, USA}
\author{M. Kang}
  \affiliation{Laboratory of Atomic and Solid State Physics, Department of Physics, Cornell University, Ithaca, NY 14853, USA}
  \affiliation{Department of Materials Science and Engineering, Cornell University, Ithaca, NY 14853, USA}
  \affiliation{Kavli Institute at Cornell for Nanoscale Science, Cornell University, Ithaca, NY 14853, USA}
\author{Z. Arthur}
  \affiliation{Canadian Light Source, Inc., 44 Innovation Boulevard, Saskatoon, SK S7N 2V3, Canada}
\author{T. M. Pedersen}
  \affiliation{Canadian Light Source, Inc., 44 Innovation Boulevard, Saskatoon, SK S7N 2V3, Canada}
\author{R. Sutarto}
  \affiliation{Canadian Light Source, Inc., 44 Innovation Boulevard, Saskatoon, SK S7N 2V3, Canada}
\author{S. Fan}
  \affiliation{National Synchrotron Light Source II, Brookhaven National Laboratory, Upton, NY 11973, USA}
\author{J. Pelliciari}
  \affiliation{National Synchrotron Light Source II, Brookhaven National Laboratory, Upton, NY 11973, USA}
\author{V. Bisogni}
  \affiliation{National Synchrotron Light Source II, Brookhaven National Laboratory, Upton, NY 11973, USA}
\author{G. Herranz}
  \affiliation{Institut de Ciència de Materials de Barcelona (ICMAB-CSIC), Campus UAB Bellaterra 08193, Spain}
\author{A. B. Georgescu}
  \affiliation{Department of Chemistry, Indiana University, Bloomington, IN 47405, USA}
\author{D. G. Hawthorn}
  \affiliation{Department of Physics and Astronomy, University of Waterloo, Waterloo ON N2L 3G1, Canada}
\author{L. F. Kourkoutis}
  \affiliation{School of Applied and Engineering Physics, Cornell University, Ithaca, NY 14853, USA} 
  \affiliation{Kavli Institute at Cornell for Nanoscale Science, Cornell University, Ithaca, NY 14853, USA}
\author{D. A. Muller}
  \affiliation{School of Applied and Engineering Physics, Cornell University, Ithaca, NY 14853, USA}
  \affiliation{Kavli Institute at Cornell for Nanoscale Science, Cornell University, Ithaca, NY 14853, USA}
\author{D. G. Schlom}
  \affiliation{Department of Materials Science and Engineering, Cornell University, Ithaca, NY 14853, USA}
  \affiliation{Kavli Institute at Cornell for Nanoscale Science, Cornell University, Ithaca, NY 14853, USA}
  \affiliation{Leibniz-Institut f{\"u}r Kristallz{\"u}chtung, Max-Born-Stra{\ss}e 2, 12489 Berlin, Germany}
\author{K. M. Shen}
  \affiliation{Laboratory of Atomic and Solid State Physics, Department of Physics, Cornell University, Ithaca, NY 14853, USA}
  \affiliation{Kavli Institute at Cornell for Nanoscale Science, Cornell University, Ithaca, NY 14853, USA}
  \affiliation{Institut de Ciència de Materials de Barcelona (ICMAB-CSIC), Campus UAB Bellaterra 08193, Spain}

\begin{abstract}
We report evidence for superconductivity with onset temperatures up to 11 K in thin films of the infinite-layer nickelate parent compound NdNiO$_2$.  A combination of oxide molecular-beam epitaxy and atomic hydrogen reduction yields samples with high crystallinity and low residual resistivities, a substantial fraction of which exhibit  superconducting transitions. We survey a large series of samples with a variety of techniques, including electrical transport, scanning transmission electron microscopy, x-ray absorption spectroscopy, and resonant inelastic x-ray scattering, to investigate the possible origins of superconductivity. We propose that superconductivity could be intrinsic to the undoped infinite-layer nickelates but suppressed by disorder due to its nodal order parameter, a finding which would necessitate a reconsideration of the nickelate phase diagram. Another possible hypothesis is that the parent materials can be hole doped from randomly dispersed apical oxygen atoms, which would suggest an alternative pathway for achieving superconductivity.
\end{abstract}
\maketitle

\begin{figure*}[ht]
  \resizebox{17.2 cm}{!}{\includegraphics{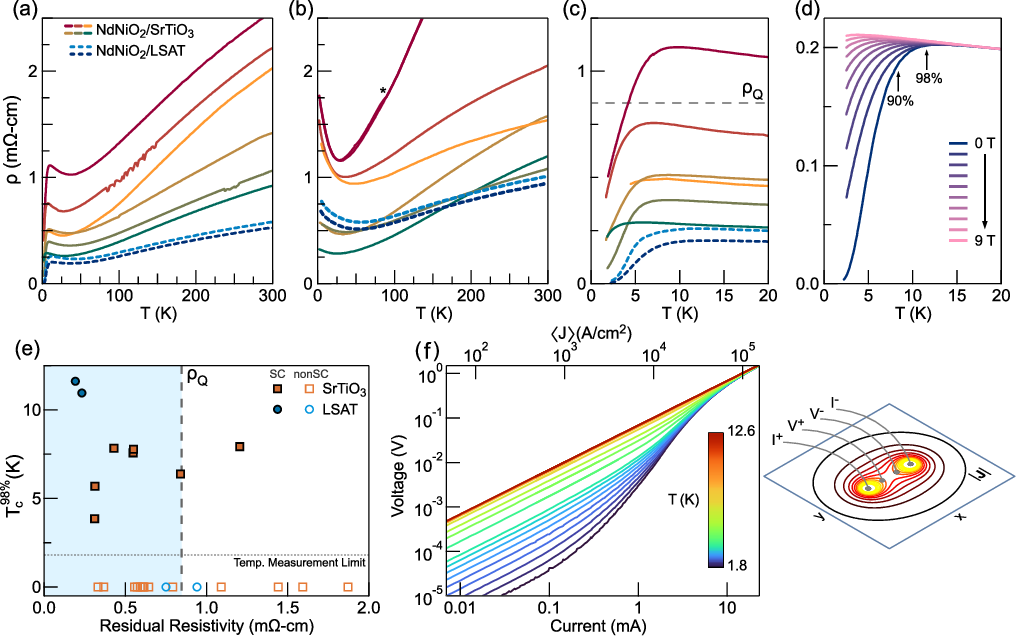}}
  \caption{\label{fig:transport} Electrical transport measurements of 20 u.c. thick \NNO~ films grown on (001) \STO~ and (001) (LaAlO$_3$)$_{0.3}$(Sr$_2$AlTaO$_6$)$_{0.7}$ (LSAT) capped with 2-3 u.c. of \STO. (a) Resistivity of a series of films which display partial superconducting transitions. (b) Resistivity of a representative series of films which do not exhibit signs of superconductivity. The onset of a small hysteretic transition in a high resistance sample at $\sim 90$~K is marked with a star. (c) Zoom in on the transition region for traces in (a). Dashed line represents the resistivity corresponding to the quantum sheet resistance per NiO$_2$ plane, $\rho_Q$. (d) Superconducting transition of an \NNO~ film on LSAT under applied magnetic fields, in 1 T steps. Arrows mark where the zero-field resistance drops to 90\% and 98\% of the normal state resistance. (e) Extracted onset \Tcon~(defined by a 2\% drop below the normal state resistance) plotted versus the film residual resistivity, \rhores~ (defined as the extrapolated normal state resistance to 0 K using high field data for superconducting samples)  (f) Pulsed IV measurements of the same film as in panel (d) measured between equally spaced (500 $\mu$m) leads in the pictured geometry. The estimated average current density between the voltage leads, $\langle J \rangle$, assuming a uniform and isotropic resistivity, is provided on the top axis.
  }
\end{figure*}

\section{Introduction}
A defining characteristic of many unconventional high-temperature superconductors, including the cuprates, Fe-based, and now the \ILN~is the dome-shaped dependence of their superconducting transition temperature ($T_c$) on hole or electron doping \cite{Li2019b}.  Sharing the motif of a $d^9$ transition metal in a square plaquette of oxygen ions \cite{Crespin1983a,Hayward1999,Hayward2003a,Ikeda2016}, the \ILN~mirror the cuprates in numerous ways, including a strange metal near optimal doping and Fermi liquid behaviour in the overdoped regime \cite{Lee2023}. On the other hand, the nickelates and cuprates also display notable differences. While the undoped cuprates are antiferromagnetic charge-transfer insulators with a $\sim 2$~ eV gap, undoped nickelates are metallic, presumably from the self-doping of the NiO$_2$ planes by rare-earth $5d$ states \cite{Jiang2019a,Botana2020,Kitatani2020,Karp2020,Sakakibara2020,DiCataldo2023}, and possess short-ranged magnetic fluctuations \cite{Lu2021,Cui2021,Fowlie2022,Rossi2024} in lieu of long-range N\'eel order.  While the cuprates are well-described by a single-band model, the rare-earth $5d$ and $3d_{z^2}$ orbitals may also play a role in the low-energy electronic structure of the nickelates \cite{Lechermann2020a}.  Calculations also suggest important differences in the $p-d$ hybridization and relative strength of correlations in the Ni$^{1+}$ system \cite{Lee2004,Jiang2020,Jiang2022}.  

These deviations from canonical `cuprate-like' behavior motivate a detailed and systematic investigation of the nature of the parent \ILN. There are, however, significant challenges associated with the synthesis of the pristine parent phase, as evidenced by the fact that their reported resistivities vary by orders of magnitude depending on the rare-earth \cite{Ikeda2016,Li2019b,Osada2020,Zeng2021,Osada2021} and reduction procedure used\cite{Kawai2009,Ikeda2013,Ikeda2014,Wei2023,Parzyck2024b}. This is likely due to the challenge of fully reducing the parent compounds versus the doped compounds, since hole doping increases the targeted Ni valence, thereby easing the reduction process \cite{Wang2020}. Possibly as a result, many recent studies focus almost exclusively on the doped compounds, and the undoped parent phase is notably absent from recent doping-dependent studies \cite{Zeng2020,Lee2023}. While superconductivity is typically restricted to a window of hole doping between 12.5 and 27.5\%, one study has reported evidence of superconductivity in undoped, CaH$_2$~ reduced, LaNiO$_2$ ($T_c < 5$ K) \cite{Osada2021} -- if the parent nickelates are indeed superconducting, it would necessitate a reconsideration of the nickelate phase diagram.  

Here, we report superconductivity in the infinite-layer parent compound NdNiO$_2$ with onset $T_c$'s up to 11 K in thin films prepared using a combination of molecular-beam epitaxy (MBE) and atomic hydrogen reduction. A series of 20 unit cell (u.c.) thick samples, capped with 2-3 u.c. of \STO, were synthesized under nearly identical conditions, enabling a highly systematic study employing a combination of electrical transport, high resolution scanning transmission electron microscopy (STEM), x-ray absorption spectroscopy (XAS), and resonant inelastic x-ray scattering (RIXS).  Among these nominally identically prepared samples, roughly one in three exhibit partial superconducting transitions while the rest display metallic behaviour--without systematic differences between the two sets in lab-based x-ray diffraction measurements \cite{SIcite}.  We conclude that the most likely scenario is that superconductivity is intrinsic to the clean limit of the undoped parent nickelates, but is readily suppressed by disorder. We also speculate on the possibility of hole doping of the NiO$_2$ planes from randomly dispersed apical oxygen atoms. Either of the two scenarios would have important implications: if superconductivity is intrinsic to the parent compounds, this would suggest that the current phase diagram should be revisited. If excess oxygen is responsible, this would indicate a powerful alternative route for hole doping and achieving superconductivity. 

\section{Results}
We show resistivity measurements on a series of \NNO~samples, where a significant fraction show partial superconducting transitions, Fig.
\hyperref[fig:transport]{\ref*{fig:transport}(a)}, while others exhibit metallic behavior, Fig. \hyperref[fig:transport]{\ref*{fig:transport}(b)}.  Residual resistivities, $\rho_{\mathrm{res}}$, were extrapolated from the normal state resistance and typically lie in the 300-700 \microohmcm~ range, significantly lower than other undoped \NNO~films in the literature ($\rho_{\mathrm{res}} \approx$ 2500 - 6000 \microohmcm) \cite{Li2020a,Tam2022,Wei2023}.  For superconducting samples, the transition region is shown in greater detail in Fig. \hyperref[fig:transport]{\ref*{fig:transport}(c)}, where the onset is apparent between 5 and 11 K. Although none of the samples exhibit a zero-resistance state down to 1.8 K, the most complete transitions are those of films grown on (LaAlO$_3$)$_{0.3}$(Sr$_2$AlTaO$_6$)$_{0.7}$ (LSAT), which also have the highest onset $T_c$'s. The superconducting nature of the transition is confirmed by application of a magnetic field in Fig. \hyperref[fig:transport]{\ref*{fig:transport}(d)}, which fully suppresses the transition by $\sim 7$~T for films on \STO~ and $\sim 9$~T for films on LSAT. 

We define the onset transition temperature, $T_c^{\mathrm{on}}$, as the temperature at which the resistance decreases by 2\% from its normal state value, as determined from high-field measurements; in Fig. \hyperref[fig:transport]{\ref*{fig:transport}(e)}, we plot $T_c^{\mathrm{on}}$ vs. $\rho_{\mathrm{res}}$ for all samples, illustrating that \rhores~of nearly all superconducting samples are clustered below the quantum of resistance per NiO$_2$ plane ($\rho_{\mathrm{res}} < \rho_Q= hd/e^2 \approx 850$ ~\microohmcm, where $d=3.285$~\AA), similar to previous reports where a downturn was reported in doped and undoped (La,Sr)NiO$_2$ \cite{Osada2021}. This suggests that disorder could play a key role in suppressing superconductivity in the undoped nickelates and would be consistent with evidence from THz \cite{Cheng2024} and penetration depth measurements \cite{Harvey2022} that the \ILN~possess a nodal, possibly $d$-wave, superconducting order parameter, since all impurities in nodal superconductors should be pair-breaking \cite{Millis1988a,Radtke1993}. This is supported by the fact that samples grown on LSAT exhibit  higher $T_c^{\mathrm{on}}$, lower $\rho_{\mathrm{res}}$, and more complete superconducting transitions, likely attributable to the improved crystalline quality and reduced propensity for defects due to the better lattice match between NdNiO$_3$ and LSAT \cite{Lee2023,Parzyck2024b}. 

The broad nature of the transitions and lack of a zero resistance state suggests that the  superconductivity is likely inhomogeneous. This could arise from a number of possibilities, including a non-uniform distribution of defects resulting in regions where superconductivity is quenched, chemical inhomogeneities, or that the superconductivity could originate from tiny amounts of a filamentary secondary phase. To rule out the latter, we have performed pulsed I-V measurements, Fig. \hyperref[fig:transport]{\ref*{fig:transport}(f)}, which show that the sample accommodates currents as high as 5-10 mA, resulting in an estimated critical current density ($J_c$) in the $10^4$~ A/cm$^2$ range, before fully returning to the normal state. This measurement was performed in an unconstrained geometry, with $J_c$ estimated using a simple analytic model \cite{Smits1958}. Though this model likely overestimates $J_c$, the extracted values are fairly comparable to prior measurements of optimally doped samples ($J_c \approx 1.0-3.5\times 10^{5}$~A/cm$^2$) \cite{Osada2020,Wei2023a}. This suggests that superconductivity exists throughout a substantial fraction of the sample, since if only a small volume fraction of the sample were superconducting, $J_c$ should be far lower than the observed values. Nevertheless, a significant number of samples with $\rho_{\mathrm{res}} < \rho_Q$ do not superconduct, raising the question about whether disorder is the only relevant variable. Some alternative scenarios include the hole doping from residual apical oxygens left behind during the reduction process, or from Sr diffusion, resulting in Nd$_{1-x}$Sr$_{x}$NiO$_{2}$ inclusions.

\begin{figure}
  \resizebox{8.6 cm}{!}{\includegraphics{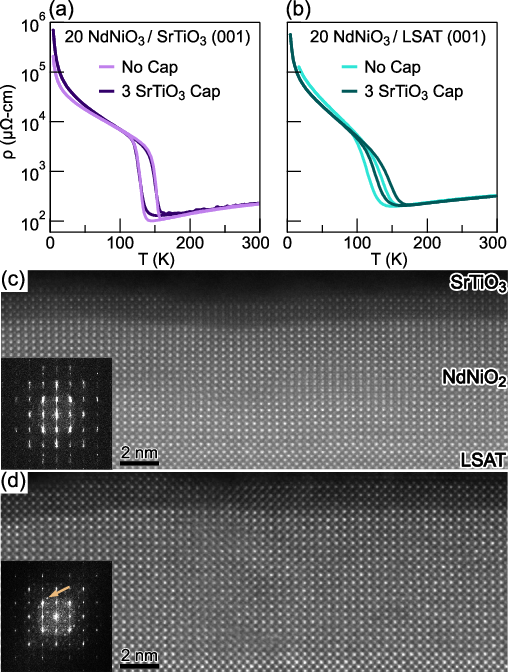}}
  \caption{\label{fig:capping} Investigation of Sr interdiffusion into \NNO. (a-b) Electrical resistivity of 20 u.c. thick, undoped, unreduced perovskite NdNiO$_3$ films grown on \STO~ and LSAT, with and without SrTiO$_3$ capping layers.  (c) ADF-STEM measurements of a superconducting NdNiO$_2$/LSAT film showing a region without $3a_0$~oxygen ordering. (d) Image of a different region of the same film showing $3a_0$~ order. Fourier transforms of the ADF images are inset with the fractional $\nicefrac{1}{3}$ order peak highlighted by an arrow.}
\end{figure}

\begin{figure*}[t]
  \resizebox{17.2 cm}{!}{\includegraphics{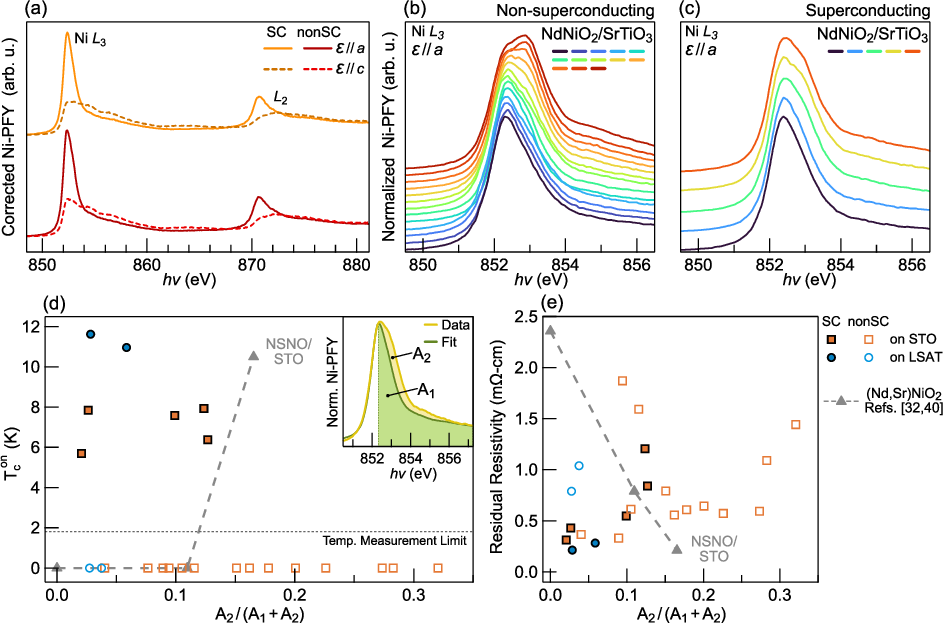}}
  \caption{\label{fig:xas} X-ray absorption spectroscopy (XAS) measurements of \NNO. (a) Self-absorption corrected Ni $L$-edge partial fluorescence yield (PFY) measurements of superconducting and non-superconducting films on SrTiO$_3$, collected at an incidence angle of $\theta=20^{\circ}$ in $\sigma$~ and $\pi$~ polarization such that $\varepsilon\parallel a$ in $\sigma$~ polarization.  $\varepsilon\parallel c$ signal is determined by decomposing the $\pi$-polarized data, $I=(I_{\pi}-I_{\sigma}\sin^2(20^{\circ})) / \cos^2(20^{\circ})$. (b) Ni $L_3$ PFY measurements of \NNO / \STO~ films not exhibiting superconducting transitions. (c) The same, for films exhibiting signatures of superconductivity. Traces in (b) and (c) have been normalized to their maximum value and offset for clarity. (d) \Tcon~versus excess spectral weight above the undoped \NNO~lineshape, as defined in the inset, for samples on SrTiO$_3$~ (squares) and LSAT (circles) substrates. (e) Comparison of \rhores~versus excess spectral weight for \NNO~and data from  Nd$_{1-x}$Sr$_x$NiO$_2$, extracted from Refs. \cite{Rossi2021a,Li2020a}.}
\end{figure*}

We first rule out the possibility of Sr interdiffusion from either the substrate or capping layer through a combination of electrical transport and STEM measurements. Sr diffusion would be most likely during film synthesis ($T\sim 500-550~^{\circ}$C), rather than the reduction stage, which is brief and occurs at lower temperatures ($T\sim 300~^{\circ}$C). In perovskite NdNiO$_3$, it is well established that small amounts of hole doping ($x \leq 0.06$) completely suppress the low-temperature insulating state \cite{Cheong1994,Alonso1995b,Yang2021,Patel2022}. In contrast, all of our unreduced NdNiO$_3$ samples, whether capped or uncapped, exhibit sharp metal-to-insulator transitions, Figs. \hyperref[fig:capping]{\ref*{fig:capping}(a)} and \hyperref[fig:capping]{\ref*{fig:capping}(b)}, with a nearly 10$^4$ increase in resistivity, excluding the possibility of Sr doping. Furthermore, annular dark field (ADF) STEM measurements on a reduced, superconducting NdNiO$_2$ / LSAT film show that the film-substrate and film-cap interfaces are sharp, Figs. \hyperref[fig:capping]{\ref*{fig:capping}(c)} and \hyperref[fig:capping]{\ref*{fig:capping}(d)}, without noticeable mosaicity or planar fault formation in the SrTiO$_3$ capping layer, which occurs when the Sr non-stoichiometry exceeds $10$~\% \cite{Brooks2009a,Brooks2015}. Given how thin the \STO~ cap is (3 u.c.) relative to the \NNO~ film (20 u.c.), this rules out significant Sr doping, since doping even 25\% of the nickelate film into the superconducting dome ($x\sim 13$\%) would require more than 20\% Sr loss from the capping layer.

\begin{figure}
  \resizebox{8.6 cm}{!}{\includegraphics{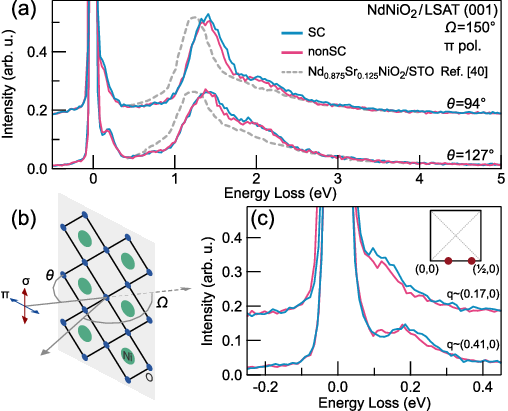}}
  \caption{\label{fig:rixs} Resonant inelastic scattering measurements of \NNO~collected in $\pi$~ polarization at 40 K and at an incident energy 100 meV below the Ni $L_3$ peak ($\sim$852.3 eV). (a) Comparison of energy loss spectra for superconducting and non-superconducting samples measured at a fixed scattering angle of $\Omega=150^{\circ}$ and incidence angles of $\theta=94^{\circ},127^{\circ}$. Data for Nd$_{0.875}$Sr$_{0.125}$NiO$_2$ from Ref. \cite{Rossi2021a} are provided for reference (dashed). (b) Experimental geometry of x-ray absorption and inelastic scattering measurements. (c) Zoom in on the region around the elastic peak showing dispersive spin excitations.}
\end{figure}

Comparing the concentration of holes in the NiO$_2$ plane across samples should be a direct way of distinguishing whether superconductivity is intrinsic to undoped \NNO, or if it arises from inadvertent additional hole doping from residual apical oxygens, or other sources. The Ni $L_3$-edge has previously been shown to be highly sensitive to hole doping \cite{Rossi2021a}, with the undoped parent compound displaying a single peak corresponding to a $2p^63d^9 \rightarrow 2p^53d^{10}$ transition, while hole doping causes the emergence of a clear shoulder $\sim 1$~ eV higher in energy which increases with hole doping, attributed to singlet holes doped into the Ni $d_{x^2-y^2}$ orbital.

In Fig. \hyperref[fig:xas]{\ref*{fig:xas}(a)} we compare representative Ni $L$-edge XAS spectra from superconducting and non-superconducting \NNO. The spectra are nearly identical, both exhibiting a sharp, single Ni $L_3$~ absorption peak at 852.4 eV and a strong dichroic response, consistent with previous measurements of undoped \NNO~ \cite{Rossi2021a}. In Figs. \hyperref[fig:xas]{\ref*{fig:xas}(b)} and \hyperref[fig:xas]{\ref*{fig:xas}(c)}, we show a comparison of the Ni $L_3$-edge from a large number of samples, with the superconducting samples exhibiting consistent, single peak lineshapes, whereas the non-superconducting samples, Fig. \hyperref[fig:xas]{\ref*{fig:xas}(b)}, display far more variability, with additional shoulders and peaks at higher energy. To compare the lineshapes quantitatively, we fit each of the spectra with a canonical undoped \NNO~lineshape (extracted from Ref. \cite{Rossi2021a} and shown in the inset of Fig. \hyperref[fig:xas]{\ref*{fig:xas}(d)} in green, and extract the excess spectral weight above the undoped \NNO~lineshape ($A_2$, in shaded yellow). We define the excess spectral weight, $y = A_2 / (A_1 + A_2)$, as $A_2$ normalized by the total XAS intensity integrated between 852.4 and 857 eV ($A_1 + A_2$). By this definition, undoped \NNO~ would have $y = 0$, while hole doping will increase $y$ through the increasing shoulder at $\approx$ 853.2 eV. We note that partially reduced insulating phases such as Nd$_3$Ni$_3$O$_8$ also exhibit peaks in the same energy range \cite{Parzyck2024a,Krieger2024}, so this definition of $y$ does not distinguish between hole doping versus the presence of other impurity phases.

In Fig. \hyperref[fig:xas]{\ref*{fig:xas}(d)}, we plot \Tcon~ versus $y$. It is notable that the majority of superconducting samples lie close to $y = 0$, resembling undoped \NNO~and suggesting that they have minimal hole doping or impurity phases, whereas the non-superconducting samples exhibit a much wider degree of variability. To address whether the excess spectral weight $y$ corresponds to doped holes or impurity phases, we note that prior studies show that \rhores~ decreases markedly with increasing hole doping \cite{Zeng2020,Li2020a,Lee2023}, so if $y$ corresponds to doped mobile holes, then \rhores~ should decrease monotonically with increasing $y$, as it does for Nd$_{1-x}$Sr$_x$NiO$_2$ (shown in dashed grey). In contrast, we find that the samples with the lowest \rhores~also have the lowest values of $y$, indicating that increasing $y$ does not correspond to the doping of mobile holes into the NiO$_2$ plane, but more likely the presence of inclusions of insulating, partially reduced phases such as Nd$_3$Ni$_3$O$_8$. This conclusion is supported by a careful analysis of the STEM images which show that while the majority of the sample is in the infinite layer phase, there still exist small clusters where excess apical oxygens remain and form oxygen ordered structures, as evidenced by $\nicefrac{1}{3}$ order peaks in the Fourier transform of the ADF images, shown in Fig. \hyperref[fig:capping]{\ref*{fig:capping}(d)} and the Supplemental Materials \cite{SIcite}. Prior work has identified these intermediate, oxygen ordered structures to be electrically insulating \cite{Parzyck2024a} and should not be the origin of the superconductivity. Taken as a whole, the XAS and STEM data clearly indicate the presence of excess oxygen atoms in the samples, but suggest they form insulating, ordered structures rather than doping mobile holes. 

In addition, prior RIXS measurements on Nd$_{1-x}$Sr$_x$NiO$_2$ have shown that the \textit{d-d} excitations and magnons soften considerably with hole doping \cite{Rossi2021a,Lu2021,Rossi2024}. In Figs. \hyperref[fig:rixs]{\ref*{fig:rixs}(a)} and \hyperref[fig:rixs]{\ref*{fig:rixs}(c)} we compare RIXS spectra from superconducting and non-superconducting \NNO/LSAT samples. The measured energies of the \textit{d-d} excitations (1.36 eV, $d_{xy}$ and 1.82 eV, $d_{xz/yz}$), as well as the magnon energy ($\sim 160$~meV) and its dispersion from $H = 0.41$ to $0.17$ not only match well with prior reports on undoped \NNO, but are also clearly distinct from doped Nd$_{0.875}$Sr$_{0.125}$NiO$_2$ (shown in dashed grey, from Ref. \cite{Rossi2021a}). Moreover, the spectra from the two samples are identical to within experimental uncertainty, indicating no observable difference in hole doping. In summary, a comparison of XAS and RIXS measurements on superconducting and non-superconducting \NNO~ show no evidence for hole doping, with superconducting samples exhibiting absorption and inelastic excitations consistent with undoped \NNO.

\begin{figure}
  \resizebox{8.6 cm}{!}{\includegraphics{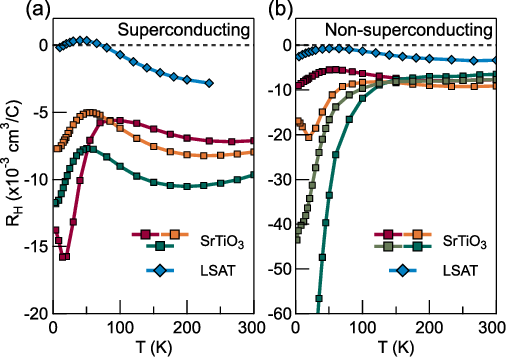}}
  \caption{\label{fig:Hall} Temperature dependence of the Hall coefficient, $R_H$, for selected 20 u.c. thick NdNiO$_2$ films synthesized on (001) \STO (squares) and (001) LSAT (diamonds) substrates.  (a) Hall coefficient measurements for films exhibiting partial superconducting transitions. (b) Same, for films which do not exhibit any signs of superconductivity.}
\end{figure}

In Fig. \ref{fig:Hall}, we present measurements of the Hall coefficient, $R_H$, since Nd$_{1-x}$Sr$_x$NiO$_2$ and the other \ILN~exhibit a marked dependence of $R_H$ on hole doping \cite{Li2020a,Zeng2020,Osada2020a,Zeng2021,Osada2021}. In underdoped, non-superconducting samples, $R_H$ remains negative at all temperatures, but upon traversing the superconducting dome, $R_H$~ changes sign from negative to positive around 100 K. Temperature dependent Hall measurements are shown in Figs. \hyperref[fig:Hall]{\ref*{fig:Hall}(a)} and \hyperref[fig:Hall]{\ref*{fig:Hall}(b)} which indicate that for nearly all samples, $R_H$ remains strictly negative at all temperatures, with the exception of superconducting samples grown on LSAT. The magnitude of $R_H$~ as $T \rightarrow 0$~ for samples grown on LSAT are also smaller ($-4.7 < R_H < 0.4$~ $\times 10^{-3}$~cm$^3$/C) than those grown on \STO, consistent with prior measurements \cite{Lee2023}.  While the values of $R_H (300 K)$ are generally consistent, the low temperature values of $R_H$ vary widely, similar to reports in the literature (e.g. from -40 to -6.5 $\times 10^{-3}$~cm$^3$/C \cite{Zeng2020,Li2020a}).  These large sample-to-sample variations makes drawing clear conclusions from the Hall coefficient difficult, but we note that $|R_H|$ for the superconducting samples are generally smaller than for non-superconducting ones. Furthermore, $R_H$ appears to be uncorrelated with $\rho_{\textrm{res}}$, suggesting that the low resistivities observed in these films are due to reduced disorder, rather than hole doping consistent with the XAS measurements in Fig. \hyperref[fig:xas]{\ref*{fig:xas}(e)}.

\begin{figure}
  \resizebox{8.6 cm}{!}{\includegraphics{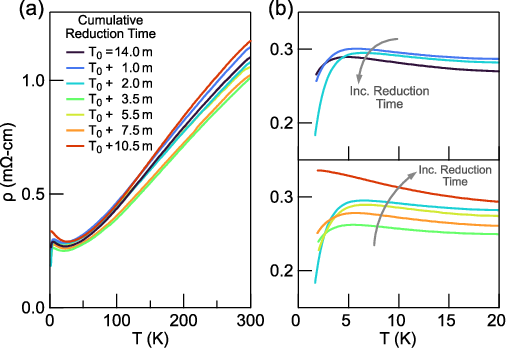}}
  \caption{\label{fig:annealing} Progressive reduction of an \NNO/\STO~ film. (a) Temperature dependent resistivity of a reduced film which is subsequently exposed to additional atomic hydrogen in short intervals, reduction conditions are detailed in the Supplemental Materials \cite{SIcite}. (b) Zoom-in on the superconducting transition region of the same data.}
\end{figure}

If superconductivity arises from hole doping from residual apical oxygens, then further reduction of a superconducting sample should suppress superconductivity. In Fig. \ref{fig:annealing} we show the resistivity of a superconducting \NNO/SrTiO$_3$ sample, following a sequence of short additional reduction steps. The originally reduced sample (reduction time $T_0 = 14$ min) shows a small downturn in the resistivity around 5 K. Subsequent reductions up to $T_0 + 3.5$ min \emph{increase} \Tcon, but as the sample is further reduced,  superconductivity is eventually suppressed at $T > T_0 + 7.5$ min. During this sequence, no change in the x-ray diffraction pattern is observed. At face value, the suppression of $T_c^{\textrm{on}}$ with reduction might appear to support the possibility of hole doping from excess oxygens, but \rhores~ also increases slightly with reduction time, suggesting that reduction beyond the optimal level results in increased pair-breaking disorder. Without a method to disentangle the effects of disorder and doping from reduction, it is difficult to draw clear conclusions from this measurement. Nevertheless, the strong sensitivity of samples to the reduction process likely explain why superconductivity is not observed in all samples with $\rho_{\mathrm{res}} < \rho_Q$, since the optimal reduction conditions for different films depend on subtle extrinsic factors, \textit{e.g.} the density of defects in the NdNiO$_3$~ precursor. While this experiment shows that superconductivity can be suppressed by further reduction, the superconducting transitions are robust over long times ($\sim 1$~ year) when films are stored under ambient conditions. 

\section{Discussion \& Conclusions}
In the hole-doped cuprates, superconductivity emerges when a number of conditions are achieved: 1) holes are doped into the CuO$_2$ plane, 2) the doped holes suppress long-range antiferromagnetic order, leaving short-ranged antiferromagnetic fluctuations, and 3) the residual resistivity drops below approximately the quantum of resistance per CuO$_2$ plane \cite{Ando2001,Ando2004}. It is noteworthy that the undoped parent nickelates  fulfill all of the above conditions without the need for additional hole doping, lending credence to the possibility that superconductivity could be intrinsic to the parent compounds. Given the mounting evidence that the nickelates have a nodal order parameter, pair-breaking from impurity scattering should have a substantial impact on $T_c$. This is consistent with the observation that nearly all superconducting \NNO~samples have $\rho_{\mathrm{res}} < \rho_Q$, and is also supported by our XAS and RIXS measurements which show that the electronic structure and excitation spectra of superconducting samples closely match those from undoped \NNO. Furthermore, nearly all samples maintain a strictly negative Hall coefficient, consistent with the behavior of undoped \ILN. Together with observations of superconductivity in undoped \LNO~\cite{Osada2021} and recently \PNO~\cite{Preziosi2024}, these pieces of evidence suggest that superconductivity could be intrinsic to the clean limit of the parent \ILN. This raises the possibility that the superconducting dome could extend all the way to $x = 0$, reminiscent of FeSe, where the undoped stoichiometric compound exhibits superconductivity without long-range antiferromagnetic order, and $T_c$ is gradually increased with electron doping. 

XAS and STEM measurements provide strong evidence for the presence of residual apical oxygens, but also indicate that they tend to self-organize into ordered vacancy structures such as Nd$_3$Ni$_3$O$_8$ which do not donate mobile holes \cite{Parzyck2024a}. Indeed, none of the measurements reported here are obviously consistent with the presence of mobile, doped holes in \NNO. Instead, both \Tcon~and \rhores~appear to be anti-correlated with the amount of excess oxygen, suggesting that the presence of excess oxygens is actually deleterious for superconductivity, as clusters of impurity phases should act as scattering centers. In the related $T'$ and infinite-layer cuprates, excess apical oxygens do not donate mobile holes to the CuO$_2$ plane and rather act as strong scattering centers for low-energy quasiparticles \cite{Richard2007,Song2012}, and their removal by post-growth reduction is essential to achieve superconductivity.

Nevertheless, it remains difficult to unequivocally rule out the possibility of hole doping from randomly dispersed apical oxygen atoms. Indeed, the possibility that excess apical oxygens could donate mobile holes to the NiO$_2$ plane in the infinite layer nickelates would be a novel finding. Controlling doping through oxygen non-stoichiometry has played an crucial role in the cuprate superconductors, enabling traversal of the phase diagrams of numerous materials such as Bi$_2$Sr$_2$CaCu$_2$O$_{8+\delta}$, Tl$_2$Ba$_2$CuO$_{6+\delta}$, and HgBa$_2$CuO$_{4+\delta}$, as well as facilitating the observation of quantum oscillations in ortho-ordered phases of YBa$_2$Cu$_3$O$_{7-\delta}$ \cite{DoironLeyraud2007}. The possibility to employ a similar tuning knob could likewise enable new approaches for realizing or enhancing superconductivity in the \ILN. 

In conclusion, we report the observation of superconductivity with $T_c$ onsets up to 11 K in undoped \NNO~thin films with high crystallinity and low residual resistivities. Superconductivity appears to emerge when \rhores~falls below the quantum of resistance per NiO$_2$ sheet. We propose that superconductivity could be intrinsic to the clean limit of undoped \NNO, as it already possesses key ingredients necessary for cuprate superconductivity, including holes in the NiO$_2$ plane (self-doped from rare-earth $5d$ orbitals) and short-ranged magnetic fluctuations. On the other hand, we cannot entirely rule out the possibility that residual apical oxygens could provide doped mobile holes to the NiO$_2$ plane, which  would present a new pathway to superconductivity in the \ILN.

\section{Methods}
\subsection{Sample Synthesis and Characterization}
Thin films of NdNiO$_3$ were grown on (001)-oriented SrTiO$_3$ and LSAT substrates using reactive-oxide molecular-beam epitaxy in a Veeco GEN10 system and reduced using an atomic hydrogen beam produced by flowing hydrogen gas through a thermal cracker.  Details of the film growth and atomic hydrogen reduction are provided in Ref. \cite{Parzyck2024b}. Structural quality and phase purity of the thin film samples were determined by Cu K$\alpha_1$ x-ray diffraction measurements performed on a PANalytical Empyrean X-ray diffractometer. Electrical transport measurements were performed using both a custom built LHe-cooled four-point probe measurement station and a Quantum Design physical property measurement system and geometric factors accounted for using the methods of Ref. \cite{Miccoli2015}.  Ohmic contacts were prepared either by ultrasonic aluminum wire bonding or by placing a small indium dot on the sample surface underneath a gold press contact.  Hall effect measurements were performed either on the entire $10\times 10$~mm sample in a square, four-point Van der Pauw configuration or using a 6-wire Hall bar geometry on a diced $5\times 10$~mm piece.

\subsection{Pulsed IV Measurements}
Pulsed IV measurements were performed using a Keithley 6221 current source and 2182A voltmeter using a 500 $\mu$s pulse width and a variable duty cycle to maintain a fixed, low average power such that heating effects were not observable.  Contacts were placed in a linear four-point geometry near the sample center with equally spaced contacts 500 $\mu$m apart.  The current density, $J(x,y) = \left| \mathbf{J} \right|$, can be estimated for leads placed at $x=\pm \delta$ for an infinite two-dimensional slab as described in Refs. \cite{Smits1958,Miccoli2015}.  The relevant quantity is then the average current density between voltage leads (spaced by $2\delta/3$):
$\langle J \rangle = \frac{3}{2\delta} \int_{-\delta/3}^{\delta/3} \left|J(x,0)\right| dx$.
This model assumes that $\rho$ 1) is uniform and isotropic on a macro-scale and 2) does not depend implicitly on $|\mathbf{J}|$.  Both of these conditions are satisfied in the high and low current limits (when superconductivity is either unaffected or completely suppressed); however in the intermediate regime the true value of $\langle J \rangle$ is likely smaller then predicted by this simple model.  As superconductivity is quenched in the higher density `on-axis' region the current will spread further off-axis to compensate, reducing $|\mathbf{J}|$ in the region near the leads. So, the measurements in Fig. \hyperref[fig:transport]{\ref*{fig:transport}(f)} provide reasonable bounds for $J_c$ between $5\times10^3$ and $5\times10^4$ A/cm$^2$, but not a precise value.

\subsection{Scanning Transmission Electron Microscopy}
STEM characterization was performed on a cross-sectional lamella prepared with the standard focused ion beam (FIB) lift-out procedure using a Thermo Fisher Helios G4 UX FIB. ADF imaging was performed on Cs-corrected Thermo Fisher Scientific (TSF) Spectra 300 X-CFEG or aberration-corrected TFS Titan Themis 300 operating at 300 kV and 30 mrad probe convergence semi-angle. For high-precision structural measurements, a series of 40 rapid-frame images were acquired and subsequently realigned and averaged by a method of rigid registration optimized to prevent lattice hops \cite{Savitzky2018} resulting in high signal-to-noise ratio image of atomic lattice.

\subsection{X-ray Absorption Spectroscopy}
Nickel $L$-edge XAS Measurements were performed at the SGM and REIXS beamlines of the Canadian Light Source on a tilt stage at room temperature under high vacuum; the nominal photon flux and energy resolution were $I_0 = 2\times 10^{11}$~ photons/sec and $ E /\Delta E \sim 8600$, respectively with a spot size of $ < 30\times 50$~ $\mu$m.  The incoming x-ray polarization was selected to be in either the $\sigma$ ($\varepsilon \parallel$ to the sample $a$-axis) or $\pi$~ ($\varepsilon$~ in the sample $bc$-plane) configuration and the fluorescence signal was measured using an array of four silicon drift detectors (SDDs) with an individual angular acceptance of $\sim 8.02^{\circ}$ and energy resolution of $> 100$~ eV allowing for removal of the substantial oxygen fluorescence background produced by the substrates. PFY measurements from different detectors were corrected for self-absorption effects using the methods of \cite{Eisebitt1993,Achkar2011} and spectra are normalized to match in the pre- (845 eV) and post-edge (885 eV) regions. For O $K$-edge spectra, and Ni $L$-edge spectra of samples on LSAT, substantial contributions from the substrate prevent accurate measurement of the x-ray fluorescence so total electron yield (TEY) measurements were performed instead. The incident photon energy was calibrated to match prior measurements \cite{Parzyck2024a} performed on common samples.

\subsection{Resonant Inelastic X-ray Scattering}
Resonant inelastic x-ray scattering (RIXS) measurements at the Ni $L_3$-edge were performed at the 2ID-SIX beamline at NSLS-II, Brookhaven National Laboratory (USA) at 40 K in a fixed $\pi$~ polarization, to maximize the intensity of spin excitations, using the experimental geometry sketched in Fig. \hyperref[fig:rixs]{\ref*{fig:rixs}(b)}. The scattering angle $\Omega$~ is defined as the angle between the incident and scattered x-rays and was fixed at 150$^{\circ}$~ to maximize the momentum transfer. The grazing angle, $\theta$, is defined as the angle between the incident x-ray and the sample surface.  The selection of the resonant energy for the RIXS measurement is 0.1 eV below the main Ni absorption peak of the Ni $L_3$-edge, where the intensities of the spin excitations are the strongest. The energy resolution, determined by estimating the full width at half maximum of the elastic peak of a multilayer reference sample, is found to be $\sim 40$~ meV at the Ni $L_3$-edge.

\section*{Acknowledgements}
The authors would like to thank Y. Birkholzer, J. Fontcuberta, H.Y. Hwang, and D. Preziosi for enlightening discussions and sharing unpublished results, as well as S. Palmer and S. Button for their assistance in substrate preparation. This work was primarily supported by the Air Force Office of Scientific Research through Grants No. FA9550-21-1-0168 and FA9550-23-1-0161, and also by the US Department of Energy (DOE), Office of Basic Energy Sciences, under contract no. DE-SC0019414, and the National Science Foundation through Grants No. DMR-2104427 and the Platform for the Accelerated Realization, Analysis, and Discovery of Interface Materials (PARADIM) under Cooperative Agreement No. DMR-2039380. Additional support for materials synthesis was provided by the Gordon and Betty Moore Foundation’s EPiQS Initiative through Grant Nos. GBMF3850 and GBMF9073. G.H. acknowledges support from projects Severo Ochoa MaTrans42 (no.CEX2023-001263-S) and PID2023-152225NB-I00 of the Spanish Agencia Estatal de Investigaci\'{o}n and by the Generalitat de Catalunya (2021 SGR 00445). A.B.G. was supported in part by Lilly Endowment, Inc., through its support for the Indiana University Pervasive Technology Institute. This research uses Beamline 2-ID of the National Synchrotron Light Source II, a DOE Office of Science User Facility operated for the DOE Office of Science by Brookhaven National Laboratory under Contract No. DESC0012704. Part of the research described in this paper was performed using the REIXS and SGM beamlines at the Canadian Light Source, a national research facility of the University of Saskatchewan, which is supported by the Canada Foundation for Innovation (CFI), the Natural Sciences and Engineering Research Council (NSERC), the National Research Council (NRC), the Canadian Institutes of Health Research (CIHR), the Government of Saskatchewan, and the University of Saskatchewan. This work made use of the Cornell Center for Materials Research shared instrument facilities. The microscopy work was supported by NSF PARADIM (DMR-2039380), with additional support from Cornell University, the Weill Institute and the Kavli Institute at Cornell. L.B. and L.F.K. acknowledge support from Packard foundation. The DFT calculations shown in this work have been performed on the Big Red 200 and Quartz supercomputer clusters at Indiana University. Substrate preparation was performed at the Cornell NanoScale Facility, a member of the National Nanotechnology Coordinated Infrastructure (NNCI), which is supported by the National Science Foundation (Grant NNCI-2025233).

\section*{Author Contributions}
C.T.P and K.M.S conceived the research and designed the experiment. Sample synthesis and characterization were performed by C.T.P. and Y.W. under the supervision of D.G.S. and K.M.S. X-ray absorption measurements were performed by C.T.P, M.K., Z.A., T.M.P, and R.S., and analyzed by C.T.P. and D.G.H.  Resonant inelastic scattering measurements were performed by S.F., J.P., and V.B.  Electron microscopy measurements were performed by L.B. under the supervision of L.F.K. and D.A.M. The results were analyzed and interpreted by C.T.P., L.B., Y.W., M.K., D.G.H., G.H., D.A.M., D.G.S., and K.M.S. Density functional theory calculations were performed by A.B.G.  C.T.P. and K.M.S. wrote the manuscript with input from all authors.

\section*{Data Availability}
The data that support the findings of this study are available within the paper and Supplementary Materials \cite{SIcite}.

\bibliography{Nickelates}

\begin{thebibliography}{59}%
\makeatletter
\providecommand \@ifxundefined [1]{%
 \@ifx{#1\undefined}
}%
\providecommand \@ifnum [1]{%
 \ifnum #1\expandafter \@firstoftwo
 \else \expandafter \@secondoftwo
 \fi
}%
\providecommand \@ifx [1]{%
 \ifx #1\expandafter \@firstoftwo
 \else \expandafter \@secondoftwo
 \fi
}%
\providecommand \natexlab [1]{#1}%
\providecommand \enquote  [1]{``#1''}%
\providecommand \bibnamefont  [1]{#1}%
\providecommand \bibfnamefont [1]{#1}%
\providecommand \citenamefont [1]{#1}%
\providecommand \href@noop [0]{\@secondoftwo}%
\providecommand \href [0]{\begingroup \@sanitize@url \@href}%
\providecommand \@href[1]{\@@startlink{#1}\@@href}%
\providecommand \@@href[1]{\endgroup#1\@@endlink}%
\providecommand \@sanitize@url [0]{\catcode `\\12\catcode `\$12\catcode
  `\&12\catcode `\#12\catcode `\^12\catcode `\_12\catcode `\%12\relax}%
\providecommand \@@startlink[1]{}%
\providecommand \@@endlink[0]{}%
\providecommand \url  [0]{\begingroup\@sanitize@url \@url }%
\providecommand \@url [1]{\endgroup\@href {#1}{\urlprefix }}%
\providecommand \urlprefix  [0]{URL }%
\providecommand \Eprint [0]{\href }%
\providecommand \doibase [0]{https://doi.org/}%
\providecommand \selectlanguage [0]{\@gobble}%
\providecommand \bibinfo  [0]{\@secondoftwo}%
\providecommand \bibfield  [0]{\@secondoftwo}%
\providecommand \translation [1]{[#1]}%
\providecommand \BibitemOpen [0]{}%
\providecommand \bibitemStop [0]{}%
\providecommand \bibitemNoStop [0]{.\EOS\space}%
\providecommand \EOS [0]{\spacefactor3000\relax}%
\providecommand \BibitemShut  [1]{\csname bibitem#1\endcsname}%
\let\auto@bib@innerbib\@empty
\bibitem [{\citenamefont {Li}\ \emph {et~al.}(2019)\citenamefont {Li},
  \citenamefont {Lee}, \citenamefont {Wang}, \citenamefont {Osada},
  \citenamefont {Crossley}, \citenamefont {Lee}, \citenamefont {Cui},
  \citenamefont {Hikita},\ and\ \citenamefont {Hwang}}]{Li2019b}%
  \BibitemOpen
  \bibfield  {author} {\bibinfo {author} {\bibfnamefont {D.}~\bibnamefont
  {Li}}, \bibinfo {author} {\bibfnamefont {K.}~\bibnamefont {Lee}}, \bibinfo
  {author} {\bibfnamefont {B.~Y.}\ \bibnamefont {Wang}}, \bibinfo {author}
  {\bibfnamefont {M.}~\bibnamefont {Osada}}, \bibinfo {author} {\bibfnamefont
  {S.}~\bibnamefont {Crossley}}, \bibinfo {author} {\bibfnamefont {H.~R.}\
  \bibnamefont {Lee}}, \bibinfo {author} {\bibfnamefont {Y.}~\bibnamefont
  {Cui}}, \bibinfo {author} {\bibfnamefont {Y.}~\bibnamefont {Hikita}},\ and\
  \bibinfo {author} {\bibfnamefont {H.~Y.}\ \bibnamefont {Hwang}},\ }\bibfield
  {title} {\bibinfo {title} {{Superconductivity in an infinite-layer
  nickelate}},\ }\href {https://doi.org/10.1038/s41586-019-1496-5} {\bibfield
  {journal} {\bibinfo  {journal} {Nature}\ }\textbf {\bibinfo {volume} {572}},\
  \bibinfo {pages} {624} (\bibinfo {year} {2019})}\BibitemShut {NoStop}%
\bibitem [{\citenamefont {Crespin}\ \emph {et~al.}(1983)\citenamefont
  {Crespin}, \citenamefont {Levitz},\ and\ \citenamefont
  {Gatineau}}]{Crespin1983a}%
  \BibitemOpen
  \bibfield  {author} {\bibinfo {author} {\bibfnamefont {M.}~\bibnamefont
  {Crespin}}, \bibinfo {author} {\bibfnamefont {B.~Y.~P.}\ \bibnamefont
  {Levitz}},\ and\ \bibinfo {author} {\bibfnamefont {L.}~\bibnamefont
  {Gatineau}},\ }\bibfield  {title} {\bibinfo {title} {{Reduced Forms of
  LaNiO$_3$ Perovskite}},\ }\href {https://doi.org/10.1039/F29837901181}
  {\bibfield  {journal} {\bibinfo  {journal} {Journal of the Chemical Society,
  Faraday Transactions II}\ }\textbf {\bibinfo {volume} {79}},\ \bibinfo
  {pages} {1181} (\bibinfo {year} {1983})}\BibitemShut {NoStop}%
\bibitem [{\citenamefont {Hayward}\ \emph {et~al.}(1999)\citenamefont
  {Hayward}, \citenamefont {Green}, \citenamefont {Rosseinsky},\ and\
  \citenamefont {Sloan}}]{Hayward1999}%
  \BibitemOpen
  \bibfield  {author} {\bibinfo {author} {\bibfnamefont {M.~A.}\ \bibnamefont
  {Hayward}}, \bibinfo {author} {\bibfnamefont {M.~A.}\ \bibnamefont {Green}},
  \bibinfo {author} {\bibfnamefont {M.~J.}\ \bibnamefont {Rosseinsky}},\ and\
  \bibinfo {author} {\bibfnamefont {J.}~\bibnamefont {Sloan}},\ }\bibfield
  {title} {\bibinfo {title} {{Sodium Hydride as a Powerful Reducing Agent for
  Topotactic Oxide Deintercalation: Synthesis and Characterization of the
  Nickel(I) Oxide LaNiO$_2$}},\ }\href {https://doi.org/10.1021/ja991573i}
  {\bibfield  {journal} {\bibinfo  {journal} {Journal of the American Chemical
  Society}\ }\textbf {\bibinfo {volume} {121}},\ \bibinfo {pages} {8843}
  (\bibinfo {year} {1999})}\BibitemShut {NoStop}%
\bibitem [{\citenamefont {Hayward}\ and\ \citenamefont
  {Rosseinsky}(2003)}]{Hayward2003a}%
  \BibitemOpen
  \bibfield  {author} {\bibinfo {author} {\bibfnamefont {M.}~\bibnamefont
  {Hayward}}\ and\ \bibinfo {author} {\bibfnamefont {M.}~\bibnamefont
  {Rosseinsky}},\ }\bibfield  {title} {\bibinfo {title} {{Synthesis of the
  infinite layer Ni(I) phase NdNiO$_{2+x}$ by low temperature reduction of
  NdNiO$_3$ with sodium hydride}},\ }\href
  {https://doi.org/10.1016/S1293-2558(03)00111-0} {\bibfield  {journal}
  {\bibinfo  {journal} {Solid State Sciences}\ }\textbf {\bibinfo {volume}
  {5}},\ \bibinfo {pages} {839} (\bibinfo {year} {2003})}\BibitemShut {NoStop}%
\bibitem [{\citenamefont {Ikeda}\ \emph {et~al.}(2016)\citenamefont {Ikeda},
  \citenamefont {Krockenberger}, \citenamefont {Irie}, \citenamefont {Naito},\
  and\ \citenamefont {Yamamoto}}]{Ikeda2016}%
  \BibitemOpen
  \bibfield  {author} {\bibinfo {author} {\bibfnamefont {A.}~\bibnamefont
  {Ikeda}}, \bibinfo {author} {\bibfnamefont {Y.}~\bibnamefont
  {Krockenberger}}, \bibinfo {author} {\bibfnamefont {H.}~\bibnamefont {Irie}},
  \bibinfo {author} {\bibfnamefont {M.}~\bibnamefont {Naito}},\ and\ \bibinfo
  {author} {\bibfnamefont {H.}~\bibnamefont {Yamamoto}},\ }\bibfield  {title}
  {\bibinfo {title} {{Direct observation of infinite NiO$_2$ planes in
  LaNiO$_2$ films}},\ }\href {https://doi.org/10.7567/APEX.9.061101} {\bibfield
   {journal} {\bibinfo  {journal} {Applied Physics Express}\ }\textbf {\bibinfo
  {volume} {9}},\ \bibinfo {pages} {061101} (\bibinfo {year}
  {2016})}\BibitemShut {NoStop}%
\bibitem [{\citenamefont {Lee}\ \emph {et~al.}(2023)\citenamefont {Lee},
  \citenamefont {Wang}, \citenamefont {Osada}, \citenamefont {Goodge},
  \citenamefont {Wang}, \citenamefont {Lee}, \citenamefont {Harvey},
  \citenamefont {Kim}, \citenamefont {Yu}, \citenamefont {Murthy},
  \citenamefont {Raghu}, \citenamefont {Kourkoutis},\ and\ \citenamefont
  {Hwang}}]{Lee2023}%
  \BibitemOpen
  \bibfield  {author} {\bibinfo {author} {\bibfnamefont {K.}~\bibnamefont
  {Lee}}, \bibinfo {author} {\bibfnamefont {B.~Y.}\ \bibnamefont {Wang}},
  \bibinfo {author} {\bibfnamefont {M.}~\bibnamefont {Osada}}, \bibinfo
  {author} {\bibfnamefont {B.~H.}\ \bibnamefont {Goodge}}, \bibinfo {author}
  {\bibfnamefont {T.~C.}\ \bibnamefont {Wang}}, \bibinfo {author}
  {\bibfnamefont {Y.}~\bibnamefont {Lee}}, \bibinfo {author} {\bibfnamefont
  {S.}~\bibnamefont {Harvey}}, \bibinfo {author} {\bibfnamefont {W.~J.}\
  \bibnamefont {Kim}}, \bibinfo {author} {\bibfnamefont {Y.}~\bibnamefont
  {Yu}}, \bibinfo {author} {\bibfnamefont {C.}~\bibnamefont {Murthy}}, \bibinfo
  {author} {\bibfnamefont {S.}~\bibnamefont {Raghu}}, \bibinfo {author}
  {\bibfnamefont {L.~F.}\ \bibnamefont {Kourkoutis}},\ and\ \bibinfo {author}
  {\bibfnamefont {H.~Y.}\ \bibnamefont {Hwang}},\ }\bibfield  {title} {\bibinfo
  {title} {{Linear-in-temperature resistivity for optimally superconducting
  (Nd,Sr)NiO$_2$}},\ }\href {https://doi.org/10.1038/s41586-023-06129-x}
  {\bibfield  {journal} {\bibinfo  {journal} {Nature}\ }\textbf {\bibinfo
  {volume} {619}},\ \bibinfo {pages} {288} (\bibinfo {year}
  {2023})}\BibitemShut {NoStop}%
\bibitem [{\citenamefont {Jiang}\ \emph {et~al.}(2019)\citenamefont {Jiang},
  \citenamefont {Si}, \citenamefont {Liao},\ and\ \citenamefont
  {Zhong}}]{Jiang2019a}%
  \BibitemOpen
  \bibfield  {author} {\bibinfo {author} {\bibfnamefont {P.}~\bibnamefont
  {Jiang}}, \bibinfo {author} {\bibfnamefont {L.}~\bibnamefont {Si}}, \bibinfo
  {author} {\bibfnamefont {Z.}~\bibnamefont {Liao}},\ and\ \bibinfo {author}
  {\bibfnamefont {Z.}~\bibnamefont {Zhong}},\ }\bibfield  {title} {\bibinfo
  {title} {{Electronic structure of rare-earth infinite-layer RNiO$_2$ (R=La,
  Nd)}},\ }\href {https://doi.org/10.1103/PhysRevB.100.201106} {\bibfield
  {journal} {\bibinfo  {journal} {Physical Review B}\ }\textbf {\bibinfo
  {volume} {100}},\ \bibinfo {pages} {201106} (\bibinfo {year}
  {2019})}\BibitemShut {NoStop}%
\bibitem [{\citenamefont {Botana}\ and\ \citenamefont
  {Norman}(2020)}]{Botana2020}%
  \BibitemOpen
  \bibfield  {author} {\bibinfo {author} {\bibfnamefont {A.~S.}\ \bibnamefont
  {Botana}}\ and\ \bibinfo {author} {\bibfnamefont {M.~R.}\ \bibnamefont
  {Norman}},\ }\bibfield  {title} {\bibinfo {title} {{Similarities and
  Differences between LaNiO$_2$~ and CaCuO$_2$~ and Implications for
  Superconductivity}},\ }\href {https://doi.org/10.1103/PhysRevX.10.011024}
  {\bibfield  {journal} {\bibinfo  {journal} {Physical Review X}\ }\textbf
  {\bibinfo {volume} {10}},\ \bibinfo {pages} {011024} (\bibinfo {year}
  {2020})}\BibitemShut {NoStop}%
\bibitem [{\citenamefont {Kitatani}\ \emph {et~al.}(2020)\citenamefont
  {Kitatani}, \citenamefont {Si}, \citenamefont {Janson}, \citenamefont
  {Arita}, \citenamefont {Zhong},\ and\ \citenamefont {Held}}]{Kitatani2020}%
  \BibitemOpen
  \bibfield  {author} {\bibinfo {author} {\bibfnamefont {M.}~\bibnamefont
  {Kitatani}}, \bibinfo {author} {\bibfnamefont {L.}~\bibnamefont {Si}},
  \bibinfo {author} {\bibfnamefont {O.}~\bibnamefont {Janson}}, \bibinfo
  {author} {\bibfnamefont {R.}~\bibnamefont {Arita}}, \bibinfo {author}
  {\bibfnamefont {Z.}~\bibnamefont {Zhong}},\ and\ \bibinfo {author}
  {\bibfnamefont {K.}~\bibnamefont {Held}},\ }\bibfield  {title} {\bibinfo
  {title} {{Nickelate superconductors—a renaissance of the one-band Hubbard
  model}},\ }\href {https://doi.org/10.1038/s41535-020-00260-y} {\bibfield
  {journal} {\bibinfo  {journal} {npj Quantum Materials}\ }\textbf {\bibinfo
  {volume} {5}},\ \bibinfo {pages} {59} (\bibinfo {year} {2020})}\BibitemShut
  {NoStop}%
\bibitem [{\citenamefont {Karp}\ \emph {et~al.}(2020)\citenamefont {Karp},
  \citenamefont {Botana}, \citenamefont {Norman}, \citenamefont {Park},
  \citenamefont {Zingl},\ and\ \citenamefont {Millis}}]{Karp2020}%
  \BibitemOpen
  \bibfield  {author} {\bibinfo {author} {\bibfnamefont {J.}~\bibnamefont
  {Karp}}, \bibinfo {author} {\bibfnamefont {A.~S.}\ \bibnamefont {Botana}},
  \bibinfo {author} {\bibfnamefont {M.~R.}\ \bibnamefont {Norman}}, \bibinfo
  {author} {\bibfnamefont {H.}~\bibnamefont {Park}}, \bibinfo {author}
  {\bibfnamefont {M.}~\bibnamefont {Zingl}},\ and\ \bibinfo {author}
  {\bibfnamefont {A.}~\bibnamefont {Millis}},\ }\bibfield  {title} {\bibinfo
  {title} {{Many-Body Electronic Structure of NdNiO$_2$ and CaCuO$_2$}},\
  }\href {https://doi.org/10.1103/PhysRevX.10.021061} {\bibfield  {journal}
  {\bibinfo  {journal} {Physical Review X}\ }\textbf {\bibinfo {volume} {10}},\
  \bibinfo {pages} {021061} (\bibinfo {year} {2020})}\BibitemShut {NoStop}%
\bibitem [{\citenamefont {Sakakibara}\ \emph {et~al.}(2020)\citenamefont
  {Sakakibara}, \citenamefont {Usui}, \citenamefont {Suzuki}, \citenamefont
  {Kotani}, \citenamefont {Aoki},\ and\ \citenamefont
  {Kuroki}}]{Sakakibara2020}%
  \BibitemOpen
  \bibfield  {author} {\bibinfo {author} {\bibfnamefont {H.}~\bibnamefont
  {Sakakibara}}, \bibinfo {author} {\bibfnamefont {H.}~\bibnamefont {Usui}},
  \bibinfo {author} {\bibfnamefont {K.}~\bibnamefont {Suzuki}}, \bibinfo
  {author} {\bibfnamefont {T.}~\bibnamefont {Kotani}}, \bibinfo {author}
  {\bibfnamefont {H.}~\bibnamefont {Aoki}},\ and\ \bibinfo {author}
  {\bibfnamefont {K.}~\bibnamefont {Kuroki}},\ }\bibfield  {title} {\bibinfo
  {title} {{Model Construction and a Possibility of Cuprate like Pairing in a
  New $d^9$ Nickelate Superconductor (Nd,Sr)NiO$_2$}},\ }\href
  {https://doi.org/10.1103/PhysRevLett.125.077003} {\bibfield  {journal}
  {\bibinfo  {journal} {Physical Review Letters}\ }\textbf {\bibinfo {volume}
  {125}},\ \bibinfo {pages} {077003} (\bibinfo {year} {2020})}\BibitemShut
  {NoStop}%
\bibitem [{\citenamefont {{Di Cataldo}}\ \emph {et~al.}(2024)\citenamefont {{Di
  Cataldo}}, \citenamefont {Worm}, \citenamefont {Tomczak}, \citenamefont
  {Si},\ and\ \citenamefont {Held}}]{DiCataldo2023}%
  \BibitemOpen
  \bibfield  {author} {\bibinfo {author} {\bibfnamefont {S.}~\bibnamefont {{Di
  Cataldo}}}, \bibinfo {author} {\bibfnamefont {P.}~\bibnamefont {Worm}},
  \bibinfo {author} {\bibfnamefont {J.~M.}\ \bibnamefont {Tomczak}}, \bibinfo
  {author} {\bibfnamefont {L.}~\bibnamefont {Si}},\ and\ \bibinfo {author}
  {\bibfnamefont {K.}~\bibnamefont {Held}},\ }\bibfield  {title} {\bibinfo
  {title} {{Unconventional superconductivity without doping in infinite-layer
  nickelates under pressure}},\ }\href
  {https://doi.org/10.1038/s41467-024-48169-5} {\bibfield  {journal} {\bibinfo
  {journal} {Nature Communications}\ }\textbf {\bibinfo {volume} {15}},\
  \bibinfo {pages} {3952} (\bibinfo {year} {2024})}\BibitemShut {NoStop}%
\bibitem [{\citenamefont {Lu}\ \emph {et~al.}(2021)\citenamefont {Lu},
  \citenamefont {Rossi}, \citenamefont {Nag}, \citenamefont {Osada},
  \citenamefont {Li}, \citenamefont {Lee}, \citenamefont {Wang}, \citenamefont
  {Garcia-Fernandez}, \citenamefont {Agrestini}, \citenamefont {Shen},
  \citenamefont {Been}, \citenamefont {Moritz}, \citenamefont {Devereaux},
  \citenamefont {Zaanen}, \citenamefont {Hwang}, \citenamefont {Zhou},\ and\
  \citenamefont {Lee}}]{Lu2021}%
  \BibitemOpen
  \bibfield  {author} {\bibinfo {author} {\bibfnamefont {H.}~\bibnamefont
  {Lu}}, \bibinfo {author} {\bibfnamefont {M.}~\bibnamefont {Rossi}}, \bibinfo
  {author} {\bibfnamefont {A.}~\bibnamefont {Nag}}, \bibinfo {author}
  {\bibfnamefont {M.}~\bibnamefont {Osada}}, \bibinfo {author} {\bibfnamefont
  {D.~F.}\ \bibnamefont {Li}}, \bibinfo {author} {\bibfnamefont
  {K.}~\bibnamefont {Lee}}, \bibinfo {author} {\bibfnamefont {B.~Y.}\
  \bibnamefont {Wang}}, \bibinfo {author} {\bibfnamefont {M.}~\bibnamefont
  {Garcia-Fernandez}}, \bibinfo {author} {\bibfnamefont {S.}~\bibnamefont
  {Agrestini}}, \bibinfo {author} {\bibfnamefont {Z.~X.}\ \bibnamefont {Shen}},
  \bibinfo {author} {\bibfnamefont {E.~M.}\ \bibnamefont {Been}}, \bibinfo
  {author} {\bibfnamefont {B.}~\bibnamefont {Moritz}}, \bibinfo {author}
  {\bibfnamefont {T.~P.}\ \bibnamefont {Devereaux}}, \bibinfo {author}
  {\bibfnamefont {J.}~\bibnamefont {Zaanen}}, \bibinfo {author} {\bibfnamefont
  {H.~Y.}\ \bibnamefont {Hwang}}, \bibinfo {author} {\bibfnamefont {K.-j.}\
  \bibnamefont {Zhou}},\ and\ \bibinfo {author} {\bibfnamefont {W.~S.}\
  \bibnamefont {Lee}},\ }\bibfield  {title} {\bibinfo {title} {{Magnetic
  excitations in infinite-layer nickelates}},\ }\href
  {https://doi.org/10.1126/science.abd7726} {\bibfield  {journal} {\bibinfo
  {journal} {Science}\ }\textbf {\bibinfo {volume} {373}},\ \bibinfo {pages}
  {213} (\bibinfo {year} {2021})}\BibitemShut {NoStop}%
\bibitem [{\citenamefont {Cui}\ \emph {et~al.}(2021)\citenamefont {Cui},
  \citenamefont {Li}, \citenamefont {Li}, \citenamefont {Zhu}, \citenamefont
  {Hu}, \citenamefont {Yang}, \citenamefont {Zhang}, \citenamefont {Yu},
  \citenamefont {Wen},\ and\ \citenamefont {Yu}}]{Cui2021}%
  \BibitemOpen
  \bibfield  {author} {\bibinfo {author} {\bibfnamefont {Y.}~\bibnamefont
  {Cui}}, \bibinfo {author} {\bibfnamefont {C.}~\bibnamefont {Li}}, \bibinfo
  {author} {\bibfnamefont {Q.}~\bibnamefont {Li}}, \bibinfo {author}
  {\bibfnamefont {X.}~\bibnamefont {Zhu}}, \bibinfo {author} {\bibfnamefont
  {Z.}~\bibnamefont {Hu}}, \bibinfo {author} {\bibfnamefont {Y.~F.}\
  \bibnamefont {Yang}}, \bibinfo {author} {\bibfnamefont {J.}~\bibnamefont
  {Zhang}}, \bibinfo {author} {\bibfnamefont {R.}~\bibnamefont {Yu}}, \bibinfo
  {author} {\bibfnamefont {H.~H.}\ \bibnamefont {Wen}},\ and\ \bibinfo {author}
  {\bibfnamefont {W.}~\bibnamefont {Yu}},\ }\bibfield  {title} {\bibinfo
  {title} {{NMR Evidence of Antiferromagnetic Spin Fluctuations in
  Nd$_{0.85}$Sr$_{0.15}$NiO$_2$}},\ }\href
  {https://doi.org/10.1088/0256-307X/38/6/067401} {\bibfield  {journal}
  {\bibinfo  {journal} {Chinese Physics Letters}\ }\textbf {\bibinfo {volume}
  {38}},\ \bibinfo {pages} {0} (\bibinfo {year} {2021})}\BibitemShut {NoStop}%
\bibitem [{\citenamefont {Fowlie}\ \emph {et~al.}(2022)\citenamefont {Fowlie},
  \citenamefont {Hadjimichael}, \citenamefont {Martins}, \citenamefont {Li},
  \citenamefont {Osada}, \citenamefont {Wang}, \citenamefont {Lee},
  \citenamefont {Lee}, \citenamefont {Salman}, \citenamefont {Prokscha},
  \citenamefont {Triscone}, \citenamefont {Hwang},\ and\ \citenamefont
  {Suter}}]{Fowlie2022}%
  \BibitemOpen
  \bibfield  {author} {\bibinfo {author} {\bibfnamefont {J.}~\bibnamefont
  {Fowlie}}, \bibinfo {author} {\bibfnamefont {M.}~\bibnamefont
  {Hadjimichael}}, \bibinfo {author} {\bibfnamefont {M.~M.}\ \bibnamefont
  {Martins}}, \bibinfo {author} {\bibfnamefont {D.}~\bibnamefont {Li}},
  \bibinfo {author} {\bibfnamefont {M.}~\bibnamefont {Osada}}, \bibinfo
  {author} {\bibfnamefont {B.~Y.}\ \bibnamefont {Wang}}, \bibinfo {author}
  {\bibfnamefont {K.}~\bibnamefont {Lee}}, \bibinfo {author} {\bibfnamefont
  {Y.}~\bibnamefont {Lee}}, \bibinfo {author} {\bibfnamefont {Z.}~\bibnamefont
  {Salman}}, \bibinfo {author} {\bibfnamefont {T.}~\bibnamefont {Prokscha}},
  \bibinfo {author} {\bibfnamefont {J.-M.}\ \bibnamefont {Triscone}}, \bibinfo
  {author} {\bibfnamefont {H.~Y.}\ \bibnamefont {Hwang}},\ and\ \bibinfo
  {author} {\bibfnamefont {A.}~\bibnamefont {Suter}},\ }\bibfield  {title}
  {\bibinfo {title} {{Intrinsic magnetism in superconducting infinite-layer
  nickelates}},\ }\href {https://www.nature.com/articles/s41567-022-01684-y}
  {\bibfield  {journal} {\bibinfo  {journal} {Nature Physics}\ }\textbf
  {\bibinfo {volume} {18}},\ \bibinfo {pages} {1043} (\bibinfo {year}
  {2022})}\BibitemShut {NoStop}%
\bibitem [{\citenamefont {Rossi}\ \emph {et~al.}(2024)\citenamefont {Rossi},
  \citenamefont {Lu}, \citenamefont {Lee}, \citenamefont {Goodge},
  \citenamefont {Choi}, \citenamefont {Osada}, \citenamefont {Lee},
  \citenamefont {Li}, \citenamefont {Wang}, \citenamefont {Jost}, \citenamefont
  {Agrestini}, \citenamefont {Garcia-Fernandez}, \citenamefont {Shen},
  \citenamefont {Zhou}, \citenamefont {Been}, \citenamefont {Moritz},
  \citenamefont {Kourkoutis}, \citenamefont {Devereaux}, \citenamefont
  {Hwang},\ and\ \citenamefont {Lee}}]{Rossi2024}%
  \BibitemOpen
  \bibfield  {author} {\bibinfo {author} {\bibfnamefont {M.}~\bibnamefont
  {Rossi}}, \bibinfo {author} {\bibfnamefont {H.}~\bibnamefont {Lu}}, \bibinfo
  {author} {\bibfnamefont {K.}~\bibnamefont {Lee}}, \bibinfo {author}
  {\bibfnamefont {B.~H.}\ \bibnamefont {Goodge}}, \bibinfo {author}
  {\bibfnamefont {J.}~\bibnamefont {Choi}}, \bibinfo {author} {\bibfnamefont
  {M.}~\bibnamefont {Osada}}, \bibinfo {author} {\bibfnamefont
  {Y.}~\bibnamefont {Lee}}, \bibinfo {author} {\bibfnamefont {D.}~\bibnamefont
  {Li}}, \bibinfo {author} {\bibfnamefont {B.~Y.}\ \bibnamefont {Wang}},
  \bibinfo {author} {\bibfnamefont {D.}~\bibnamefont {Jost}}, \bibinfo {author}
  {\bibfnamefont {S.}~\bibnamefont {Agrestini}}, \bibinfo {author}
  {\bibfnamefont {M.}~\bibnamefont {Garcia-Fernandez}}, \bibinfo {author}
  {\bibfnamefont {Z.~X.}\ \bibnamefont {Shen}}, \bibinfo {author}
  {\bibfnamefont {K.~J.}\ \bibnamefont {Zhou}}, \bibinfo {author}
  {\bibfnamefont {E.}~\bibnamefont {Been}}, \bibinfo {author} {\bibfnamefont
  {B.}~\bibnamefont {Moritz}}, \bibinfo {author} {\bibfnamefont {L.~F.}\
  \bibnamefont {Kourkoutis}}, \bibinfo {author} {\bibfnamefont {T.~P.}\
  \bibnamefont {Devereaux}}, \bibinfo {author} {\bibfnamefont {H.~Y.}\
  \bibnamefont {Hwang}},\ and\ \bibinfo {author} {\bibfnamefont {W.~S.}\
  \bibnamefont {Lee}},\ }\bibfield  {title} {\bibinfo {title} {{Universal
  orbital and magnetic structures in infinite-layer nickelates}},\ }\bibfield
  {journal} {\bibinfo  {journal} {Physical Review B}\ }\textbf {\bibinfo
  {volume} {109}},\ \href {https://doi.org/10.1103/PhysRevB.109.024512}
  {10.1103/PhysRevB.109.024512} (\bibinfo {year} {2024})\BibitemShut {NoStop}%
\bibitem [{\citenamefont {Lechermann}(2020)}]{Lechermann2020a}%
  \BibitemOpen
  \bibfield  {author} {\bibinfo {author} {\bibfnamefont {F.}~\bibnamefont
  {Lechermann}},\ }\bibfield  {title} {\bibinfo {title} {{Late transition metal
  oxides with infinite-layer structure: Nickelates versus cuprates}},\ }\href
  {https://doi.org/10.1103/PhysRevB.101.081110} {\bibfield  {journal} {\bibinfo
   {journal} {Physical Review B}\ }\textbf {\bibinfo {volume} {101}},\ \bibinfo
  {pages} {081110} (\bibinfo {year} {2020})}\BibitemShut {NoStop}%
\bibitem [{\citenamefont {Lee}\ and\ \citenamefont {Pickett}(2004)}]{Lee2004}%
  \BibitemOpen
  \bibfield  {author} {\bibinfo {author} {\bibfnamefont {K.-W.}\ \bibnamefont
  {Lee}}\ and\ \bibinfo {author} {\bibfnamefont {W.~E.}\ \bibnamefont
  {Pickett}},\ }\bibfield  {title} {\bibinfo {title} {{Infinite-layer
  LaNiO$_2$: Ni$^{1+}$ is not Cu$^{2+}$}},\ }\href
  {https://doi.org/10.1103/PhysRevB.70.165109} {\bibfield  {journal} {\bibinfo
  {journal} {Physical Review B}\ }\textbf {\bibinfo {volume} {70}},\ \bibinfo
  {pages} {165109} (\bibinfo {year} {2004})}\BibitemShut {NoStop}%
\bibitem [{\citenamefont {Jiang}\ \emph {et~al.}(2020)\citenamefont {Jiang},
  \citenamefont {Berciu},\ and\ \citenamefont {Sawatzky}}]{Jiang2020}%
  \BibitemOpen
  \bibfield  {author} {\bibinfo {author} {\bibfnamefont {M.}~\bibnamefont
  {Jiang}}, \bibinfo {author} {\bibfnamefont {M.}~\bibnamefont {Berciu}},\ and\
  \bibinfo {author} {\bibfnamefont {G.~A.}\ \bibnamefont {Sawatzky}},\
  }\bibfield  {title} {\bibinfo {title} {{Critical Nature of the Ni Spin State
  in Doped NdNiO$_2$}},\ }\href
  {https://doi.org/10.1103/PhysRevLett.124.207004} {\bibfield  {journal}
  {\bibinfo  {journal} {Physical Review Letters}\ }\textbf {\bibinfo {volume}
  {124}},\ \bibinfo {pages} {207004} (\bibinfo {year} {2020})}\BibitemShut
  {NoStop}%
\bibitem [{\citenamefont {Jiang}\ \emph {et~al.}(2022)\citenamefont {Jiang},
  \citenamefont {Berciu},\ and\ \citenamefont {Sawatzky}}]{Jiang2022}%
  \BibitemOpen
  \bibfield  {author} {\bibinfo {author} {\bibfnamefont {M.}~\bibnamefont
  {Jiang}}, \bibinfo {author} {\bibfnamefont {M.}~\bibnamefont {Berciu}},\ and\
  \bibinfo {author} {\bibfnamefont {G.~A.}\ \bibnamefont {Sawatzky}},\
  }\bibfield  {title} {\bibinfo {title} {{Stabilization of singlet hole-doped
  state in infinite-layer nickelate superconductors}},\ }\href
  {https://doi.org/10.1103/PhysRevB.106.115150} {\bibfield  {journal} {\bibinfo
   {journal} {Physical Review B}\ }\textbf {\bibinfo {volume} {106}},\ \bibinfo
  {pages} {115150} (\bibinfo {year} {2022})}\BibitemShut {NoStop}%
\bibitem [{\citenamefont {Osada}\ \emph
  {et~al.}(2020{\natexlab{a}})\citenamefont {Osada}, \citenamefont {Wang},
  \citenamefont {Goodge}, \citenamefont {Lee}, \citenamefont {Yoon},
  \citenamefont {Sakuma}, \citenamefont {Li}, \citenamefont {Miura},
  \citenamefont {Kourkoutis},\ and\ \citenamefont {Hwang}}]{Osada2020}%
  \BibitemOpen
  \bibfield  {author} {\bibinfo {author} {\bibfnamefont {M.}~\bibnamefont
  {Osada}}, \bibinfo {author} {\bibfnamefont {B.~Y.}\ \bibnamefont {Wang}},
  \bibinfo {author} {\bibfnamefont {B.~H.}\ \bibnamefont {Goodge}}, \bibinfo
  {author} {\bibfnamefont {K.}~\bibnamefont {Lee}}, \bibinfo {author}
  {\bibfnamefont {H.}~\bibnamefont {Yoon}}, \bibinfo {author} {\bibfnamefont
  {K.}~\bibnamefont {Sakuma}}, \bibinfo {author} {\bibfnamefont
  {D.}~\bibnamefont {Li}}, \bibinfo {author} {\bibfnamefont {M.}~\bibnamefont
  {Miura}}, \bibinfo {author} {\bibfnamefont {L.~F.}\ \bibnamefont
  {Kourkoutis}},\ and\ \bibinfo {author} {\bibfnamefont {H.~Y.}\ \bibnamefont
  {Hwang}},\ }\bibfield  {title} {\bibinfo {title} {{A Superconducting
  Praseodymium Nickelate with Infinite Layer Structure}},\ }\href
  {https://pubs.acs.org/doi/10.1021/acs.nanolett.0c01392} {\bibfield  {journal}
  {\bibinfo  {journal} {Nano Letters}\ }\textbf {\bibinfo {volume} {20}},\
  \bibinfo {pages} {5735} (\bibinfo {year} {2020}{\natexlab{a}})}\BibitemShut
  {NoStop}%
\bibitem [{\citenamefont {Zeng}\ \emph {et~al.}(2022)\citenamefont {Zeng},
  \citenamefont {Li}, \citenamefont {Chow}, \citenamefont {Cao}, \citenamefont
  {Zhang}, \citenamefont {Tang}, \citenamefont {Yin}, \citenamefont {Lim},
  \citenamefont {Hu}, \citenamefont {Yang},\ and\ \citenamefont
  {Ariando}}]{Zeng2021}%
  \BibitemOpen
  \bibfield  {author} {\bibinfo {author} {\bibfnamefont {S.}~\bibnamefont
  {Zeng}}, \bibinfo {author} {\bibfnamefont {C.}~\bibnamefont {Li}}, \bibinfo
  {author} {\bibfnamefont {L.~E.}\ \bibnamefont {Chow}}, \bibinfo {author}
  {\bibfnamefont {Y.}~\bibnamefont {Cao}}, \bibinfo {author} {\bibfnamefont
  {Z.}~\bibnamefont {Zhang}}, \bibinfo {author} {\bibfnamefont {C.~S.}\
  \bibnamefont {Tang}}, \bibinfo {author} {\bibfnamefont {X.}~\bibnamefont
  {Yin}}, \bibinfo {author} {\bibfnamefont {Z.~S.}\ \bibnamefont {Lim}},
  \bibinfo {author} {\bibfnamefont {J.}~\bibnamefont {Hu}}, \bibinfo {author}
  {\bibfnamefont {P.}~\bibnamefont {Yang}},\ and\ \bibinfo {author}
  {\bibfnamefont {A.}~\bibnamefont {Ariando}},\ }\bibfield  {title} {\bibinfo
  {title} {{Superconductivity in infinite-layer nickelate
  La$_{1-x}$Ca$_x$NiO$_2$ thin films}},\ }\href
  {https://doi.org/10.1126/sciadv.abl9927} {\bibfield  {journal} {\bibinfo
  {journal} {Science Advances}\ }\textbf {\bibinfo {volume} {8}},\ \bibinfo
  {pages} {eabl9927} (\bibinfo {year} {2022})}\BibitemShut {NoStop}%
\bibitem [{\citenamefont {Osada}\ \emph {et~al.}(2021)\citenamefont {Osada},
  \citenamefont {Wang}, \citenamefont {Goodge}, \citenamefont {Harvey},
  \citenamefont {Lee}, \citenamefont {Li}, \citenamefont {Kourkoutis},\ and\
  \citenamefont {Hwang}}]{Osada2021}%
  \BibitemOpen
  \bibfield  {author} {\bibinfo {author} {\bibfnamefont {M.}~\bibnamefont
  {Osada}}, \bibinfo {author} {\bibfnamefont {B.~Y.}\ \bibnamefont {Wang}},
  \bibinfo {author} {\bibfnamefont {B.~H.}\ \bibnamefont {Goodge}}, \bibinfo
  {author} {\bibfnamefont {S.~P.}\ \bibnamefont {Harvey}}, \bibinfo {author}
  {\bibfnamefont {K.}~\bibnamefont {Lee}}, \bibinfo {author} {\bibfnamefont
  {D.}~\bibnamefont {Li}}, \bibinfo {author} {\bibfnamefont {L.~F.}\
  \bibnamefont {Kourkoutis}},\ and\ \bibinfo {author} {\bibfnamefont {H.~Y.}\
  \bibnamefont {Hwang}},\ }\bibfield  {title} {\bibinfo {title} {{Nickelate
  Superconductivity without Rare‐Earth Magnetism: (La,Sr)NiO$_2$}},\ }\href
  {https://doi.org/10.1002/adma.202104083} {\bibfield  {journal} {\bibinfo
  {journal} {Advanced Materials}\ }\textbf {\bibinfo {volume} {33}},\ \bibinfo
  {pages} {2104083} (\bibinfo {year} {2021})}\BibitemShut {NoStop}%
\bibitem [{\citenamefont {Kawai}\ \emph {et~al.}(2009)\citenamefont {Kawai},
  \citenamefont {Inoue}, \citenamefont {Mizumaki}, \citenamefont {Kawamura},
  \citenamefont {Ichikawa},\ and\ \citenamefont {Shimakawa}}]{Kawai2009}%
  \BibitemOpen
  \bibfield  {author} {\bibinfo {author} {\bibfnamefont {M.}~\bibnamefont
  {Kawai}}, \bibinfo {author} {\bibfnamefont {S.}~\bibnamefont {Inoue}},
  \bibinfo {author} {\bibfnamefont {M.}~\bibnamefont {Mizumaki}}, \bibinfo
  {author} {\bibfnamefont {N.}~\bibnamefont {Kawamura}}, \bibinfo {author}
  {\bibfnamefont {N.}~\bibnamefont {Ichikawa}},\ and\ \bibinfo {author}
  {\bibfnamefont {Y.}~\bibnamefont {Shimakawa}},\ }\bibfield  {title} {\bibinfo
  {title} {{Reversible changes of epitaxial thin films from perovskite
  LaNiO$_3$ to infinite-layer structure LaNiO$_2$}},\ }\href
  {https://doi.org/10.1063/1.3078276} {\bibfield  {journal} {\bibinfo
  {journal} {Applied Physics Letters}\ }\textbf {\bibinfo {volume} {94}},\
  \bibinfo {pages} {082102} (\bibinfo {year} {2009})}\BibitemShut {NoStop}%
\bibitem [{\citenamefont {Ikeda}\ \emph {et~al.}(2013)\citenamefont {Ikeda},
  \citenamefont {Manabe},\ and\ \citenamefont {Naito}}]{Ikeda2013}%
  \BibitemOpen
  \bibfield  {author} {\bibinfo {author} {\bibfnamefont {A.}~\bibnamefont
  {Ikeda}}, \bibinfo {author} {\bibfnamefont {T.}~\bibnamefont {Manabe}},\ and\
  \bibinfo {author} {\bibfnamefont {M.}~\bibnamefont {Naito}},\ }\bibfield
  {title} {\bibinfo {title} {{Improved conductivity of infinite-layer LaNiO$_2$
  thin films by metal organic decomposition}},\ }\href
  {https://doi.org/10.1016/j.physc.2013.09.007} {\bibfield  {journal} {\bibinfo
   {journal} {Physica C: Superconductivity and its Applications}\ }\textbf
  {\bibinfo {volume} {495}},\ \bibinfo {pages} {134} (\bibinfo {year}
  {2013})}\BibitemShut {NoStop}%
\bibitem [{\citenamefont {Ikeda}\ \emph {et~al.}(2014)\citenamefont {Ikeda},
  \citenamefont {Manabe},\ and\ \citenamefont {Naito}}]{Ikeda2014}%
  \BibitemOpen
  \bibfield  {author} {\bibinfo {author} {\bibfnamefont {A.}~\bibnamefont
  {Ikeda}}, \bibinfo {author} {\bibfnamefont {T.}~\bibnamefont {Manabe}},\ and\
  \bibinfo {author} {\bibfnamefont {M.}~\bibnamefont {Naito}},\ }\bibfield
  {title} {\bibinfo {title} {{Comparison of reduction agents in the synthesis
  of infinite-layer LaNiO$_2$ films}},\ }\href
  {https://doi.org/10.1016/j.physc.2014.09.002} {\bibfield  {journal} {\bibinfo
   {journal} {Physica C: Superconductivity and its Applications}\ }\textbf
  {\bibinfo {volume} {506}},\ \bibinfo {pages} {83} (\bibinfo {year}
  {2014})}\BibitemShut {NoStop}%
\bibitem [{\citenamefont {Wei}\ \emph {et~al.}(2023{\natexlab{a}})\citenamefont
  {Wei}, \citenamefont {Shin}, \citenamefont {Hong}, \citenamefont {Shin},
  \citenamefont {Thind}, \citenamefont {Yang}, \citenamefont {Klie},
  \citenamefont {Walker},\ and\ \citenamefont {Ahn}}]{Wei2023}%
  \BibitemOpen
  \bibfield  {author} {\bibinfo {author} {\bibfnamefont {W.}~\bibnamefont
  {Wei}}, \bibinfo {author} {\bibfnamefont {K.}~\bibnamefont {Shin}}, \bibinfo
  {author} {\bibfnamefont {H.}~\bibnamefont {Hong}}, \bibinfo {author}
  {\bibfnamefont {Y.}~\bibnamefont {Shin}}, \bibinfo {author} {\bibfnamefont
  {A.~S.}\ \bibnamefont {Thind}}, \bibinfo {author} {\bibfnamefont
  {Y.}~\bibnamefont {Yang}}, \bibinfo {author} {\bibfnamefont {R.~F.}\
  \bibnamefont {Klie}}, \bibinfo {author} {\bibfnamefont {F.~J.}\ \bibnamefont
  {Walker}},\ and\ \bibinfo {author} {\bibfnamefont {C.~H.}\ \bibnamefont
  {Ahn}},\ }\bibfield  {title} {\bibinfo {title} {{Solid state reduction of
  nickelate thin films}},\ }\href
  {https://doi.org/10.1103/PhysRevMaterials.7.013802} {\bibfield  {journal}
  {\bibinfo  {journal} {Physical Review Materials}\ }\textbf {\bibinfo {volume}
  {7}},\ \bibinfo {pages} {013802} (\bibinfo {year}
  {2023}{\natexlab{a}})}\BibitemShut {NoStop}%
\bibitem [{\citenamefont {Parzyck}\ \emph
  {et~al.}(2024{\natexlab{a}})\citenamefont {Parzyck}, \citenamefont {Anil},
  \citenamefont {Wu}, \citenamefont {Goodge}, \citenamefont {Roddy},
  \citenamefont {Kourkoutis}, \citenamefont {Schlom},\ and\ \citenamefont
  {Shen}}]{Parzyck2024b}%
  \BibitemOpen
  \bibfield  {author} {\bibinfo {author} {\bibfnamefont {C.~T.}\ \bibnamefont
  {Parzyck}}, \bibinfo {author} {\bibfnamefont {V.}~\bibnamefont {Anil}},
  \bibinfo {author} {\bibfnamefont {Y.}~\bibnamefont {Wu}}, \bibinfo {author}
  {\bibfnamefont {B.~H.}\ \bibnamefont {Goodge}}, \bibinfo {author}
  {\bibfnamefont {M.}~\bibnamefont {Roddy}}, \bibinfo {author} {\bibfnamefont
  {L.~F.}\ \bibnamefont {Kourkoutis}}, \bibinfo {author} {\bibfnamefont
  {D.~G.}\ \bibnamefont {Schlom}},\ and\ \bibinfo {author} {\bibfnamefont
  {K.~M.}\ \bibnamefont {Shen}},\ }\bibfield  {title} {\bibinfo {title}
  {{Synthesis of thin film infinite-layer nickelates by atomic hydrogen
  reduction: Clarifying the role of the capping layer}},\ }\href
  {https://doi.org/10.1063/5.0197304} {\bibfield  {journal} {\bibinfo
  {journal} {APL Materials}\ }\textbf {\bibinfo {volume} {12}},\ \bibinfo
  {pages} {031132} (\bibinfo {year} {2024}{\natexlab{a}})}\BibitemShut
  {NoStop}%
\bibitem [{\citenamefont {Wang}\ \emph {et~al.}(2020)\citenamefont {Wang},
  \citenamefont {Zheng}, \citenamefont {Krivyakina}, \citenamefont {Chmaissem},
  \citenamefont {Lopes}, \citenamefont {Lynn}, \citenamefont {Gallington},
  \citenamefont {Ren}, \citenamefont {Rosenkranz}, \citenamefont {Mitchell},\
  and\ \citenamefont {Phelan}}]{Wang2020}%
  \BibitemOpen
  \bibfield  {author} {\bibinfo {author} {\bibfnamefont {B.-X.}\ \bibnamefont
  {Wang}}, \bibinfo {author} {\bibfnamefont {H.}~\bibnamefont {Zheng}},
  \bibinfo {author} {\bibfnamefont {E.}~\bibnamefont {Krivyakina}}, \bibinfo
  {author} {\bibfnamefont {O.}~\bibnamefont {Chmaissem}}, \bibinfo {author}
  {\bibfnamefont {P.~P.}\ \bibnamefont {Lopes}}, \bibinfo {author}
  {\bibfnamefont {J.~W.}\ \bibnamefont {Lynn}}, \bibinfo {author}
  {\bibfnamefont {L.~C.}\ \bibnamefont {Gallington}}, \bibinfo {author}
  {\bibfnamefont {Y.}~\bibnamefont {Ren}}, \bibinfo {author} {\bibfnamefont
  {S.}~\bibnamefont {Rosenkranz}}, \bibinfo {author} {\bibfnamefont {J.~F.}\
  \bibnamefont {Mitchell}},\ and\ \bibinfo {author} {\bibfnamefont
  {D.}~\bibnamefont {Phelan}},\ }\bibfield  {title} {\bibinfo {title}
  {{Synthesis and characterization of bulk Nd$_{1-x}$Sr$_x$NiO$_2$ and
  Nd$_{1-x}$Sr$_x$NiO$_3$}},\ }\href
  {https://doi.org/10.1103/PhysRevMaterials.4.084409} {\bibfield  {journal}
  {\bibinfo  {journal} {Physical Review Materials}\ }\textbf {\bibinfo {volume}
  {4}},\ \bibinfo {pages} {084409} (\bibinfo {year} {2020})}\BibitemShut
  {NoStop}%
\bibitem [{\citenamefont {Zeng}\ \emph {et~al.}(2020)\citenamefont {Zeng},
  \citenamefont {Tang}, \citenamefont {Yin}, \citenamefont {Li}, \citenamefont
  {Li}, \citenamefont {Huang}, \citenamefont {Hu}, \citenamefont {Liu},
  \citenamefont {Omar}, \citenamefont {Jani}, \citenamefont {Lim},
  \citenamefont {Han}, \citenamefont {Wan}, \citenamefont {Yang}, \citenamefont
  {Pennycook}, \citenamefont {Wee},\ and\ \citenamefont {Ariando}}]{Zeng2020}%
  \BibitemOpen
  \bibfield  {author} {\bibinfo {author} {\bibfnamefont {S.}~\bibnamefont
  {Zeng}}, \bibinfo {author} {\bibfnamefont {C.~S.}\ \bibnamefont {Tang}},
  \bibinfo {author} {\bibfnamefont {X.}~\bibnamefont {Yin}}, \bibinfo {author}
  {\bibfnamefont {C.}~\bibnamefont {Li}}, \bibinfo {author} {\bibfnamefont
  {M.}~\bibnamefont {Li}}, \bibinfo {author} {\bibfnamefont {Z.}~\bibnamefont
  {Huang}}, \bibinfo {author} {\bibfnamefont {J.}~\bibnamefont {Hu}}, \bibinfo
  {author} {\bibfnamefont {W.}~\bibnamefont {Liu}}, \bibinfo {author}
  {\bibfnamefont {G.~J.}\ \bibnamefont {Omar}}, \bibinfo {author}
  {\bibfnamefont {H.}~\bibnamefont {Jani}}, \bibinfo {author} {\bibfnamefont
  {Z.~S.}\ \bibnamefont {Lim}}, \bibinfo {author} {\bibfnamefont
  {K.}~\bibnamefont {Han}}, \bibinfo {author} {\bibfnamefont {D.}~\bibnamefont
  {Wan}}, \bibinfo {author} {\bibfnamefont {P.}~\bibnamefont {Yang}}, \bibinfo
  {author} {\bibfnamefont {S.~J.}\ \bibnamefont {Pennycook}}, \bibinfo {author}
  {\bibfnamefont {A.~T.~S.}\ \bibnamefont {Wee}},\ and\ \bibinfo {author}
  {\bibfnamefont {A.}~\bibnamefont {Ariando}},\ }\bibfield  {title} {\bibinfo
  {title} {{Phase Diagram and Superconducting Dome of Infinite-Layer
  Nd$_{1-x}$Sr$_x$NiO$_2$ Thin Films}},\ }\href
  {https://doi.org/10.1103/PhysRevLett.125.147003} {\bibfield  {journal}
  {\bibinfo  {journal} {Physical Review Letters}\ }\textbf {\bibinfo {volume}
  {125}},\ \bibinfo {pages} {147003} (\bibinfo {year} {2020})}\BibitemShut
  {NoStop}%
\bibitem [{SIc()}]{SIcite}%
  \BibitemOpen
  \href@noop {} {}\bibinfo {howpublished} {See the Supplemental Materials for
  additional resistivity and hall effect measurements of superconducting and
  non-superconducting samples, lab-based x-ray diffraction measurements,
  additional x-ray absorbtion spectroscopy data, and further STEM and DFT
  analysis of the effects of residual oxygen.}\BibitemShut {Stop}%
\bibitem [{\citenamefont {Li}\ \emph {et~al.}(2020)\citenamefont {Li},
  \citenamefont {Wang}, \citenamefont {Lee}, \citenamefont {Harvey},
  \citenamefont {Osada}, \citenamefont {Goodge}, \citenamefont {Kourkoutis},\
  and\ \citenamefont {Hwang}}]{Li2020a}%
  \BibitemOpen
  \bibfield  {author} {\bibinfo {author} {\bibfnamefont {D.}~\bibnamefont
  {Li}}, \bibinfo {author} {\bibfnamefont {B.~Y.}\ \bibnamefont {Wang}},
  \bibinfo {author} {\bibfnamefont {K.}~\bibnamefont {Lee}}, \bibinfo {author}
  {\bibfnamefont {S.~P.}\ \bibnamefont {Harvey}}, \bibinfo {author}
  {\bibfnamefont {M.}~\bibnamefont {Osada}}, \bibinfo {author} {\bibfnamefont
  {B.~H.}\ \bibnamefont {Goodge}}, \bibinfo {author} {\bibfnamefont {L.~F.}\
  \bibnamefont {Kourkoutis}},\ and\ \bibinfo {author} {\bibfnamefont {H.~Y.}\
  \bibnamefont {Hwang}},\ }\bibfield  {title} {\bibinfo {title}
  {{Superconducting Dome in Nd$_{1-x}$Sr$_x$NiO$_2$ Infinite Layer Films}},\
  }\href {https://doi.org/10.1103/PhysRevLett.125.027001} {\bibfield  {journal}
  {\bibinfo  {journal} {Physical Review Letters}\ }\textbf {\bibinfo {volume}
  {125}},\ \bibinfo {pages} {027001} (\bibinfo {year} {2020})}\BibitemShut
  {NoStop}%
\bibitem [{\citenamefont {Tam}\ \emph {et~al.}(2022)\citenamefont {Tam},
  \citenamefont {Choi}, \citenamefont {Ding}, \citenamefont {Agrestini},
  \citenamefont {Nag}, \citenamefont {Wu}, \citenamefont {Huang}, \citenamefont
  {Luo}, \citenamefont {Gao}, \citenamefont {Garc{\'{i}}a-Fern{\'{a}}ndez},
  \citenamefont {Qiao},\ and\ \citenamefont {Zhou}}]{Tam2022}%
  \BibitemOpen
  \bibfield  {author} {\bibinfo {author} {\bibfnamefont {C.~C.}\ \bibnamefont
  {Tam}}, \bibinfo {author} {\bibfnamefont {J.}~\bibnamefont {Choi}}, \bibinfo
  {author} {\bibfnamefont {X.}~\bibnamefont {Ding}}, \bibinfo {author}
  {\bibfnamefont {S.}~\bibnamefont {Agrestini}}, \bibinfo {author}
  {\bibfnamefont {A.}~\bibnamefont {Nag}}, \bibinfo {author} {\bibfnamefont
  {M.}~\bibnamefont {Wu}}, \bibinfo {author} {\bibfnamefont {B.}~\bibnamefont
  {Huang}}, \bibinfo {author} {\bibfnamefont {H.}~\bibnamefont {Luo}}, \bibinfo
  {author} {\bibfnamefont {P.}~\bibnamefont {Gao}}, \bibinfo {author}
  {\bibfnamefont {M.}~\bibnamefont {Garc{\'{i}}a-Fern{\'{a}}ndez}}, \bibinfo
  {author} {\bibfnamefont {L.}~\bibnamefont {Qiao}},\ and\ \bibinfo {author}
  {\bibfnamefont {K.-J.}\ \bibnamefont {Zhou}},\ }\bibfield  {title} {\bibinfo
  {title} {{Charge density waves in infinite-layer NdNiO$_2$ nickelates}},\
  }\href {https://doi.org/10.1038/s41563-022-01330-1} {\bibfield  {journal}
  {\bibinfo  {journal} {Nature Materials}\ }\textbf {\bibinfo {volume} {21}},\
  \bibinfo {pages} {1116} (\bibinfo {year} {2022})}\BibitemShut {NoStop}%
\bibitem [{\citenamefont {Cheng}\ \emph {et~al.}(2024)\citenamefont {Cheng},
  \citenamefont {Cheng}, \citenamefont {Lee}, \citenamefont {Luo},
  \citenamefont {Chen}, \citenamefont {Lee}, \citenamefont {Wang},
  \citenamefont {Mootz}, \citenamefont {Perakis}, \citenamefont {Shen},
  \citenamefont {Hwang},\ and\ \citenamefont {Wang}}]{Cheng2024}%
  \BibitemOpen
  \bibfield  {author} {\bibinfo {author} {\bibfnamefont {B.}~\bibnamefont
  {Cheng}}, \bibinfo {author} {\bibfnamefont {D.}~\bibnamefont {Cheng}},
  \bibinfo {author} {\bibfnamefont {K.}~\bibnamefont {Lee}}, \bibinfo {author}
  {\bibfnamefont {L.}~\bibnamefont {Luo}}, \bibinfo {author} {\bibfnamefont
  {Z.}~\bibnamefont {Chen}}, \bibinfo {author} {\bibfnamefont {Y.}~\bibnamefont
  {Lee}}, \bibinfo {author} {\bibfnamefont {B.~Y.}\ \bibnamefont {Wang}},
  \bibinfo {author} {\bibfnamefont {M.}~\bibnamefont {Mootz}}, \bibinfo
  {author} {\bibfnamefont {I.~E.}\ \bibnamefont {Perakis}}, \bibinfo {author}
  {\bibfnamefont {Z.-X.}\ \bibnamefont {Shen}}, \bibinfo {author}
  {\bibfnamefont {H.~Y.}\ \bibnamefont {Hwang}},\ and\ \bibinfo {author}
  {\bibfnamefont {J.}~\bibnamefont {Wang}},\ }\bibfield  {title} {\bibinfo
  {title} {{Evidence for $d$-wave superconductivity of infinite-layer
  nickelates from low-energy electrodynamics}},\ }\href
  {https://doi.org/10.1038/s41563-023-01766-z} {\bibfield  {journal} {\bibinfo
  {journal} {Nature Materials}\ }\textbf {\bibinfo {volume} {23}},\ \bibinfo
  {pages} {775} (\bibinfo {year} {2024})}\BibitemShut {NoStop}%
\bibitem [{\citenamefont {Harvey}\ \emph {et~al.}(2022)\citenamefont {Harvey},
  \citenamefont {Wang}, \citenamefont {Fowlie}, \citenamefont {Osada},
  \citenamefont {Lee}, \citenamefont {Lee}, \citenamefont {Li},\ and\
  \citenamefont {Hwang}}]{Harvey2022}%
  \BibitemOpen
  \bibfield  {author} {\bibinfo {author} {\bibfnamefont {S.~P.}\ \bibnamefont
  {Harvey}}, \bibinfo {author} {\bibfnamefont {B.~Y.}\ \bibnamefont {Wang}},
  \bibinfo {author} {\bibfnamefont {J.}~\bibnamefont {Fowlie}}, \bibinfo
  {author} {\bibfnamefont {M.}~\bibnamefont {Osada}}, \bibinfo {author}
  {\bibfnamefont {K.}~\bibnamefont {Lee}}, \bibinfo {author} {\bibfnamefont
  {Y.}~\bibnamefont {Lee}}, \bibinfo {author} {\bibfnamefont {D.}~\bibnamefont
  {Li}},\ and\ \bibinfo {author} {\bibfnamefont {H.~Y.}\ \bibnamefont
  {Hwang}},\ }\href@noop {} {\bibinfo {title} {{Evidence for nodal
  superconductivity in infinite-layer nickelates}}} (\bibinfo {year} {2022}),\
  \Eprint {https://arxiv.org/abs/2201.12971} {arXiv:2201.12971
  [cond-mat.supr-con]} \BibitemShut {NoStop}%
\bibitem [{\citenamefont {Millis}\ \emph {et~al.}(1988)\citenamefont {Millis},
  \citenamefont {Sachdev},\ and\ \citenamefont {Varma}}]{Millis1988a}%
  \BibitemOpen
  \bibfield  {author} {\bibinfo {author} {\bibfnamefont {A.~J.}\ \bibnamefont
  {Millis}}, \bibinfo {author} {\bibfnamefont {S.}~\bibnamefont {Sachdev}},\
  and\ \bibinfo {author} {\bibfnamefont {C.~M.}\ \bibnamefont {Varma}},\
  }\bibfield  {title} {\bibinfo {title} {{Inelastic scattering and pair
  breaking in anisotropic and isotropic superconductors}},\ }\href
  {https://doi.org/10.1103/PhysRevB.37.4975} {\bibfield  {journal} {\bibinfo
  {journal} {Physical Review B}\ }\textbf {\bibinfo {volume} {37}},\ \bibinfo
  {pages} {4975} (\bibinfo {year} {1988})}\BibitemShut {NoStop}%
\bibitem [{\citenamefont {Radtke}\ \emph {et~al.}(1993)\citenamefont {Radtke},
  \citenamefont {Levin}, \citenamefont {Sch{\"{u}}ttler},\ and\ \citenamefont
  {Norman}}]{Radtke1993}%
  \BibitemOpen
  \bibfield  {author} {\bibinfo {author} {\bibfnamefont {R.~J.}\ \bibnamefont
  {Radtke}}, \bibinfo {author} {\bibfnamefont {K.}~\bibnamefont {Levin}},
  \bibinfo {author} {\bibfnamefont {H.-B.}\ \bibnamefont {Sch{\"{u}}ttler}},\
  and\ \bibinfo {author} {\bibfnamefont {M.~R.}\ \bibnamefont {Norman}},\
  }\bibfield  {title} {\bibinfo {title} {{Predictions for impurity-induced
  T$_c$ suppression in the high-temperature superconductors}},\ }\href
  {https://doi.org/10.1103/PhysRevB.48.653} {\bibfield  {journal} {\bibinfo
  {journal} {Physical Review B}\ }\textbf {\bibinfo {volume} {48}},\ \bibinfo
  {pages} {653} (\bibinfo {year} {1993})}\BibitemShut {NoStop}%
\bibitem [{\citenamefont {Smits}(1958)}]{Smits1958}%
  \BibitemOpen
  \bibfield  {author} {\bibinfo {author} {\bibfnamefont {F.~M.}\ \bibnamefont
  {Smits}},\ }\bibfield  {title} {\bibinfo {title} {{Measurement of Sheet
  Resistivities with the Four-Point Probe}},\ }\href
  {https://doi.org/10.1002/j.1538-7305.1958.tb03883.x} {\bibfield  {journal}
  {\bibinfo  {journal} {Bell System Technical Journal}\ }\textbf {\bibinfo
  {volume} {37}},\ \bibinfo {pages} {711} (\bibinfo {year} {1958})}\BibitemShut
  {NoStop}%
\bibitem [{\citenamefont {Wei}\ \emph {et~al.}(2023{\natexlab{b}})\citenamefont
  {Wei}, \citenamefont {Vu}, \citenamefont {Zhang}, \citenamefont {Walker},\
  and\ \citenamefont {Ahn}}]{Wei2023a}%
  \BibitemOpen
  \bibfield  {author} {\bibinfo {author} {\bibfnamefont {W.}~\bibnamefont
  {Wei}}, \bibinfo {author} {\bibfnamefont {D.}~\bibnamefont {Vu}}, \bibinfo
  {author} {\bibfnamefont {Z.}~\bibnamefont {Zhang}}, \bibinfo {author}
  {\bibfnamefont {F.~J.}\ \bibnamefont {Walker}},\ and\ \bibinfo {author}
  {\bibfnamefont {C.~H.}\ \bibnamefont {Ahn}},\ }\bibfield  {title} {\bibinfo
  {title} {{Superconducting Nd$_{1-x}$Eu$_x$NiO$_2$ thin films using in situ
  synthesis}},\ }\href {https://doi.org/10.1126/sciadv.adh3327} {\bibfield
  {journal} {\bibinfo  {journal} {Science Advances}\ }\textbf {\bibinfo
  {volume} {9}},\ \bibinfo {pages} {eadh3327} (\bibinfo {year}
  {2023}{\natexlab{b}})}\BibitemShut {NoStop}%
\bibitem [{\citenamefont {Rossi}\ \emph {et~al.}(2021)\citenamefont {Rossi},
  \citenamefont {Lu}, \citenamefont {Nag}, \citenamefont {Li}, \citenamefont
  {Osada}, \citenamefont {Lee}, \citenamefont {Wang}, \citenamefont
  {Agrestini}, \citenamefont {Garcia-Fernandez}, \citenamefont {Kas},
  \citenamefont {Chuang}, \citenamefont {Shen}, \citenamefont {Hwang},
  \citenamefont {Moritz}, \citenamefont {Zhou}, \citenamefont {Devereaux},\
  and\ \citenamefont {Lee}}]{Rossi2021a}%
  \BibitemOpen
  \bibfield  {author} {\bibinfo {author} {\bibfnamefont {M.}~\bibnamefont
  {Rossi}}, \bibinfo {author} {\bibfnamefont {H.}~\bibnamefont {Lu}}, \bibinfo
  {author} {\bibfnamefont {A.}~\bibnamefont {Nag}}, \bibinfo {author}
  {\bibfnamefont {D.}~\bibnamefont {Li}}, \bibinfo {author} {\bibfnamefont
  {M.}~\bibnamefont {Osada}}, \bibinfo {author} {\bibfnamefont
  {K.}~\bibnamefont {Lee}}, \bibinfo {author} {\bibfnamefont {B.~Y.}\
  \bibnamefont {Wang}}, \bibinfo {author} {\bibfnamefont {S.}~\bibnamefont
  {Agrestini}}, \bibinfo {author} {\bibfnamefont {M.}~\bibnamefont
  {Garcia-Fernandez}}, \bibinfo {author} {\bibfnamefont {J.~J.}\ \bibnamefont
  {Kas}}, \bibinfo {author} {\bibfnamefont {Y.-D.}\ \bibnamefont {Chuang}},
  \bibinfo {author} {\bibfnamefont {Z.~X.}\ \bibnamefont {Shen}}, \bibinfo
  {author} {\bibfnamefont {H.~Y.}\ \bibnamefont {Hwang}}, \bibinfo {author}
  {\bibfnamefont {B.}~\bibnamefont {Moritz}}, \bibinfo {author} {\bibfnamefont
  {K.-J.}\ \bibnamefont {Zhou}}, \bibinfo {author} {\bibfnamefont {T.~P.}\
  \bibnamefont {Devereaux}},\ and\ \bibinfo {author} {\bibfnamefont {W.~S.}\
  \bibnamefont {Lee}},\ }\bibfield  {title} {\bibinfo {title} {{Orbital and
  spin character of doped carriers in infinite-layer nickelates}},\ }\href
  {https://doi.org/10.1103/PhysRevB.104.L220505} {\bibfield  {journal}
  {\bibinfo  {journal} {Physical Review B}\ }\textbf {\bibinfo {volume}
  {104}},\ \bibinfo {pages} {L220505} (\bibinfo {year} {2021})}\BibitemShut
  {NoStop}%
\bibitem [{\citenamefont {Cheong}\ \emph {et~al.}(1994)\citenamefont {Cheong},
  \citenamefont {Hwang}, \citenamefont {Batlogg}, \citenamefont {Cooper},\ and\
  \citenamefont {Canfield}}]{Cheong1994}%
  \BibitemOpen
  \bibfield  {author} {\bibinfo {author} {\bibfnamefont {S.-W.}\ \bibnamefont
  {Cheong}}, \bibinfo {author} {\bibfnamefont {H.}~\bibnamefont {Hwang}},
  \bibinfo {author} {\bibfnamefont {B.}~\bibnamefont {Batlogg}}, \bibinfo
  {author} {\bibfnamefont {A.}~\bibnamefont {Cooper}},\ and\ \bibinfo {author}
  {\bibfnamefont {P.}~\bibnamefont {Canfield}},\ }\bibfield  {title} {\bibinfo
  {title} {{Electron-hole doping of the metal-insulator transition compound
  RENiO$_3$}},\ }\href {https://doi.org/10.1016/0921-4526(94)90873-7}
  {\bibfield  {journal} {\bibinfo  {journal} {Physica B: Condensed Matter}\
  }\textbf {\bibinfo {volume} {194-196}},\ \bibinfo {pages} {1087} (\bibinfo
  {year} {1994})}\BibitemShut {NoStop}%
\bibitem [{\citenamefont {Alonso}\ \emph {et~al.}(1995)\citenamefont {Alonso},
  \citenamefont {Mart{\'{i}}nez-Lope},\ and\ \citenamefont
  {Hidalgo}}]{Alonso1995b}%
  \BibitemOpen
  \bibfield  {author} {\bibinfo {author} {\bibfnamefont {J.~A.}\ \bibnamefont
  {Alonso}}, \bibinfo {author} {\bibfnamefont {M.~J.}\ \bibnamefont
  {Mart{\'{i}}nez-Lope}},\ and\ \bibinfo {author} {\bibfnamefont {M.~A.}\
  \bibnamefont {Hidalgo}},\ }\bibfield  {title} {\bibinfo {title} {{Hole and
  electron doping of RNiO$_3$ (R = La, Nd)}},\ }\href
  {https://doi.org/10.1006/jssc.1995.1196} {\bibfield  {journal} {\bibinfo
  {journal} {Journal of Solid State Chemistry}\ }\textbf {\bibinfo {volume}
  {116}},\ \bibinfo {pages} {146} (\bibinfo {year} {1995})}\BibitemShut
  {NoStop}%
\bibitem [{\citenamefont {Yang}\ \emph {et~al.}(2021)\citenamefont {Yang},
  \citenamefont {Wen}, \citenamefont {Cui}, \citenamefont {Chen},\ and\
  \citenamefont {Zhao}}]{Yang2021}%
  \BibitemOpen
  \bibfield  {author} {\bibinfo {author} {\bibfnamefont {H.}~\bibnamefont
  {Yang}}, \bibinfo {author} {\bibfnamefont {Z.}~\bibnamefont {Wen}}, \bibinfo
  {author} {\bibfnamefont {Y.}~\bibnamefont {Cui}}, \bibinfo {author}
  {\bibfnamefont {Y.}~\bibnamefont {Chen}},\ and\ \bibinfo {author}
  {\bibfnamefont {Y.}~\bibnamefont {Zhao}},\ }\bibfield  {title} {\bibinfo
  {title} {{The Preparation, Structure, and Metal–Insulator Transition in
  Bulk Nd$_{1-x}$Ca$_x$NiO$_3$ (x = $0 \sim 0.3$)}},\ }\href
  {https://doi.org/10.1007/s10948-021-05929-4} {\bibfield  {journal} {\bibinfo
  {journal} {Journal of Superconductivity and Novel Magnetism}\ }\textbf
  {\bibinfo {volume} {34}},\ \bibinfo {pages} {2339} (\bibinfo {year}
  {2021})}\BibitemShut {NoStop}%
\bibitem [{\citenamefont {Patel}\ \emph {et~al.}(2022)\citenamefont {Patel},
  \citenamefont {Patra}, \citenamefont {Ojha}, \citenamefont {Kumar},
  \citenamefont {Sarkar}, \citenamefont {Saha}, \citenamefont {Bhattacharya},
  \citenamefont {Freeland}, \citenamefont {Kim}, \citenamefont {Ryan},
  \citenamefont {Mahadevan},\ and\ \citenamefont {Middey}}]{Patel2022}%
  \BibitemOpen
  \bibfield  {author} {\bibinfo {author} {\bibfnamefont {R.~K.}\ \bibnamefont
  {Patel}}, \bibinfo {author} {\bibfnamefont {K.}~\bibnamefont {Patra}},
  \bibinfo {author} {\bibfnamefont {S.~K.}\ \bibnamefont {Ojha}}, \bibinfo
  {author} {\bibfnamefont {S.}~\bibnamefont {Kumar}}, \bibinfo {author}
  {\bibfnamefont {S.}~\bibnamefont {Sarkar}}, \bibinfo {author} {\bibfnamefont
  {A.}~\bibnamefont {Saha}}, \bibinfo {author} {\bibfnamefont {N.}~\bibnamefont
  {Bhattacharya}}, \bibinfo {author} {\bibfnamefont {J.~W.}\ \bibnamefont
  {Freeland}}, \bibinfo {author} {\bibfnamefont {J.~W.}\ \bibnamefont {Kim}},
  \bibinfo {author} {\bibfnamefont {P.~J.}\ \bibnamefont {Ryan}}, \bibinfo
  {author} {\bibfnamefont {P.}~\bibnamefont {Mahadevan}},\ and\ \bibinfo
  {author} {\bibfnamefont {S.}~\bibnamefont {Middey}},\ }\bibfield  {title}
  {\bibinfo {title} {{Hole doping in a negative charge transfer insulator}},\
  }\href {https://doi.org/10.1038/s42005-022-00993-1} {\bibfield  {journal}
  {\bibinfo  {journal} {Communications Physics}\ }\textbf {\bibinfo {volume}
  {5}},\ \bibinfo {pages} {1} (\bibinfo {year} {2022})}\BibitemShut {NoStop}%
\bibitem [{\citenamefont {Brooks}\ \emph {et~al.}(2009)\citenamefont {Brooks},
  \citenamefont {Kourkoutis}, \citenamefont {Heeg}, \citenamefont {Schubert},
  \citenamefont {Muller},\ and\ \citenamefont {Schlom}}]{Brooks2009a}%
  \BibitemOpen
  \bibfield  {author} {\bibinfo {author} {\bibfnamefont {C.~M.}\ \bibnamefont
  {Brooks}}, \bibinfo {author} {\bibfnamefont {L.~F.}\ \bibnamefont
  {Kourkoutis}}, \bibinfo {author} {\bibfnamefont {T.}~\bibnamefont {Heeg}},
  \bibinfo {author} {\bibfnamefont {J.}~\bibnamefont {Schubert}}, \bibinfo
  {author} {\bibfnamefont {D.~A.}\ \bibnamefont {Muller}},\ and\ \bibinfo
  {author} {\bibfnamefont {D.~G.}\ \bibnamefont {Schlom}},\ }\bibfield  {title}
  {\bibinfo {title} {{Growth of homoepitaxial SrTiO$_3$ thin films by
  molecular-beam epitaxy}},\ }\href {https://doi.org/10.1063/1.3117365}
  {\bibfield  {journal} {\bibinfo  {journal} {Applied Physics Letters}\
  }\textbf {\bibinfo {volume} {94}},\ \bibinfo {pages} {3} (\bibinfo {year}
  {2009})}\BibitemShut {NoStop}%
\bibitem [{\citenamefont {Brooks}\ \emph {et~al.}(2015)\citenamefont {Brooks},
  \citenamefont {Wilson}, \citenamefont {Schäfer}, \citenamefont {Mundy},
  \citenamefont {Holtz}, \citenamefont {Muller}, \citenamefont {Schubert},
  \citenamefont {Cahill},\ and\ \citenamefont {Schlom}}]{Brooks2015}%
  \BibitemOpen
  \bibfield  {author} {\bibinfo {author} {\bibfnamefont {C.~M.}\ \bibnamefont
  {Brooks}}, \bibinfo {author} {\bibfnamefont {R.~B.}\ \bibnamefont {Wilson}},
  \bibinfo {author} {\bibfnamefont {A.}~\bibnamefont {Schäfer}}, \bibinfo
  {author} {\bibfnamefont {J.~A.}\ \bibnamefont {Mundy}}, \bibinfo {author}
  {\bibfnamefont {M.~E.}\ \bibnamefont {Holtz}}, \bibinfo {author}
  {\bibfnamefont {D.~A.}\ \bibnamefont {Muller}}, \bibinfo {author}
  {\bibfnamefont {J.}~\bibnamefont {Schubert}}, \bibinfo {author}
  {\bibfnamefont {D.~G.}\ \bibnamefont {Cahill}},\ and\ \bibinfo {author}
  {\bibfnamefont {D.~G.}\ \bibnamefont {Schlom}},\ }\bibfield  {title}
  {\bibinfo {title} {{Tuning thermal conductivity in homoepitaxial SrTiO$_3$
  films via defects}},\ }\href {https://doi.org/10.1063/1.4927200} {\bibfield
  {journal} {\bibinfo  {journal} {Applied Physics Letters}\ }\textbf {\bibinfo
  {volume} {107}},\ \bibinfo {pages} {051902} (\bibinfo {year}
  {2015})}\BibitemShut {NoStop}%
\bibitem [{\citenamefont {Parzyck}\ \emph
  {et~al.}(2024{\natexlab{b}})\citenamefont {Parzyck}, \citenamefont {Gupta},
  \citenamefont {Wu}, \citenamefont {Anil}, \citenamefont {Bhatt},
  \citenamefont {Bouliane}, \citenamefont {Gong}, \citenamefont {Gregory},
  \citenamefont {Luo}, \citenamefont {Sutarto}, \citenamefont {He},
  \citenamefont {Chuang}, \citenamefont {Zhou}, \citenamefont {Herranz},
  \citenamefont {Kourkoutis}, \citenamefont {Singer}, \citenamefont {Schlom},
  \citenamefont {Hawthorn},\ and\ \citenamefont {Shen}}]{Parzyck2024a}%
  \BibitemOpen
  \bibfield  {author} {\bibinfo {author} {\bibfnamefont {C.~T.}\ \bibnamefont
  {Parzyck}}, \bibinfo {author} {\bibfnamefont {N.~K.}\ \bibnamefont {Gupta}},
  \bibinfo {author} {\bibfnamefont {Y.}~\bibnamefont {Wu}}, \bibinfo {author}
  {\bibfnamefont {V.}~\bibnamefont {Anil}}, \bibinfo {author} {\bibfnamefont
  {L.}~\bibnamefont {Bhatt}}, \bibinfo {author} {\bibfnamefont
  {M.}~\bibnamefont {Bouliane}}, \bibinfo {author} {\bibfnamefont
  {R.}~\bibnamefont {Gong}}, \bibinfo {author} {\bibfnamefont {B.~Z.}\
  \bibnamefont {Gregory}}, \bibinfo {author} {\bibfnamefont {A.}~\bibnamefont
  {Luo}}, \bibinfo {author} {\bibfnamefont {R.}~\bibnamefont {Sutarto}},
  \bibinfo {author} {\bibfnamefont {F.}~\bibnamefont {He}}, \bibinfo {author}
  {\bibfnamefont {Y.-D.}\ \bibnamefont {Chuang}}, \bibinfo {author}
  {\bibfnamefont {T.}~\bibnamefont {Zhou}}, \bibinfo {author} {\bibfnamefont
  {G.}~\bibnamefont {Herranz}}, \bibinfo {author} {\bibfnamefont {L.~F.}\
  \bibnamefont {Kourkoutis}}, \bibinfo {author} {\bibfnamefont
  {A.}~\bibnamefont {Singer}}, \bibinfo {author} {\bibfnamefont {D.~G.}\
  \bibnamefont {Schlom}}, \bibinfo {author} {\bibfnamefont {D.~G.}\
  \bibnamefont {Hawthorn}},\ and\ \bibinfo {author} {\bibfnamefont {K.~M.}\
  \bibnamefont {Shen}},\ }\bibfield  {title} {\bibinfo {title} {{Absence of
  $3a_0$ charge density wave order in the infinite-layer nickelate
  NdNiO$_2$}},\ }\href {https://doi.org/10.1038/s41563-024-01797-0} {\bibfield
  {journal} {\bibinfo  {journal} {Nature Materials}\ }\textbf {\bibinfo
  {volume} {23}},\ \bibinfo {pages} {486} (\bibinfo {year}
  {2024}{\natexlab{b}})}\BibitemShut {NoStop}%
\bibitem [{\citenamefont {Krieger}\ \emph {et~al.}(2024)\citenamefont
  {Krieger}, \citenamefont {Sahib}, \citenamefont {Rosa}, \citenamefont {Rath},
  \citenamefont {Chen}, \citenamefont {Raji}, \citenamefont {Pinho},
  \citenamefont {Lefevre}, \citenamefont {Ghiringhelli}, \citenamefont
  {Gloter}, \citenamefont {Viart}, \citenamefont {Salluzzo},\ and\
  \citenamefont {Preziosi}}]{Krieger2024}%
  \BibitemOpen
  \bibfield  {author} {\bibinfo {author} {\bibfnamefont {G.}~\bibnamefont
  {Krieger}}, \bibinfo {author} {\bibfnamefont {H.}~\bibnamefont {Sahib}},
  \bibinfo {author} {\bibfnamefont {F.}~\bibnamefont {Rosa}}, \bibinfo {author}
  {\bibfnamefont {M.}~\bibnamefont {Rath}}, \bibinfo {author} {\bibfnamefont
  {Y.}~\bibnamefont {Chen}}, \bibinfo {author} {\bibfnamefont {A.}~\bibnamefont
  {Raji}}, \bibinfo {author} {\bibfnamefont {P.~V.~B.}\ \bibnamefont {Pinho}},
  \bibinfo {author} {\bibfnamefont {C.}~\bibnamefont {Lefevre}}, \bibinfo
  {author} {\bibfnamefont {G.}~\bibnamefont {Ghiringhelli}}, \bibinfo {author}
  {\bibfnamefont {A.}~\bibnamefont {Gloter}}, \bibinfo {author} {\bibfnamefont
  {N.}~\bibnamefont {Viart}}, \bibinfo {author} {\bibfnamefont
  {M.}~\bibnamefont {Salluzzo}},\ and\ \bibinfo {author} {\bibfnamefont
  {D.}~\bibnamefont {Preziosi}},\ }\href@noop {} {\bibinfo {title} {Signatures
  of canted antiferromagnetism in infinite-layer nickelates studied by x-ray
  magnetic dichroism}} (\bibinfo {year} {2024}),\ \Eprint
  {https://arxiv.org/abs/2403.16969} {arXiv:2403.16969 [cond-mat.mtrl-sci]}
  \BibitemShut {NoStop}%
\bibitem [{\citenamefont {Osada}\ \emph
  {et~al.}(2020{\natexlab{b}})\citenamefont {Osada}, \citenamefont {Wang},
  \citenamefont {Lee}, \citenamefont {Li},\ and\ \citenamefont
  {Hwang}}]{Osada2020a}%
  \BibitemOpen
  \bibfield  {author} {\bibinfo {author} {\bibfnamefont {M.}~\bibnamefont
  {Osada}}, \bibinfo {author} {\bibfnamefont {B.~Y.}\ \bibnamefont {Wang}},
  \bibinfo {author} {\bibfnamefont {K.}~\bibnamefont {Lee}}, \bibinfo {author}
  {\bibfnamefont {D.}~\bibnamefont {Li}},\ and\ \bibinfo {author}
  {\bibfnamefont {H.~Y.}\ \bibnamefont {Hwang}},\ }\bibfield  {title} {\bibinfo
  {title} {{Phase diagram of infinite layer praseodymium nickelate
  Pr$_{1-x}$Sr$_x$NiO$_2$ thin films}},\ }\href
  {https://doi.org/10.1103/PhysRevMaterials.4.121801} {\bibfield  {journal}
  {\bibinfo  {journal} {Phys. Rev. Mater.}\ }\textbf {\bibinfo {volume} {4}},\
  \bibinfo {pages} {121801} (\bibinfo {year} {2020}{\natexlab{b}})}\BibitemShut
  {NoStop}%
\bibitem [{\citenamefont {Ando}\ \emph {et~al.}(2001)\citenamefont {Ando},
  \citenamefont {Lavrov}, \citenamefont {Komiya}, \citenamefont {Segawa},\ and\
  \citenamefont {Sun}}]{Ando2001}%
  \BibitemOpen
  \bibfield  {author} {\bibinfo {author} {\bibfnamefont {Y.}~\bibnamefont
  {Ando}}, \bibinfo {author} {\bibfnamefont {A.~N.}\ \bibnamefont {Lavrov}},
  \bibinfo {author} {\bibfnamefont {S.}~\bibnamefont {Komiya}}, \bibinfo
  {author} {\bibfnamefont {K.}~\bibnamefont {Segawa}},\ and\ \bibinfo {author}
  {\bibfnamefont {X.~F.}\ \bibnamefont {Sun}},\ }\bibfield  {title} {\bibinfo
  {title} {{Mobility of the Doped Holes and the Antiferromagnetic Correlations
  in Underdoped High-$T_c$ Cuprates}},\ }\href
  {https://doi.org/10.1103/PhysRevLett.87.017001} {\bibfield  {journal}
  {\bibinfo  {journal} {Physical Review Letters}\ }\textbf {\bibinfo {volume}
  {87}},\ \bibinfo {pages} {017001} (\bibinfo {year} {2001})}\BibitemShut
  {NoStop}%
\bibitem [{\citenamefont {Ando}\ \emph {et~al.}(2004)\citenamefont {Ando},
  \citenamefont {Komiya}, \citenamefont {Segawa}, \citenamefont {Ono},\ and\
  \citenamefont {Kurita}}]{Ando2004}%
  \BibitemOpen
  \bibfield  {author} {\bibinfo {author} {\bibfnamefont {Y.}~\bibnamefont
  {Ando}}, \bibinfo {author} {\bibfnamefont {S.}~\bibnamefont {Komiya}},
  \bibinfo {author} {\bibfnamefont {K.}~\bibnamefont {Segawa}}, \bibinfo
  {author} {\bibfnamefont {S.}~\bibnamefont {Ono}},\ and\ \bibinfo {author}
  {\bibfnamefont {Y.}~\bibnamefont {Kurita}},\ }\bibfield  {title} {\bibinfo
  {title} {{Electronic Phase Diagram of High-$T_c$ Cuprate Superconductors from
  a Mapping of the In-Plane Resistivity Curvature}},\ }\href
  {https://doi.org/10.1103/PhysRevLett.93.267001} {\bibfield  {journal}
  {\bibinfo  {journal} {Physical Review Letters}\ }\textbf {\bibinfo {volume}
  {93}},\ \bibinfo {pages} {267001} (\bibinfo {year} {2004})}\BibitemShut
  {NoStop}%
\bibitem [{\citenamefont {Preziosi}()}]{Preziosi2024}%
  \BibitemOpen
  \bibfield  {author} {\bibinfo {author} {\bibfnamefont {D.}~\bibnamefont
  {Preziosi}},\ }\href@noop {} {}\bibinfo {howpublished} {private
  communication}\BibitemShut {NoStop}%
\bibitem [{\citenamefont {Richard}\ \emph {et~al.}(2007)\citenamefont
  {Richard}, \citenamefont {Neupane}, \citenamefont {Xu}, \citenamefont
  {Fournier}, \citenamefont {Li}, \citenamefont {Dai}, \citenamefont {Wang},\
  and\ \citenamefont {Ding}}]{Richard2007}%
  \BibitemOpen
  \bibfield  {author} {\bibinfo {author} {\bibfnamefont {P.}~\bibnamefont
  {Richard}}, \bibinfo {author} {\bibfnamefont {M.}~\bibnamefont {Neupane}},
  \bibinfo {author} {\bibfnamefont {Y.-M.}\ \bibnamefont {Xu}}, \bibinfo
  {author} {\bibfnamefont {P.}~\bibnamefont {Fournier}}, \bibinfo {author}
  {\bibfnamefont {S.}~\bibnamefont {Li}}, \bibinfo {author} {\bibfnamefont
  {P.}~\bibnamefont {Dai}}, \bibinfo {author} {\bibfnamefont {Z.}~\bibnamefont
  {Wang}},\ and\ \bibinfo {author} {\bibfnamefont {H.}~\bibnamefont {Ding}},\
  }\bibfield  {title} {\bibinfo {title} {{Competition between
  Antiferromagnetism and Superconductivity in the Electron-Doped Cuprates
  Triggered by Oxygen Reduction}},\ }\href
  {https://doi.org/10.1103/PhysRevLett.99.157002} {\bibfield  {journal}
  {\bibinfo  {journal} {Physical Review Letters}\ }\textbf {\bibinfo {volume}
  {99}},\ \bibinfo {pages} {157002} (\bibinfo {year} {2007})}\BibitemShut
  {NoStop}%
\bibitem [{\citenamefont {Song}\ \emph {et~al.}(2012)\citenamefont {Song},
  \citenamefont {Park}, \citenamefont {Kim}, \citenamefont {Kim}, \citenamefont
  {Leem}, \citenamefont {Choi}, \citenamefont {Jung}, \citenamefont {Koh},
  \citenamefont {Han}, \citenamefont {Yoshida}, \citenamefont {Eisaki},
  \citenamefont {Lu}, \citenamefont {Shen},\ and\ \citenamefont
  {Kim}}]{Song2012}%
  \BibitemOpen
  \bibfield  {author} {\bibinfo {author} {\bibfnamefont {D.}~\bibnamefont
  {Song}}, \bibinfo {author} {\bibfnamefont {S.~R.}\ \bibnamefont {Park}},
  \bibinfo {author} {\bibfnamefont {C.}~\bibnamefont {Kim}}, \bibinfo {author}
  {\bibfnamefont {Y.}~\bibnamefont {Kim}}, \bibinfo {author} {\bibfnamefont
  {C.}~\bibnamefont {Leem}}, \bibinfo {author} {\bibfnamefont {S.}~\bibnamefont
  {Choi}}, \bibinfo {author} {\bibfnamefont {W.}~\bibnamefont {Jung}}, \bibinfo
  {author} {\bibfnamefont {Y.}~\bibnamefont {Koh}}, \bibinfo {author}
  {\bibfnamefont {G.}~\bibnamefont {Han}}, \bibinfo {author} {\bibfnamefont
  {Y.}~\bibnamefont {Yoshida}}, \bibinfo {author} {\bibfnamefont
  {H.}~\bibnamefont {Eisaki}}, \bibinfo {author} {\bibfnamefont {D.~H.}\
  \bibnamefont {Lu}}, \bibinfo {author} {\bibfnamefont {Z.-X.}\ \bibnamefont
  {Shen}},\ and\ \bibinfo {author} {\bibfnamefont {C.}~\bibnamefont {Kim}},\
  }\bibfield  {title} {\bibinfo {title} {{Oxygen-content-dependent electronic
  structures of electron-doped cuprates}},\ }\href
  {https://doi.org/10.1103/PhysRevB.86.144520} {\bibfield  {journal} {\bibinfo
  {journal} {Physical Review B}\ }\textbf {\bibinfo {volume} {86}},\ \bibinfo
  {pages} {144520} (\bibinfo {year} {2012})}\BibitemShut {NoStop}%
\bibitem [{\citenamefont {Doiron-Leyraud}\ \emph {et~al.}(2007)\citenamefont
  {Doiron-Leyraud}, \citenamefont {Proust}, \citenamefont {LeBoeuf},
  \citenamefont {Levallois}, \citenamefont {Bonnemaison}, \citenamefont
  {Liang}, \citenamefont {Bonn}, \citenamefont {Hardy},\ and\ \citenamefont
  {Taillefer}}]{DoironLeyraud2007}%
  \BibitemOpen
  \bibfield  {author} {\bibinfo {author} {\bibfnamefont {N.}~\bibnamefont
  {Doiron-Leyraud}}, \bibinfo {author} {\bibfnamefont {C.}~\bibnamefont
  {Proust}}, \bibinfo {author} {\bibfnamefont {D.}~\bibnamefont {LeBoeuf}},
  \bibinfo {author} {\bibfnamefont {J.}~\bibnamefont {Levallois}}, \bibinfo
  {author} {\bibfnamefont {J.}~\bibnamefont {Bonnemaison}}, \bibinfo {author}
  {\bibfnamefont {R.}~\bibnamefont {Liang}}, \bibinfo {author} {\bibfnamefont
  {D.}~\bibnamefont {Bonn}}, \bibinfo {author} {\bibfnamefont {W.}~\bibnamefont
  {Hardy}},\ and\ \bibinfo {author} {\bibfnamefont {L.}~\bibnamefont
  {Taillefer}},\ }\bibfield  {title} {\bibinfo {title} {{Quantum oscillations
  and the Fermi surface in an underdoped high-$T_c$ superconductor}},\
  }\href@noop {} {\bibfield  {journal} {\bibinfo  {journal} {Nature}\ }\textbf
  {\bibinfo {volume} {447}},\ \bibinfo {pages} {565} (\bibinfo {year}
  {2007})}\BibitemShut {NoStop}%
\bibitem [{\citenamefont {Miccoli}\ \emph {et~al.}(2015)\citenamefont
  {Miccoli}, \citenamefont {Edler}, \citenamefont {Pfn{\"{u}}r},\ and\
  \citenamefont {Tegenkamp}}]{Miccoli2015}%
  \BibitemOpen
  \bibfield  {author} {\bibinfo {author} {\bibfnamefont {I.}~\bibnamefont
  {Miccoli}}, \bibinfo {author} {\bibfnamefont {F.}~\bibnamefont {Edler}},
  \bibinfo {author} {\bibfnamefont {H.}~\bibnamefont {Pfn{\"{u}}r}},\ and\
  \bibinfo {author} {\bibfnamefont {C.}~\bibnamefont {Tegenkamp}},\ }\bibfield
  {title} {\bibinfo {title} {{The 100th anniversary of the four-point probe
  technique: The role of probe geometries in isotropic and anisotropic
  systems}},\ }\href
  {https://iopscience.iop.org/article/10.1088/0953-8984/27/22/223201}
  {\bibfield  {journal} {\bibinfo  {journal} {Journal of Physics Condensed
  Matter}\ }\textbf {\bibinfo {volume} {27}},\ \bibinfo {pages} {223201}
  (\bibinfo {year} {2015})}\BibitemShut {NoStop}%
\bibitem [{\citenamefont {Savitzky}\ \emph {et~al.}(2018)\citenamefont
  {Savitzky}, \citenamefont {{El Baggari}}, \citenamefont {Clement},
  \citenamefont {Waite}, \citenamefont {Goodge}, \citenamefont {Baek},
  \citenamefont {Sheckelton}, \citenamefont {Pasco}, \citenamefont {Nair},
  \citenamefont {Schreiber}, \citenamefont {Hoffman}, \citenamefont {Admasu},
  \citenamefont {Kim}, \citenamefont {Cheong}, \citenamefont {Bhattacharya},
  \citenamefont {Schlom}, \citenamefont {McQueen}, \citenamefont {Hovden},\
  and\ \citenamefont {Kourkoutis}}]{Savitzky2018}%
  \BibitemOpen
  \bibfield  {author} {\bibinfo {author} {\bibfnamefont {B.~H.}\ \bibnamefont
  {Savitzky}}, \bibinfo {author} {\bibfnamefont {I.}~\bibnamefont {{El
  Baggari}}}, \bibinfo {author} {\bibfnamefont {C.~B.}\ \bibnamefont
  {Clement}}, \bibinfo {author} {\bibfnamefont {E.}~\bibnamefont {Waite}},
  \bibinfo {author} {\bibfnamefont {B.~H.}\ \bibnamefont {Goodge}}, \bibinfo
  {author} {\bibfnamefont {D.~J.}\ \bibnamefont {Baek}}, \bibinfo {author}
  {\bibfnamefont {J.~P.}\ \bibnamefont {Sheckelton}}, \bibinfo {author}
  {\bibfnamefont {C.}~\bibnamefont {Pasco}}, \bibinfo {author} {\bibfnamefont
  {H.}~\bibnamefont {Nair}}, \bibinfo {author} {\bibfnamefont {N.~J.}\
  \bibnamefont {Schreiber}}, \bibinfo {author} {\bibfnamefont {J.}~\bibnamefont
  {Hoffman}}, \bibinfo {author} {\bibfnamefont {A.~S.}\ \bibnamefont {Admasu}},
  \bibinfo {author} {\bibfnamefont {J.}~\bibnamefont {Kim}}, \bibinfo {author}
  {\bibfnamefont {S.-W.}\ \bibnamefont {Cheong}}, \bibinfo {author}
  {\bibfnamefont {A.}~\bibnamefont {Bhattacharya}}, \bibinfo {author}
  {\bibfnamefont {D.~G.}\ \bibnamefont {Schlom}}, \bibinfo {author}
  {\bibfnamefont {T.~M.}\ \bibnamefont {McQueen}}, \bibinfo {author}
  {\bibfnamefont {R.}~\bibnamefont {Hovden}},\ and\ \bibinfo {author}
  {\bibfnamefont {L.~F.}\ \bibnamefont {Kourkoutis}},\ }\bibfield  {title}
  {\bibinfo {title} {{Image registration of low signal-to-noise cryo-STEM
  data}},\ }\href {https://doi.org/10.1016/j.ultramic.2018.04.008} {\bibfield
  {journal} {\bibinfo  {journal} {Ultramicroscopy}\ }\textbf {\bibinfo {volume}
  {191}},\ \bibinfo {pages} {56} (\bibinfo {year} {2018})}\BibitemShut
  {NoStop}%
\bibitem [{\citenamefont {Eisebitt}\ \emph {et~al.}(1993)\citenamefont
  {Eisebitt}, \citenamefont {B\"oske}, \citenamefont {Rubensson},\ and\
  \citenamefont {Eberhardt}}]{Eisebitt1993}%
  \BibitemOpen
  \bibfield  {author} {\bibinfo {author} {\bibfnamefont {S.}~\bibnamefont
  {Eisebitt}}, \bibinfo {author} {\bibfnamefont {T.}~\bibnamefont {B\"oske}},
  \bibinfo {author} {\bibfnamefont {J.-E.}\ \bibnamefont {Rubensson}},\ and\
  \bibinfo {author} {\bibfnamefont {W.}~\bibnamefont {Eberhardt}},\ }\bibfield
  {title} {\bibinfo {title} {Determination of absorption coefficients for
  concentrated samples by fluorescence detection},\ }\href
  {https://doi.org/10.1103/PhysRevB.47.14103} {\bibfield  {journal} {\bibinfo
  {journal} {Physical Review B}\ }\textbf {\bibinfo {volume} {47}},\ \bibinfo
  {pages} {14103} (\bibinfo {year} {1993})}\BibitemShut {NoStop}%
\bibitem [{\citenamefont {Achkar}\ \emph {et~al.}(2011)\citenamefont {Achkar},
  \citenamefont {Regier}, \citenamefont {Wadati}, \citenamefont {Kim},
  \citenamefont {Zhang},\ and\ \citenamefont {Hawthorn}}]{Achkar2011}%
  \BibitemOpen
  \bibfield  {author} {\bibinfo {author} {\bibfnamefont {A.~J.}\ \bibnamefont
  {Achkar}}, \bibinfo {author} {\bibfnamefont {T.~Z.}\ \bibnamefont {Regier}},
  \bibinfo {author} {\bibfnamefont {H.}~\bibnamefont {Wadati}}, \bibinfo
  {author} {\bibfnamefont {Y.-J.}\ \bibnamefont {Kim}}, \bibinfo {author}
  {\bibfnamefont {H.}~\bibnamefont {Zhang}},\ and\ \bibinfo {author}
  {\bibfnamefont {D.~G.}\ \bibnamefont {Hawthorn}},\ }\bibfield  {title}
  {\bibinfo {title} {Bulk sensitive x-ray absorption spectroscopy free of
  self-absorption effects},\ }\href
  {https://doi.org/10.1103/PhysRevB.83.081106} {\bibfield  {journal} {\bibinfo
  {journal} {Physical Review B}\ }\textbf {\bibinfo {volume} {83}},\ \bibinfo
  {pages} {081106} (\bibinfo {year} {2011})}\BibitemShut {NoStop}%
\end{thebibliography}%


\begin{thebibliography}{10}%
\makeatletter
\providecommand \@ifxundefined [1]{%
 \@ifx{#1\undefined}
}%
\providecommand \@ifnum [1]{%
 \ifnum #1\expandafter \@firstoftwo
 \else \expandafter \@secondoftwo
 \fi
}%
\providecommand \@ifx [1]{%
 \ifx #1\expandafter \@firstoftwo
 \else \expandafter \@secondoftwo
 \fi
}%
\providecommand \natexlab [1]{#1}%
\providecommand \enquote  [1]{``#1''}%
\providecommand \bibnamefont  [1]{#1}%
\providecommand \bibfnamefont [1]{#1}%
\providecommand \citenamefont [1]{#1}%
\providecommand \href@noop [0]{\@secondoftwo}%
\providecommand \href [0]{\begingroup \@sanitize@url \@href}%
\providecommand \@href[1]{\@@startlink{#1}\@@href}%
\providecommand \@@href[1]{\endgroup#1\@@endlink}%
\providecommand \@sanitize@url [0]{\catcode `\\12\catcode `\$12\catcode
  `\&12\catcode `\#12\catcode `\^12\catcode `\_12\catcode `\%12\relax}%
\providecommand \@@startlink[1]{}%
\providecommand \@@endlink[0]{}%
\providecommand \url  [0]{\begingroup\@sanitize@url \@url }%
\providecommand \@url [1]{\endgroup\@href {#1}{\urlprefix }}%
\providecommand \urlprefix  [0]{URL }%
\providecommand \Eprint [0]{\href }%
\providecommand \doibase [0]{https://doi.org/}%
\providecommand \selectlanguage [0]{\@gobble}%
\providecommand \bibinfo  [0]{\@secondoftwo}%
\providecommand \bibfield  [0]{\@secondoftwo}%
\providecommand \translation [1]{[#1]}%
\providecommand \BibitemOpen [0]{}%
\providecommand \bibitemStop [0]{}%
\providecommand \bibitemNoStop [0]{.\EOS\space}%
\providecommand \EOS [0]{\spacefactor3000\relax}%
\providecommand \BibitemShut  [1]{\csname bibitem#1\endcsname}%
\let\auto@bib@innerbib\@empty
\bibitem [{\citenamefont {Zeng}\ \emph {et~al.}(2020)\citenamefont {Zeng},
  \citenamefont {Tang}, \citenamefont {Yin}, \citenamefont {Li}, \citenamefont
  {Li}, \citenamefont {Huang}, \citenamefont {Hu}, \citenamefont {Liu},
  \citenamefont {Omar}, \citenamefont {Jani}, \citenamefont {Lim},
  \citenamefont {Han}, \citenamefont {Wan}, \citenamefont {Yang}, \citenamefont
  {Pennycook}, \citenamefont {Wee},\ and\ \citenamefont {Ariando}}]{Zeng2020}%
  \BibitemOpen
  \bibfield  {author} {\bibinfo {author} {\bibfnamefont {S.}~\bibnamefont
  {Zeng}}, \bibinfo {author} {\bibfnamefont {C.~S.}\ \bibnamefont {Tang}},
  \bibinfo {author} {\bibfnamefont {X.}~\bibnamefont {Yin}}, \bibinfo {author}
  {\bibfnamefont {C.}~\bibnamefont {Li}}, \bibinfo {author} {\bibfnamefont
  {M.}~\bibnamefont {Li}}, \bibinfo {author} {\bibfnamefont {Z.}~\bibnamefont
  {Huang}}, \bibinfo {author} {\bibfnamefont {J.}~\bibnamefont {Hu}}, \bibinfo
  {author} {\bibfnamefont {W.}~\bibnamefont {Liu}}, \bibinfo {author}
  {\bibfnamefont {G.~J.}\ \bibnamefont {Omar}}, \bibinfo {author}
  {\bibfnamefont {H.}~\bibnamefont {Jani}}, \bibinfo {author} {\bibfnamefont
  {Z.~S.}\ \bibnamefont {Lim}}, \bibinfo {author} {\bibfnamefont
  {K.}~\bibnamefont {Han}}, \bibinfo {author} {\bibfnamefont {D.}~\bibnamefont
  {Wan}}, \bibinfo {author} {\bibfnamefont {P.}~\bibnamefont {Yang}}, \bibinfo
  {author} {\bibfnamefont {S.~J.}\ \bibnamefont {Pennycook}}, \bibinfo {author}
  {\bibfnamefont {A.~T.~S.}\ \bibnamefont {Wee}},\ and\ \bibinfo {author}
  {\bibfnamefont {A.}~\bibnamefont {Ariando}},\ }\bibfield  {title} {\bibinfo
  {title} {{Phase Diagram and Superconducting Dome of Infinite-Layer
  Nd$_{1-x}$Sr$_x$NiO$_2$ Thin Films}},\ }\href
  {https://doi.org/10.1103/PhysRevLett.125.147003} {\bibfield  {journal}
  {\bibinfo  {journal} {Physical Review Letters}\ }\textbf {\bibinfo {volume}
  {125}},\ \bibinfo {pages} {147003} (\bibinfo {year} {2020})}\BibitemShut
  {NoStop}%
\bibitem [{\citenamefont {Li}\ \emph {et~al.}(2020)\citenamefont {Li},
  \citenamefont {Wang}, \citenamefont {Lee}, \citenamefont {Harvey},
  \citenamefont {Osada}, \citenamefont {Goodge}, \citenamefont {Kourkoutis},\
  and\ \citenamefont {Hwang}}]{Li2020a}%
  \BibitemOpen
  \bibfield  {author} {\bibinfo {author} {\bibfnamefont {D.}~\bibnamefont
  {Li}}, \bibinfo {author} {\bibfnamefont {B.~Y.}\ \bibnamefont {Wang}},
  \bibinfo {author} {\bibfnamefont {K.}~\bibnamefont {Lee}}, \bibinfo {author}
  {\bibfnamefont {S.~P.}\ \bibnamefont {Harvey}}, \bibinfo {author}
  {\bibfnamefont {M.}~\bibnamefont {Osada}}, \bibinfo {author} {\bibfnamefont
  {B.~H.}\ \bibnamefont {Goodge}}, \bibinfo {author} {\bibfnamefont {L.~F.}\
  \bibnamefont {Kourkoutis}},\ and\ \bibinfo {author} {\bibfnamefont {H.~Y.}\
  \bibnamefont {Hwang}},\ }\bibfield  {title} {\bibinfo {title}
  {{Superconducting Dome in Nd$_{1-x}$Sr$_x$NiO$_2$ Infinite Layer Films}},\
  }\href {https://doi.org/10.1103/PhysRevLett.125.027001} {\bibfield  {journal}
  {\bibinfo  {journal} {Physical Review Letters}\ }\textbf {\bibinfo {volume}
  {125}},\ \bibinfo {pages} {027001} (\bibinfo {year} {2020})}\BibitemShut
  {NoStop}%
\bibitem [{\citenamefont {Lee}\ \emph {et~al.}(2023)\citenamefont {Lee},
  \citenamefont {Wang}, \citenamefont {Osada}, \citenamefont {Goodge},
  \citenamefont {Wang}, \citenamefont {Lee}, \citenamefont {Harvey},
  \citenamefont {Kim}, \citenamefont {Yu}, \citenamefont {Murthy},
  \citenamefont {Raghu}, \citenamefont {Kourkoutis},\ and\ \citenamefont
  {Hwang}}]{Lee2023}%
  \BibitemOpen
  \bibfield  {author} {\bibinfo {author} {\bibfnamefont {K.}~\bibnamefont
  {Lee}}, \bibinfo {author} {\bibfnamefont {B.~Y.}\ \bibnamefont {Wang}},
  \bibinfo {author} {\bibfnamefont {M.}~\bibnamefont {Osada}}, \bibinfo
  {author} {\bibfnamefont {B.~H.}\ \bibnamefont {Goodge}}, \bibinfo {author}
  {\bibfnamefont {T.~C.}\ \bibnamefont {Wang}}, \bibinfo {author}
  {\bibfnamefont {Y.}~\bibnamefont {Lee}}, \bibinfo {author} {\bibfnamefont
  {S.}~\bibnamefont {Harvey}}, \bibinfo {author} {\bibfnamefont {W.~J.}\
  \bibnamefont {Kim}}, \bibinfo {author} {\bibfnamefont {Y.}~\bibnamefont
  {Yu}}, \bibinfo {author} {\bibfnamefont {C.}~\bibnamefont {Murthy}}, \bibinfo
  {author} {\bibfnamefont {S.}~\bibnamefont {Raghu}}, \bibinfo {author}
  {\bibfnamefont {L.~F.}\ \bibnamefont {Kourkoutis}},\ and\ \bibinfo {author}
  {\bibfnamefont {H.~Y.}\ \bibnamefont {Hwang}},\ }\bibfield  {title} {\bibinfo
  {title} {{Linear-in-temperature resistivity for optimally superconducting
  (Nd,Sr)NiO$_2$}},\ }\href {https://doi.org/10.1038/s41586-023-06129-x}
  {\bibfield  {journal} {\bibinfo  {journal} {Nature}\ }\textbf {\bibinfo
  {volume} {619}},\ \bibinfo {pages} {288} (\bibinfo {year}
  {2023})}\BibitemShut {NoStop}%
\bibitem [{\citenamefont {Parzyck}\ \emph
  {et~al.}(2024{\natexlab{a}})\citenamefont {Parzyck}, \citenamefont {Gupta},
  \citenamefont {Wu}, \citenamefont {Anil}, \citenamefont {Bhatt},
  \citenamefont {Bouliane}, \citenamefont {Gong}, \citenamefont {Gregory},
  \citenamefont {Luo}, \citenamefont {Sutarto}, \citenamefont {He},
  \citenamefont {Chuang}, \citenamefont {Zhou}, \citenamefont {Herranz},
  \citenamefont {Kourkoutis}, \citenamefont {Singer}, \citenamefont {Schlom},
  \citenamefont {Hawthorn},\ and\ \citenamefont {Shen}}]{Parzyck2024a}%
  \BibitemOpen
  \bibfield  {author} {\bibinfo {author} {\bibfnamefont {C.~T.}\ \bibnamefont
  {Parzyck}}, \bibinfo {author} {\bibfnamefont {N.~K.}\ \bibnamefont {Gupta}},
  \bibinfo {author} {\bibfnamefont {Y.}~\bibnamefont {Wu}}, \bibinfo {author}
  {\bibfnamefont {V.}~\bibnamefont {Anil}}, \bibinfo {author} {\bibfnamefont
  {L.}~\bibnamefont {Bhatt}}, \bibinfo {author} {\bibfnamefont
  {M.}~\bibnamefont {Bouliane}}, \bibinfo {author} {\bibfnamefont
  {R.}~\bibnamefont {Gong}}, \bibinfo {author} {\bibfnamefont {B.~Z.}\
  \bibnamefont {Gregory}}, \bibinfo {author} {\bibfnamefont {A.}~\bibnamefont
  {Luo}}, \bibinfo {author} {\bibfnamefont {R.}~\bibnamefont {Sutarto}},
  \bibinfo {author} {\bibfnamefont {F.}~\bibnamefont {He}}, \bibinfo {author}
  {\bibfnamefont {Y.-D.}\ \bibnamefont {Chuang}}, \bibinfo {author}
  {\bibfnamefont {T.}~\bibnamefont {Zhou}}, \bibinfo {author} {\bibfnamefont
  {G.}~\bibnamefont {Herranz}}, \bibinfo {author} {\bibfnamefont {L.~F.}\
  \bibnamefont {Kourkoutis}}, \bibinfo {author} {\bibfnamefont
  {A.}~\bibnamefont {Singer}}, \bibinfo {author} {\bibfnamefont {D.~G.}\
  \bibnamefont {Schlom}}, \bibinfo {author} {\bibfnamefont {D.~G.}\
  \bibnamefont {Hawthorn}},\ and\ \bibinfo {author} {\bibfnamefont {K.~M.}\
  \bibnamefont {Shen}},\ }\bibfield  {title} {\bibinfo {title} {{Absence of
  $3a_0$ charge density wave order in the infinite-layer nickelate
  NdNiO$_2$}},\ }\href {https://doi.org/10.1038/s41563-024-01797-0} {\bibfield
  {journal} {\bibinfo  {journal} {Nature Materials}\ }\textbf {\bibinfo
  {volume} {23}},\ \bibinfo {pages} {486} (\bibinfo {year}
  {2024}{\natexlab{a}})}\BibitemShut {NoStop}%
\bibitem [{\citenamefont {Parzyck}\ \emph
  {et~al.}(2024{\natexlab{b}})\citenamefont {Parzyck}, \citenamefont {Anil},
  \citenamefont {Wu}, \citenamefont {Goodge}, \citenamefont {Roddy},
  \citenamefont {Kourkoutis}, \citenamefont {Schlom},\ and\ \citenamefont
  {Shen}}]{Parzyck2024b}%
  \BibitemOpen
  \bibfield  {author} {\bibinfo {author} {\bibfnamefont {C.~T.}\ \bibnamefont
  {Parzyck}}, \bibinfo {author} {\bibfnamefont {V.}~\bibnamefont {Anil}},
  \bibinfo {author} {\bibfnamefont {Y.}~\bibnamefont {Wu}}, \bibinfo {author}
  {\bibfnamefont {B.~H.}\ \bibnamefont {Goodge}}, \bibinfo {author}
  {\bibfnamefont {M.}~\bibnamefont {Roddy}}, \bibinfo {author} {\bibfnamefont
  {L.~F.}\ \bibnamefont {Kourkoutis}}, \bibinfo {author} {\bibfnamefont
  {D.~G.}\ \bibnamefont {Schlom}},\ and\ \bibinfo {author} {\bibfnamefont
  {K.~M.}\ \bibnamefont {Shen}},\ }\bibfield  {title} {\bibinfo {title}
  {{Synthesis of thin film infinite-layer nickelates by atomic hydrogen
  reduction: Clarifying the role of the capping layer}},\ }\href
  {https://doi.org/10.1063/5.0197304} {\bibfield  {journal} {\bibinfo
  {journal} {APL Materials}\ }\textbf {\bibinfo {volume} {12}},\ \bibinfo
  {pages} {031132} (\bibinfo {year} {2024}{\natexlab{b}})}\BibitemShut
  {NoStop}%
\bibitem [{\citenamefont {Tokura}\ \emph {et~al.}(1989)\citenamefont {Tokura},
  \citenamefont {Takagi},\ and\ \citenamefont {Uchida}}]{Tokura1989}%
  \BibitemOpen
  \bibfield  {author} {\bibinfo {author} {\bibfnamefont {Y.}~\bibnamefont
  {Tokura}}, \bibinfo {author} {\bibfnamefont {H.}~\bibnamefont {Takagi}},\
  and\ \bibinfo {author} {\bibfnamefont {S.}~\bibnamefont {Uchida}},\
  }\bibfield  {title} {\bibinfo {title} {A superconducting copper oxide
  compound with electrons as the charge carriers},\ }\href
  {https://doi.org/10.1038/337345a0} {\bibfield  {journal} {\bibinfo  {journal}
  {Nature}\ }\textbf {\bibinfo {volume} {337}},\ \bibinfo {pages} {345}
  (\bibinfo {year} {1989})}\BibitemShut {NoStop}%
\bibitem [{\citenamefont {Kotiuga}\ \emph {et~al.}(2019)\citenamefont
  {Kotiuga}, \citenamefont {Zhang}, \citenamefont {Li}, \citenamefont
  {Rodolakis}, \citenamefont {Zhou}, \citenamefont {Sutarto}, \citenamefont
  {He}, \citenamefont {Wang}, \citenamefont {Sun}, \citenamefont {Wang},
  \citenamefont {Aghamiri}, \citenamefont {Hancock}, \citenamefont {Rokhinson},
  \citenamefont {Landau}, \citenamefont {Abate}, \citenamefont {Freeland},
  \citenamefont {Comin}, \citenamefont {Ramanathan},\ and\ \citenamefont
  {Rabe}}]{Kotiuga2019}%
  \BibitemOpen
  \bibfield  {author} {\bibinfo {author} {\bibfnamefont {M.}~\bibnamefont
  {Kotiuga}}, \bibinfo {author} {\bibfnamefont {Z.}~\bibnamefont {Zhang}},
  \bibinfo {author} {\bibfnamefont {J.}~\bibnamefont {Li}}, \bibinfo {author}
  {\bibfnamefont {F.}~\bibnamefont {Rodolakis}}, \bibinfo {author}
  {\bibfnamefont {H.}~\bibnamefont {Zhou}}, \bibinfo {author} {\bibfnamefont
  {R.}~\bibnamefont {Sutarto}}, \bibinfo {author} {\bibfnamefont
  {F.}~\bibnamefont {He}}, \bibinfo {author} {\bibfnamefont {Q.}~\bibnamefont
  {Wang}}, \bibinfo {author} {\bibfnamefont {Y.}~\bibnamefont {Sun}}, \bibinfo
  {author} {\bibfnamefont {Y.}~\bibnamefont {Wang}}, \bibinfo {author}
  {\bibfnamefont {N.~A.}\ \bibnamefont {Aghamiri}}, \bibinfo {author}
  {\bibfnamefont {S.~B.}\ \bibnamefont {Hancock}}, \bibinfo {author}
  {\bibfnamefont {L.~P.}\ \bibnamefont {Rokhinson}}, \bibinfo {author}
  {\bibfnamefont {D.~P.}\ \bibnamefont {Landau}}, \bibinfo {author}
  {\bibfnamefont {Y.}~\bibnamefont {Abate}}, \bibinfo {author} {\bibfnamefont
  {J.~W.}\ \bibnamefont {Freeland}}, \bibinfo {author} {\bibfnamefont
  {R.}~\bibnamefont {Comin}}, \bibinfo {author} {\bibfnamefont
  {S.}~\bibnamefont {Ramanathan}},\ and\ \bibinfo {author} {\bibfnamefont
  {K.~M.}\ \bibnamefont {Rabe}},\ }\bibfield  {title} {\bibinfo {title}
  {Carrier localization in perovskite nickelates from oxygen vacancies},\
  }\href {https://doi.org/10.1073/pnas.1910490116} {\bibfield  {journal}
  {\bibinfo  {journal} {Proceedings of the National Academy of Sciences of the
  United States of America}\ }\textbf {\bibinfo {volume} {116}},\ \bibinfo
  {pages} {21992} (\bibinfo {year} {2019})}\BibitemShut {NoStop}%
\bibitem [{\citenamefont {Giannozzi}\ \emph {et~al.}(2009)\citenamefont
  {Giannozzi}, \citenamefont {Baroni}, \citenamefont {Bonini}, \citenamefont
  {Calandra}, \citenamefont {Car}, \citenamefont {Cavazzoni}, \citenamefont
  {Ceresoli}, \citenamefont {Chiarotti}, \citenamefont {Cococcioni},
  \citenamefont {Dabo}, \citenamefont {{Dal Corso}}, \citenamefont
  {de~Gironcoli}, \citenamefont {Fabris}, \citenamefont {Fratesi},
  \citenamefont {Gebauer}, \citenamefont {Gerstmann}, \citenamefont
  {Gougoussis}, \citenamefont {Kokalj}, \citenamefont {Lazzeri}, \citenamefont
  {Martin-Samos}, \citenamefont {Marzari}, \citenamefont {Mauri}, \citenamefont
  {Mazzarello}, \citenamefont {Paolini}, \citenamefont {Pasquarello},
  \citenamefont {Paulatto}, \citenamefont {Sbraccia}, \citenamefont {Scandolo},
  \citenamefont {Sclauzero}, \citenamefont {Seitsonen}, \citenamefont
  {Smogunov}, \citenamefont {Umari},\ and\ \citenamefont
  {Wentzcovitch}}]{Giannozzi2009}%
  \BibitemOpen
  \bibfield  {author} {\bibinfo {author} {\bibfnamefont {P.}~\bibnamefont
  {Giannozzi}}, \bibinfo {author} {\bibfnamefont {S.}~\bibnamefont {Baroni}},
  \bibinfo {author} {\bibfnamefont {N.}~\bibnamefont {Bonini}}, \bibinfo
  {author} {\bibfnamefont {M.}~\bibnamefont {Calandra}}, \bibinfo {author}
  {\bibfnamefont {R.}~\bibnamefont {Car}}, \bibinfo {author} {\bibfnamefont
  {C.}~\bibnamefont {Cavazzoni}}, \bibinfo {author} {\bibfnamefont
  {D.}~\bibnamefont {Ceresoli}}, \bibinfo {author} {\bibfnamefont {G.~L.}\
  \bibnamefont {Chiarotti}}, \bibinfo {author} {\bibfnamefont {M.}~\bibnamefont
  {Cococcioni}}, \bibinfo {author} {\bibfnamefont {I.}~\bibnamefont {Dabo}},
  \bibinfo {author} {\bibfnamefont {A.}~\bibnamefont {{Dal Corso}}}, \bibinfo
  {author} {\bibfnamefont {S.}~\bibnamefont {de~Gironcoli}}, \bibinfo {author}
  {\bibfnamefont {S.}~\bibnamefont {Fabris}}, \bibinfo {author} {\bibfnamefont
  {G.}~\bibnamefont {Fratesi}}, \bibinfo {author} {\bibfnamefont
  {R.}~\bibnamefont {Gebauer}}, \bibinfo {author} {\bibfnamefont
  {U.}~\bibnamefont {Gerstmann}}, \bibinfo {author} {\bibfnamefont
  {C.}~\bibnamefont {Gougoussis}}, \bibinfo {author} {\bibfnamefont
  {A.}~\bibnamefont {Kokalj}}, \bibinfo {author} {\bibfnamefont
  {M.}~\bibnamefont {Lazzeri}}, \bibinfo {author} {\bibfnamefont
  {L.}~\bibnamefont {Martin-Samos}}, \bibinfo {author} {\bibfnamefont
  {N.}~\bibnamefont {Marzari}}, \bibinfo {author} {\bibfnamefont
  {F.}~\bibnamefont {Mauri}}, \bibinfo {author} {\bibfnamefont
  {R.}~\bibnamefont {Mazzarello}}, \bibinfo {author} {\bibfnamefont
  {S.}~\bibnamefont {Paolini}}, \bibinfo {author} {\bibfnamefont
  {A.}~\bibnamefont {Pasquarello}}, \bibinfo {author} {\bibfnamefont
  {L.}~\bibnamefont {Paulatto}}, \bibinfo {author} {\bibfnamefont
  {C.}~\bibnamefont {Sbraccia}}, \bibinfo {author} {\bibfnamefont
  {S.}~\bibnamefont {Scandolo}}, \bibinfo {author} {\bibfnamefont
  {G.}~\bibnamefont {Sclauzero}}, \bibinfo {author} {\bibfnamefont {A.~P.}\
  \bibnamefont {Seitsonen}}, \bibinfo {author} {\bibfnamefont {A.}~\bibnamefont
  {Smogunov}}, \bibinfo {author} {\bibfnamefont {P.}~\bibnamefont {Umari}},\
  and\ \bibinfo {author} {\bibfnamefont {R.~M.}\ \bibnamefont {Wentzcovitch}},\
  }\bibfield  {title} {\bibinfo {title} {{QUANTUM ESPRESSO: a modular and
  open-source software project for quantum simulations of materials}},\ }\href
  {https://doi.org/10.1088/0953-8984/21/39/395502} {\bibfield  {journal}
  {\bibinfo  {journal} {Journal of Physics: Condensed Matter}\ }\textbf
  {\bibinfo {volume} {21}},\ \bibinfo {pages} {395502} (\bibinfo {year}
  {2009})}\BibitemShut {NoStop}%
\bibitem [{\citenamefont {Giannozzi}\ \emph {et~al.}(2017)\citenamefont
  {Giannozzi}, \citenamefont {Andreussi}, \citenamefont {Brumme}, \citenamefont
  {Bunau}, \citenamefont {{Buongiorno Nardelli}}, \citenamefont {Calandra},
  \citenamefont {Car}, \citenamefont {Cavazzoni}, \citenamefont {Ceresoli},
  \citenamefont {Cococcioni}, \citenamefont {Colonna}, \citenamefont
  {Carnimeo}, \citenamefont {{Dal Corso}}, \citenamefont {de~Gironcoli},
  \citenamefont {Delugas}, \citenamefont {DiStasio}, \citenamefont {Ferretti},
  \citenamefont {Floris}, \citenamefont {Fratesi}, \citenamefont {Fugallo},
  \citenamefont {Gebauer}, \citenamefont {Gerstmann}, \citenamefont {Giustino},
  \citenamefont {Gorni}, \citenamefont {Jia}, \citenamefont {Kawamura},
  \citenamefont {Ko}, \citenamefont {Kokalj}, \citenamefont
  {K{\"{u}}{\c{c}}{\"{u}}kbenli}, \citenamefont {Lazzeri}, \citenamefont
  {Marsili}, \citenamefont {Marzari}, \citenamefont {Mauri}, \citenamefont
  {Nguyen}, \citenamefont {Nguyen}, \citenamefont {Otero-de-la Roza},
  \citenamefont {Paulatto}, \citenamefont {Ponc{\'{e}}}, \citenamefont {Rocca},
  \citenamefont {Sabatini}, \citenamefont {Santra}, \citenamefont {Schlipf},
  \citenamefont {Seitsonen}, \citenamefont {Smogunov}, \citenamefont {Timrov},
  \citenamefont {Thonhauser}, \citenamefont {Umari}, \citenamefont {Vast},
  \citenamefont {Wu},\ and\ \citenamefont {Baroni}}]{Giannozzi2017a}%
  \BibitemOpen
  \bibfield  {author} {\bibinfo {author} {\bibfnamefont {P.}~\bibnamefont
  {Giannozzi}}, \bibinfo {author} {\bibfnamefont {O.}~\bibnamefont
  {Andreussi}}, \bibinfo {author} {\bibfnamefont {T.}~\bibnamefont {Brumme}},
  \bibinfo {author} {\bibfnamefont {O.}~\bibnamefont {Bunau}}, \bibinfo
  {author} {\bibfnamefont {M.}~\bibnamefont {{Buongiorno Nardelli}}}, \bibinfo
  {author} {\bibfnamefont {M.}~\bibnamefont {Calandra}}, \bibinfo {author}
  {\bibfnamefont {R.}~\bibnamefont {Car}}, \bibinfo {author} {\bibfnamefont
  {C.}~\bibnamefont {Cavazzoni}}, \bibinfo {author} {\bibfnamefont
  {D.}~\bibnamefont {Ceresoli}}, \bibinfo {author} {\bibfnamefont
  {M.}~\bibnamefont {Cococcioni}}, \bibinfo {author} {\bibfnamefont
  {N.}~\bibnamefont {Colonna}}, \bibinfo {author} {\bibfnamefont
  {I.}~\bibnamefont {Carnimeo}}, \bibinfo {author} {\bibfnamefont
  {A.}~\bibnamefont {{Dal Corso}}}, \bibinfo {author} {\bibfnamefont
  {S.}~\bibnamefont {de~Gironcoli}}, \bibinfo {author} {\bibfnamefont
  {P.}~\bibnamefont {Delugas}}, \bibinfo {author} {\bibfnamefont {R.~A.}\
  \bibnamefont {DiStasio}}, \bibinfo {author} {\bibfnamefont {A.}~\bibnamefont
  {Ferretti}}, \bibinfo {author} {\bibfnamefont {A.}~\bibnamefont {Floris}},
  \bibinfo {author} {\bibfnamefont {G.}~\bibnamefont {Fratesi}}, \bibinfo
  {author} {\bibfnamefont {G.}~\bibnamefont {Fugallo}}, \bibinfo {author}
  {\bibfnamefont {R.}~\bibnamefont {Gebauer}}, \bibinfo {author} {\bibfnamefont
  {U.}~\bibnamefont {Gerstmann}}, \bibinfo {author} {\bibfnamefont
  {F.}~\bibnamefont {Giustino}}, \bibinfo {author} {\bibfnamefont
  {T.}~\bibnamefont {Gorni}}, \bibinfo {author} {\bibfnamefont
  {J.}~\bibnamefont {Jia}}, \bibinfo {author} {\bibfnamefont {M.}~\bibnamefont
  {Kawamura}}, \bibinfo {author} {\bibfnamefont {H.-Y.}\ \bibnamefont {Ko}},
  \bibinfo {author} {\bibfnamefont {A.}~\bibnamefont {Kokalj}}, \bibinfo
  {author} {\bibfnamefont {E.}~\bibnamefont {K{\"{u}}{\c{c}}{\"{u}}kbenli}},
  \bibinfo {author} {\bibfnamefont {M.}~\bibnamefont {Lazzeri}}, \bibinfo
  {author} {\bibfnamefont {M.}~\bibnamefont {Marsili}}, \bibinfo {author}
  {\bibfnamefont {N.}~\bibnamefont {Marzari}}, \bibinfo {author} {\bibfnamefont
  {F.}~\bibnamefont {Mauri}}, \bibinfo {author} {\bibfnamefont {N.~L.}\
  \bibnamefont {Nguyen}}, \bibinfo {author} {\bibfnamefont {H.-V.}\
  \bibnamefont {Nguyen}}, \bibinfo {author} {\bibfnamefont {A.}~\bibnamefont
  {Otero-de-la Roza}}, \bibinfo {author} {\bibfnamefont {L.}~\bibnamefont
  {Paulatto}}, \bibinfo {author} {\bibfnamefont {S.}~\bibnamefont
  {Ponc{\'{e}}}}, \bibinfo {author} {\bibfnamefont {D.}~\bibnamefont {Rocca}},
  \bibinfo {author} {\bibfnamefont {R.}~\bibnamefont {Sabatini}}, \bibinfo
  {author} {\bibfnamefont {B.}~\bibnamefont {Santra}}, \bibinfo {author}
  {\bibfnamefont {M.}~\bibnamefont {Schlipf}}, \bibinfo {author} {\bibfnamefont
  {A.~P.}\ \bibnamefont {Seitsonen}}, \bibinfo {author} {\bibfnamefont
  {A.}~\bibnamefont {Smogunov}}, \bibinfo {author} {\bibfnamefont
  {I.}~\bibnamefont {Timrov}}, \bibinfo {author} {\bibfnamefont
  {T.}~\bibnamefont {Thonhauser}}, \bibinfo {author} {\bibfnamefont
  {P.}~\bibnamefont {Umari}}, \bibinfo {author} {\bibfnamefont
  {N.}~\bibnamefont {Vast}}, \bibinfo {author} {\bibfnamefont {X.}~\bibnamefont
  {Wu}},\ and\ \bibinfo {author} {\bibfnamefont {S.}~\bibnamefont {Baroni}},\
  }\bibfield  {title} {\bibinfo {title} {{Advanced capabilities for materials
  modelling with Quantum ESPRESSO}},\ }\href
  {https://doi.org/10.1088/1361-648X/aa8f79} {\bibfield  {journal} {\bibinfo
  {journal} {Journal of Physics: Condensed Matter}\ }\textbf {\bibinfo {volume}
  {29}},\ \bibinfo {pages} {465901} (\bibinfo {year} {2017})}\BibitemShut
  {NoStop}%
\bibitem [{\citenamefont {Prandini}\ \emph {et~al.}(2018)\citenamefont
  {Prandini}, \citenamefont {Marrazzo}, \citenamefont {Castelli}, \citenamefont
  {Mounet},\ and\ \citenamefont {Marzari}}]{prandini2018precision}%
  \BibitemOpen
  \bibfield  {author} {\bibinfo {author} {\bibfnamefont {G.}~\bibnamefont
  {Prandini}}, \bibinfo {author} {\bibfnamefont {A.}~\bibnamefont {Marrazzo}},
  \bibinfo {author} {\bibfnamefont {I.~E.}\ \bibnamefont {Castelli}}, \bibinfo
  {author} {\bibfnamefont {N.}~\bibnamefont {Mounet}},\ and\ \bibinfo {author}
  {\bibfnamefont {N.}~\bibnamefont {Marzari}},\ }\bibfield  {title} {\bibinfo
  {title} {Precision and efficiency in solid-state pseudopotential
  calculations},\ }\href {https://doi.org/10.1038/s41524-018-0127-2} {\bibfield
   {journal} {\bibinfo  {journal} {npj Computational Materials}\ }\textbf
  {\bibinfo {volume} {4}},\ \bibinfo {pages} {72} (\bibinfo {year} {2018})},\
  \bibinfo {note}
  {\href{http://materialscloud.org/sssp}{http://materialscloud.org/sssp}}\BibitemShut
  {NoStop}%
\end{thebibliography}%
\end{document}


\title{Supplemental Materials for: Superconductivity in the Parent Infinite-Layer Nickelate NdNiO$_2$}
\author{C. T. Parzyck}
  \affiliation{Laboratory of Atomic and Solid State Physics, Department of Physics, Cornell University, Ithaca, NY 14853, USA}
\author{Y. Wu}
  \affiliation{Laboratory of Atomic and Solid State Physics, Department of Physics, Cornell University, Ithaca, NY 14853, USA}
\author{L. Bhatt}
  \affiliation{School of Applied and Engineering Physics, Cornell University, Ithaca, NY 14853, USA}
\author{M. Kang}
  \affiliation{Laboratory of Atomic and Solid State Physics, Department of Physics, Cornell University, Ithaca, NY 14853, USA}
  \affiliation{Department of Materials Science and Engineering, Cornell University, Ithaca, NY 14853, USA}
  \affiliation{Kavli Institute at Cornell for Nanoscale Science, Cornell University, Ithaca, NY 14853, USA}
\author{Z. Arthur}
  \affiliation{Canadian Light Source, Inc., 44 Innovation Boulevard, Saskatoon, SK S7N 2V3, Canada}
\author{T. M. Pedersen}
  \affiliation{Canadian Light Source, Inc., 44 Innovation Boulevard, Saskatoon, SK S7N 2V3, Canada}
\author{R. Sutarto}
  \affiliation{Canadian Light Source, Inc., 44 Innovation Boulevard, Saskatoon, SK S7N 2V3, Canada}
\author{S. Fan}
  \affiliation{National Synchrotron Light Source II, Brookhaven National Laboratory, Upton, NY 11973, USA}
\author{J. Pelliciari}
  \affiliation{National Synchrotron Light Source II, Brookhaven National Laboratory, Upton, NY 11973, USA}
\author{V. Bisogni}
  \affiliation{National Synchrotron Light Source II, Brookhaven National Laboratory, Upton, NY 11973, USA}
\author{G. Herranz}
  \affiliation{Institut de Ciència de Materials de Barcelona (ICMAB-CSIC), Campus UAB Bellaterra 08193, Spain}
\author{A. B. Georgescu}
  \affiliation{Department of Chemistry, Indiana University, Bloomington, IN 47405, USA}
\author{D. G. Hawthorn}
  \affiliation{Department of Physics and Astronomy, University of Waterloo, Waterloo ON N2L 3G1, Canada}
\author{L. F. Kourkoutis}
  \affiliation{School of Applied and Engineering Physics, Cornell University, Ithaca, NY 14853, USA} 
  \affiliation{Kavli Institute at Cornell for Nanoscale Science, Cornell University, Ithaca, NY 14853, USA}
\author{D. A. Muller}
  \affiliation{School of Applied and Engineering Physics, Cornell University, Ithaca, NY 14853, USA}
  \affiliation{Kavli Institute at Cornell for Nanoscale Science, Cornell University, Ithaca, NY 14853, USA}
\author{D. G. Schlom}
  \affiliation{Department of Materials Science and Engineering, Cornell University, Ithaca, NY 14853, USA}
  \affiliation{Kavli Institute at Cornell for Nanoscale Science, Cornell University, Ithaca, NY 14853, USA}
  \affiliation{Leibniz-Institut f{\"u}r Kristallz{\"u}chtung, Max-Born-Stra{\ss}e 2, 12489 Berlin, Germany}
\author{K. M. Shen}
  \affiliation{Laboratory of Atomic and Solid State Physics, Department of Physics, Cornell University, Ithaca, NY 14853, USA}
  \affiliation{Kavli Institute at Cornell for Nanoscale Science, Cornell University, Ithaca, NY 14853, USA}
  \affiliation{Institut de Ciència de Materials de Barcelona (ICMAB-CSIC), Campus UAB Bellaterra 08193, Spain}

\maketitle
\tableofcontents
\clearpage
\section{Normal State Resistivity and the Low-Temperature Upturn}
\begin{figure*}
  \resizebox{17.6 cm}{!}{\includegraphics{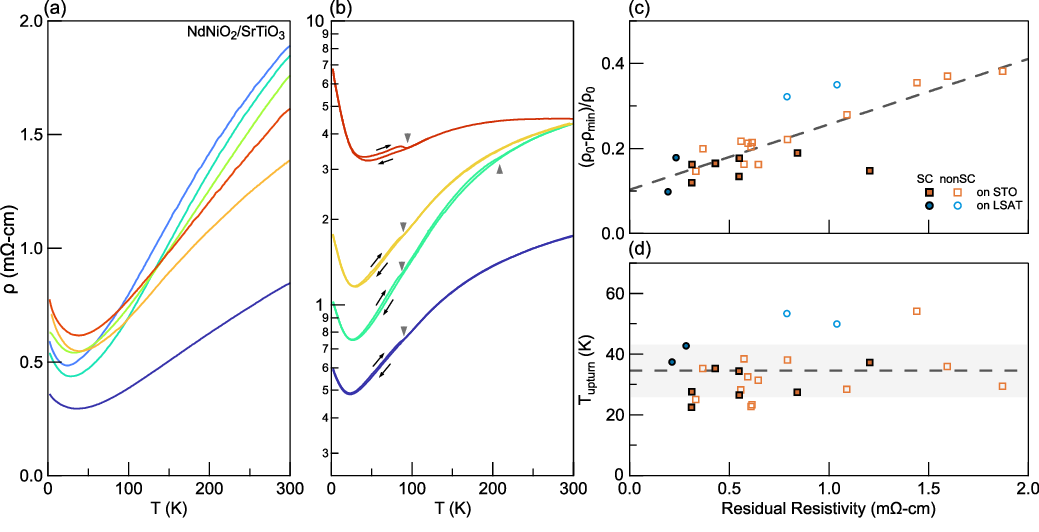}}
  \caption{\label{fig:S:additional} Temperature dependent resistivity of additional reduced \NNO~ films grown on \STO~ and capped with 2-3 unit cells of \STO~ and summary of the normal state behavior.  (a) Resistivity curves of additional `low-resistance' samples which were analyzed in the main text, displaying a non-hysteretic upturn at low temperature and no signatures of superconductivity. (b) Resistivity of samples which display hysteretic transitions, marked with grey triangles; temperature sweep directions are indicated with black arrows. (c) A metric of the magnitude of the low temperature upturn, $(\rho_{\textrm{res}} - \rho_{\textrm{min}})/\rho_{\textrm{res}}$, versus the film residual resistivity, \residual~ as well as the line of best fit to the combined data set (dashed). (d) The temperature at which the upturn occurs (\textit{i.e.} where the smallest normal state resistivity is measured), $T_{\textrm{upturn}}$, for all films plotted against \residual.  The average value of $T_{\textrm{upturn}}$, 34.5 K, is marked with a dashed line and the region bounded by one standard deviation, $\pm 9.1$~ K, is shaded.}
\end{figure*}

In this section we analyze some properties of the normal state resistance of the ensemble of samples discussed in the main text.  The resistivity of a number of additional, non-superconducting films prepared on \STO~ are  shown in Figs. \ref{fig:S:additional}(a) and \ref{fig:S:additional}(b).  Many of the films display relatively low residual resistivities, around 500 \microohmcm, with a small upturn visible at low temperature, shown in Fig. \ref{fig:S:additional}(a).  While this temperature dependence qualitatively matches prior measurements of undoped and underdoped (Nd,Sr)NiO$_2$~ \cite{Zeng2020,Li2020a,Lee2023}, some samples, shown in Fig. \ref{fig:S:additional}(b), display additional hysteretic resistive transitions, which are marked with grey triangles. Most commonly, a small hysteresis loop is visible between $\sim 25$~K and $87 - 95$~ K, though one sample exhibits a second loop closing near 210 K. The transition appears most strongly in samples with higher overall resistivity and, in particular, higher room temperature resistivity then other samples (exceeding 1.75 m\Ohm-cm).  The correlation between the magnitude of this transition and the increasing film resistivity suggests it is a result of a structural or electronic transition in another minority phase present in these sub-optimal samples.  

We next summarize some of the features observed across the array of samples presented here, including those which have been previously discussed in Refs. \cite{Parzyck2024a,Parzyck2024b}, in Figs. \ref{fig:S:additional}(c) and \ref{fig:S:additional}(d) to highlight the nature of the low temperature upturn in the resistivity.  From all of the measured resistivity curves we extract the magnitude of the upturn, defined by the normalized difference between the residual resistivity and the minimum measured resistivity, $(\rho_{\textrm{res}} - \rho_{\textrm{min}})/\rho_{\textrm{res}}$, as well as the temperature at which the minimum (normal state) resistivity is attained, $T_{\textrm{upturn}}$, and summarize them in Figs. \ref{fig:S:additional}(c) and \ref{fig:S:additional}(d), respectively.  We find that the magnitude of the upturn is clearly correlated to the residual resistivity.  Unlike the case of hole doped (Nd,Sr)NiO$_2$, however, the temperature at which the upturn occurs is uncorrelated to the film resistivity, with an average value of 34.5 K (dashed) and a spread of roughly $\pm 9.1$~K (shaded).  The strong correlation of the upturn magnitude with film residual resistivity, in a collection of nominally undoped films, suggests that the upturn observed here may be driven by disorder rather then an intrinsic mechanism.

\section{Residual Oxygen -- Additional STEM and DFT}
\begin{figure*}
  \resizebox{14.5cm}{!}{\includegraphics{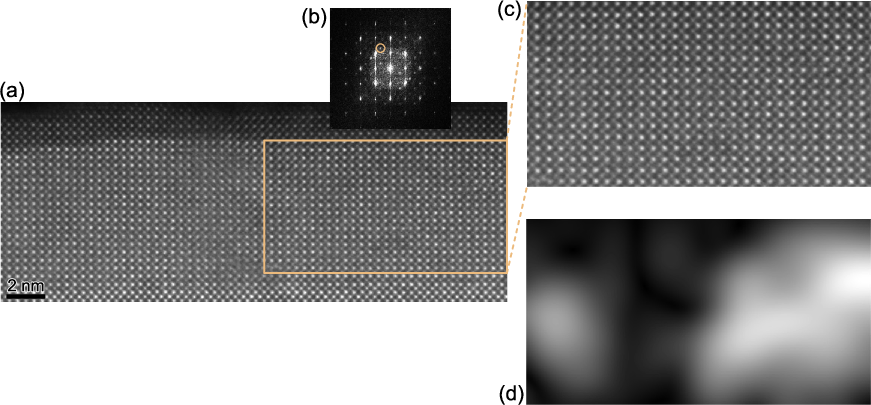}}
  \caption{\label{fig:S:tem} Ordered excess oxygen in a superconducting \NNO~ film grown on LSAT. (a) Reproduction of the ADF-STEM image from Fig. 2(d) of the main text. (b) Magnitude of the Fourier transform of this image, with a $\nicefrac{1}{3}^{\textrm{rd}}$~ order peak highlighted (orange circle). (c) Zoom in on the highlighted region of (a).  (d) Fourier filtered version of panel (c) obtained by taking the inverse transform of the circled peak in (b).}
\end{figure*} 

In Fig. 2(c) of the main text we show two images from different locations of the same reduced, superconducting, \NNO/LSAT thin film. In one region of the film no superstructure peaks are visible, in the second, however, clear $\nicefrac{1}{3}^{\textrm{rd}}$~ order peaks are visible, indicating the presence of an intermediate phase of ordered apical oxygen ions, likely Nd$_3$Ni$_3$O$_8$.  In Fig. \ref{fig:S:tem} we visualize this secondary phase in real space by Fourier filtering the image -- taking the inverse Fourier transform of the peak circled in Fig. \ref{fig:S:tem}(b), which corresponds to the $\nicefrac{1}{3}^{\textrm{rd}}$ ordering.  In Fig. \ref{fig:S:tem}(c) we see that the secondary phase appears to form clusters rather then a layer-wise or uniform distribution throughout the film.  The Fourier filtered image, showing domain distribution of an ordered intermediate phase, was obtained by taking the inverse Fourier transform of the marked superlattice peak corresponding to the $\nicefrac{1}{3}^{\textrm{rd}}$ ordering.

\begin{figure*}
  \resizebox{16cm}{!}{\includegraphics{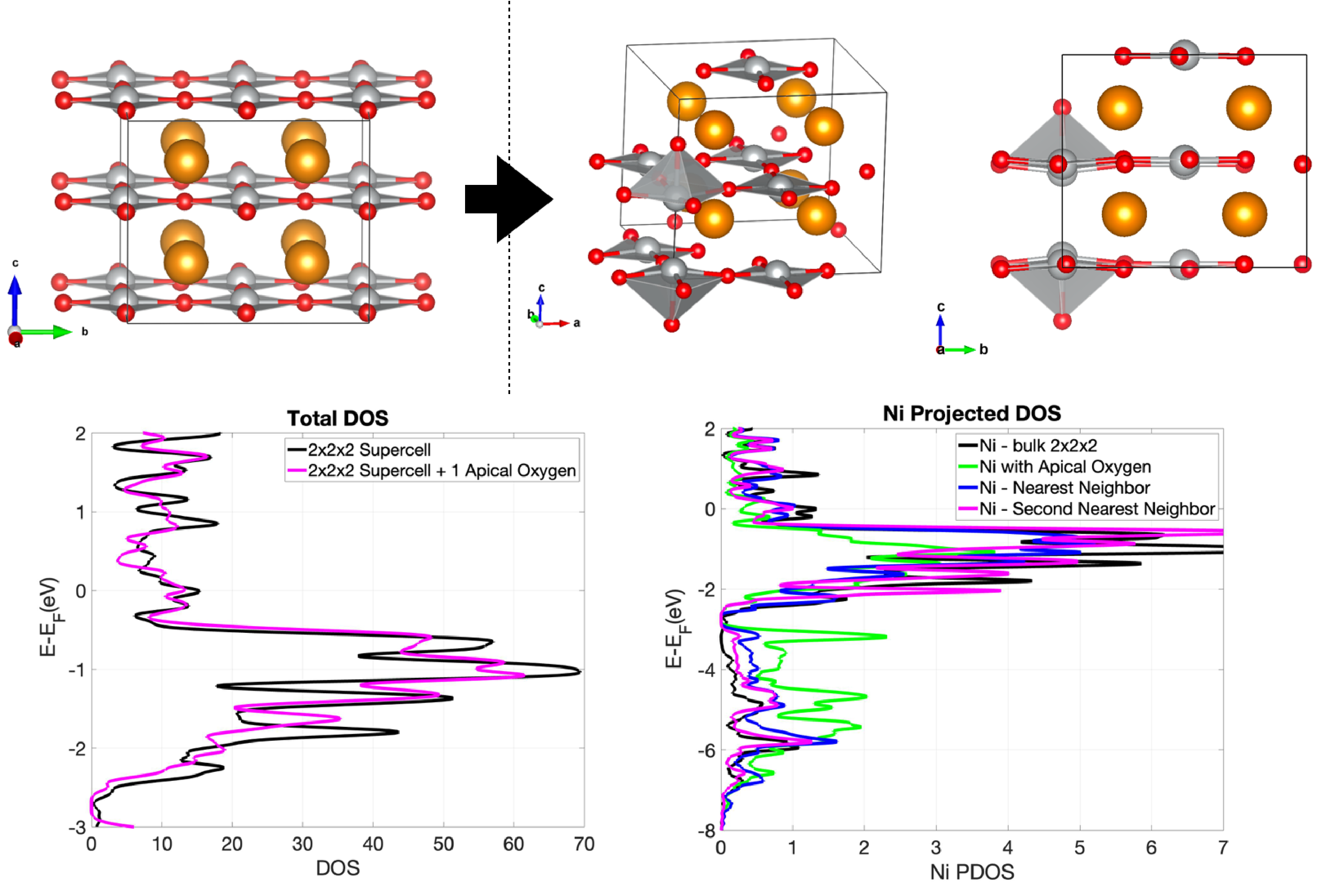}}
  \caption{\label{fig:S:DFT3a0} Top: $2\times 2\times 2$~ NdNiO$_2$ supercell (Ni-gray, Nd-golden, O-red) obtained from DFT calculations displaying Ni-O-Ni and O-Ni-O bond angles of 180$^{\circ}$ in plane. Top right: $2\times 2\times 2$~ supercell with an additional apical oxygen added, which induces local distortions. Bottom: density of states calculations for the two pictured structures.  Even though the total density of states around the Fermi level is similar for the two materials, particularly around the Fermi level, the Ni projected density of states are clearly distinct, with three inequivalent Ni sites.}
\end{figure*}

\begin{figure*}
  \resizebox{16cm}{!}{\includegraphics{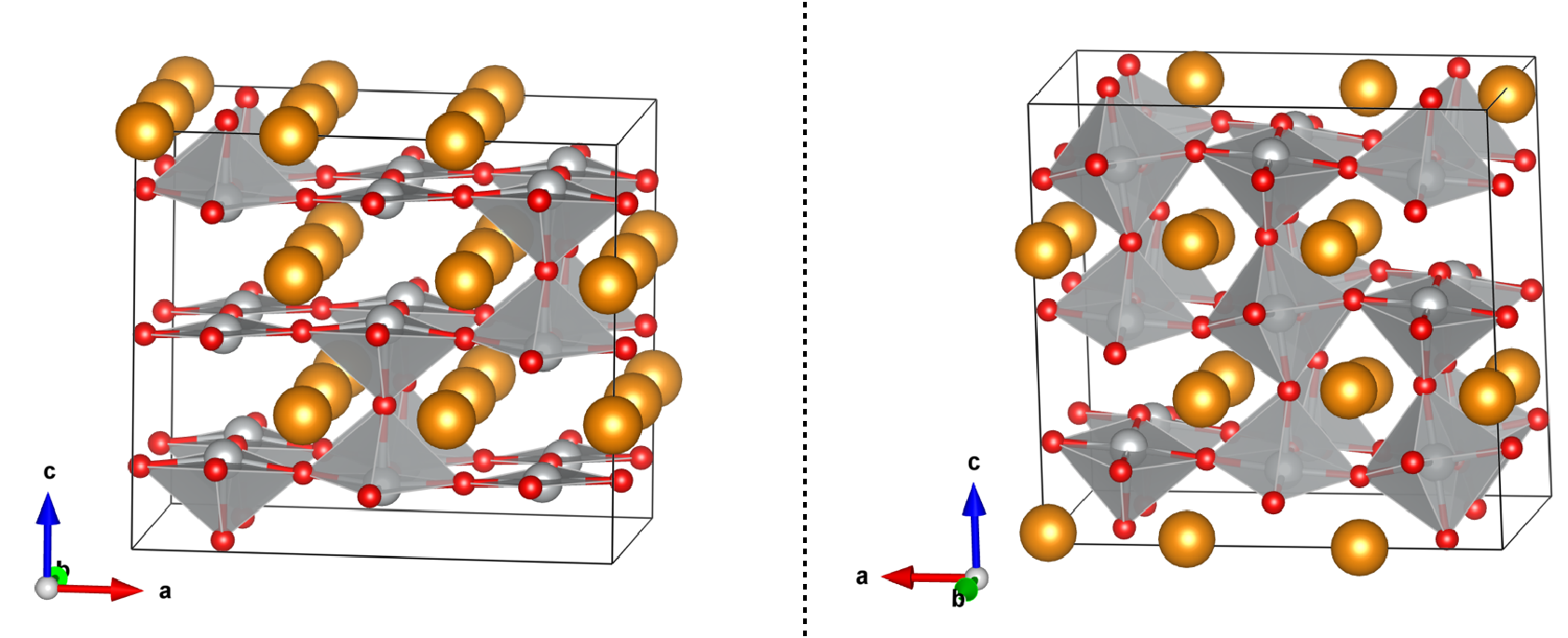}}
  \caption{\label{fig:S:DFT2} Nd$_3$Ni$_3$O$_7$ and Nd$_3$Ni$_3$O$_8$ unit cells (Ni-gray, Nd-golden, O-red) as obtained from DFT calculations. Both of these structures show strong structural distortions induced by the apical oxygen and the resulting NiO$_5$ tetrahedra and NiO$_6$}
\end{figure*}

While in some oxide compounds, most notably many cuprate superconductors, oxygen non-stoichiometry can effectively control the material doping, in others, including the $T'$~ cuprates \cite{Tokura1989} and perovskite nickelates \cite{Kotiuga2019},  excess oxygen ions do not simply contribute free carriers to the surrounding transition metal $d$ orbitals.  In Figs. \ref{fig:S:DFT3a0} and \ref{fig:S:DFT2} we present density functional theory calculations analyzing the effect of additional ordered and disordered apical oxygen ions on the structure and density of states of \NNO.  Our DFT results on infinite-layer nickelate supercells with sparse remaining apical oxygens find that they are associated with large structural distortions around the nearby Ni sites. In a relaxation calculation, large rotations (yielding Ni-O-Ni bond angles of $\sim 170$~ degrees) are observed in a $2\times 2\times 2$~ supercell with a single apical oxygen added -- disruption of the NiO$_2$ square lattice structure is expected to \textit{suppress} superconductivity.  Additionally, even though the total density of states at the Fermi level is not significantly affected by the presence of the added apical oxygen, the Ni atoms do differ in oxidation state in our calculation results. These density functional theory calculations were performed using Quantum Espresso version 6.8, ultrasoft pseudopotentials, a 540 eV energy cutoff, and a $4\times 4\times 4$~ k-mesh for the $2\times 2\times 2$~ supercell \cite{Giannozzi2009,Giannozzi2017a,prandini2018precision}. Additionally, we have performed calculations using a $3\times 3\times 3$~ supercell for Nd$_3$Ni$_3$O$_7$ and Nd$_3$Ni$_3$O$_8$. Both of these structures show strong structural distortions induced by the apical oxygen and the resulting NiO$_5$ tetrahedra and NiO$_6$ octahedra; for example the Nd$_3$Ni$_3$O$_7$ shows O-Ni-O bond angles of $\sim 165^{\circ}$~ for the tetrahedrally coordinated Ni atom, while the Ni-O-Ni bond angle along the tetrahedrally to octahedrally coordinated Ni atoms is 160$^{\circ}$.

\section{Additional Resistivity and Hall-Effect Data}
\begin{figure*}
  \resizebox{17.6cm}{!}{\includegraphics{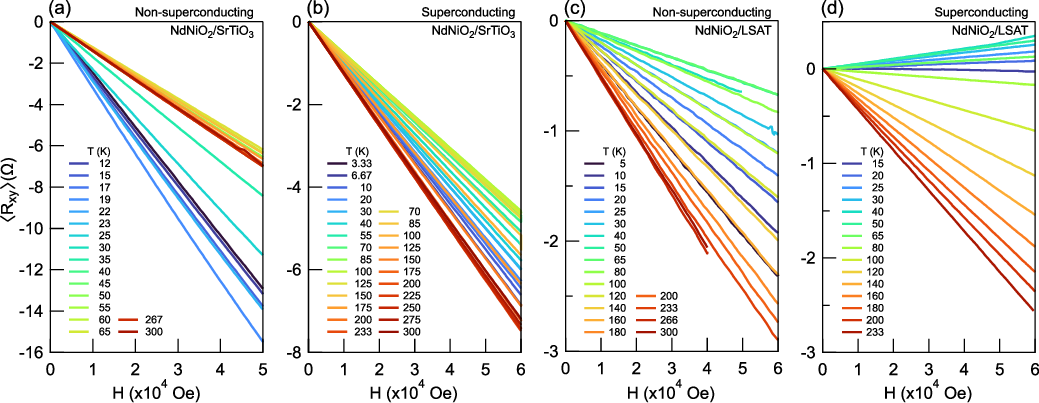}}
  \caption{\label{fig:S:halldata} Hall resistivity, as a function of applied field, for 20 u.c. NdNiO$_2$ films on \STO~ and LSAT capped with 3 u.c. of \STO.  Selected samples include (a) non-superconducting and (b) superconducting films on \STO~ as well as (c) non-superconducting and (d) superconducting films on LSAT.}
\end{figure*}

Here we present further analysis of the temperature dependent hall data described in Fig. 5 of the main text. In Fig. \ref{fig:S:halldata} we show the measured hall resistance, $\langle R_{xy}\rangle$ for a representative set of samples of each type--both superconducting and non-superconducting, on \STO~ and on LSAT.  The displayed resistance is the average of four measurements:
\begin{equation*}
    \langle R_{xy}\rangle(H) = \frac{1}{4}\left( R_{xy}(H) - R_{xy}(-H) + R_{yx}(H) - R_{yx}(-H)\right),
\end{equation*}
where $R_{xy}$ and $R_{yx}$ are taken in the same square Van der Pauw geometry but with reversed current and voltage contacts.  Here, and in all measured samples the $\langle R_{xy}\rangle$ versus $H$ relation is found to be linear to the highest measured fields and the fitted slope is extracted to measure $R_H$.

\begin{figure*}
  \resizebox{17.6cm}{!}{\includegraphics{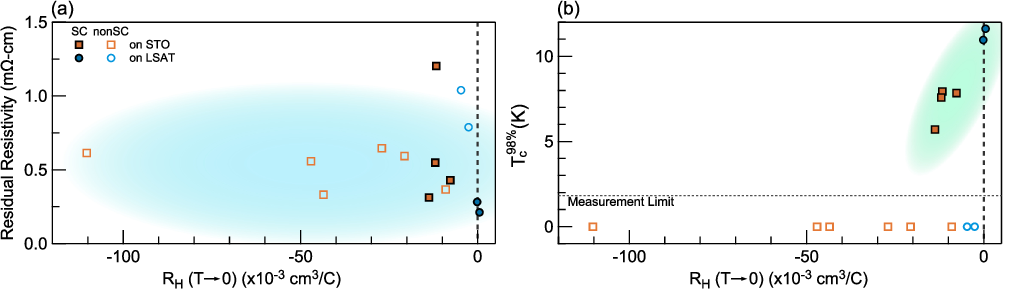}}
  \caption{\label{fig:S:hallsummary} Low temperature hall coefficients, $R_H(T\rightarrow 0)$, for additional superconducting and non-superconducting samples grown on \STO~ and LSAT. The lowest measurable temperature varies slightly from sample to sample (2-15 K) depending on measurement system configuration and the interference with the superconducting transition at the lowest temperatures. 
  (a) $R_H(T\rightarrow 0)$ as a function of the film residual resistivity, showing little-to-no correlation.  (b) $R_H(T\rightarrow 0)$ versus the onset superconducting transition temperature, $T_c$.}
\end{figure*}

In Fig. \ref{fig:S:hallsummary} we summarize the low temperature hall coefficients, $R_H(T\rightarrow 0)$ (the value of $R_H$ at the lowest accessible temperature -- 2 K for the non-superconducting films and $T_c^{\textrm{on}}$ for the superconducting ones) for all measured films, including those described in the main text. $R_H(T\rightarrow 0)$ is compared to other measurements including the film residual resistivity, onset transition temperature, $T_c^{98\%}$.  Surprisingly, films on \STO~ which show nearly identical residual resistivities ($\sim 500$~ \microohmcm) display disparate values of the hall coefficient, varying between $-0.110$~ and $-0.007$~ cm$^3$/C (though most values lie above -0.050).  As remarked on in the main text, this discrepancy is not unprecedented, as different groups have reported highly variable values of $R_H$: $-0.040$~ for Nd$_{0.92}$Sr$_{0.08}$NiO$_2$ in Ref. \cite{Zeng2020} and -0.007 for NdNiO$_2$ in Ref. \cite{Li2020a}, for instance.  Similar to Ref. \cite{Li2020a}, a large reduction in the magnitude of $R_H$ is observed when switching to substrates from \STO~ to LSAT -- the low temperature values of $R_H$ to be around -0.003 to -0.005 for (non-superconducting) film on LSAT which is close to the measured value of -0.001 for Nd$_{0.95}$Sr$_{0.05}$NiO$_2$ in \cite{Li2020a}.

We find that \residual~ and $R_H$ are essentially uncorrelated in Fig. \ref{fig:S:hallsummary}(a). Samples in which superconductivity is observed, however, do display some clustering of the low temperature hall coefficients nearer to zero and some correlation between the onset $T_c$ and the low-temperature hall coefficient, Fig. \ref{fig:S:hallsummary}(b).  However, much like \residual, having a small value of $R_H(T\rightarrow 0)$ does not guarantee the presence of a superconducting transition.  We surmise from the lack of correlation in Fig. \ref{fig:S:hallsummary}(a) that the low resistivities of our films result from reduced disorder and improved crystallinity -- not primarily from doping. 

\section{Additional Measurements of Superconducting Films}
\begin{figure*}
  \resizebox{17.6cm}{!}{\includegraphics{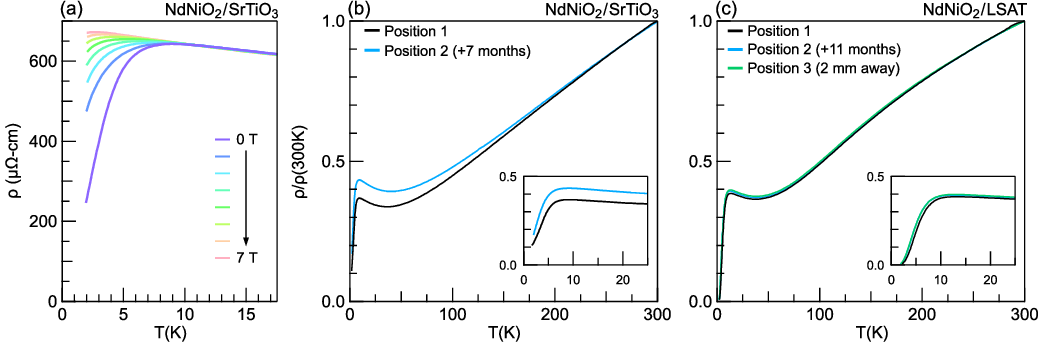}}
  \caption{\label{fig:S:moreRvsT} Additional resistivity data for superconducting \NNO~ films. (a) Resistance of a representative \NNO/\STO~ film under magnetic field. (b) Resistivity of the same \NNO/\STO~ sample measured initially after reduction, and again 7 months later, at two different positions near the sample center. (c) Resistivity of the same \NNO/LSAT sample measured after reduction, and again 11 months later, at two spatially separated positions on the sample surface.}
\end{figure*}

In Fig. \ref{fig:S:moreRvsT} we present some additional measurements of superconducting films grown on SrTiO$_3$ and LSAT.  As a comparison to Fig. 1(d) of the main text, which shows the magnetic field dependence of the superconducting transition for a film grown on LSAT, we illustrate the complementary measurement for a film grown on SrTiO$_3$ in Fig. \ref{fig:S:moreRvsT}(a).  As can be seen, the transition is quickly suppressed with the addition of a perpendicular magnetic field and nearly completely suppressed down to a temperature of 1.8 K under a 7 T applied field.  In Fig. \ref{fig:S:moreRvsT}(b) we illustrate the resistance of the same NdNiO$_2$/SrTiO$_3$ film measured at two nearby positions (within a few mm from the sample center) seven months apart (in the interim period the sample was stored at room temperature in a nitrogen purged desiccator).  Following the aging period the superconducting transition appears slightly different in shape but remains similar in onset temperature and in magnitude of the resistive drop.  A similar aging experiment was performed on an NdNiO$_2$/LSAT film and the results are reported in Fig. \ref{fig:S:moreRvsT}(c).  For the film on LSAT, an even smaller deviation in the transition shape is observed after 11 months of aging, though a small decreases in the onset $T_c$ and magnitude of the resistive drop are discernible.  Additionally, this sample was simultaneously measured at two separate positions, at the sample center and roughly 2 mm away, showing a nearly identical resistive drop at both locations.  These results indicate that the superconductivity, while sensitive to the reducing environment conditions, is stable under ambient conditions and reasonably uniform on the macro-scale.

\section{X-ray Absorption Measurements}
\begin{figure*}
  \resizebox{17.6cm}{!}{\includegraphics{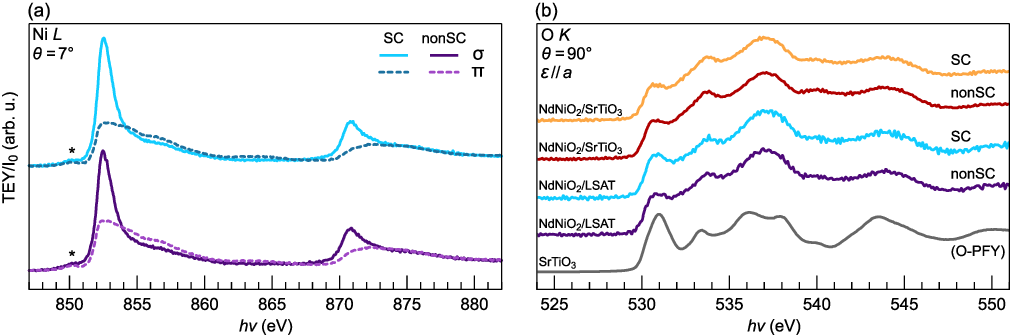}}
  \caption{\label{fig:S:moreXAS} Additional x-ray absorption spectroscopy measurements. (a) Polarization dependent XAS measurements at the Ni $L$-edge for superconducting and non-superconducting NdNiO$_2$ films on LSAT, reported as the total electron yield (TEY) normalized to the incident beam intensity ($I_0$). The contribution of the La $M_4$-edge from the substrate is marked with a star (*). Measurements were performed at an incidence angle of $\theta=7^{\circ}$ such that $\varepsilon\parallel a$~ in $\sigma$~ polarization and $\varepsilon$ is nearly parallel to $c$~ in $\pi$ polarization. Spectra have been normalized to match in the pre- and post-edge regions and offset for clarity. (b) Oxygen $K$-edge TEY spectra for \NNO~ films shown in (a) and in Fig. 3(a) of the main text, taken at normal incidence such that $\varepsilon\parallel a$.  Traces have been offset for clarity and the partial-fluorescence yield (PFY) spectra of a bare \STO~ substrate is included for reference. }
\end{figure*}

In Fig. \ref{fig:S:moreXAS} we present additional total electron yield (TEY) x-ray absorption measurements of NdNiO$_2$ samples on SrTiO$_3$ and LSAT.  Figure \ref{fig:S:moreXAS}(a) displays x-ray absorption data taken at the nickel $L$-edge on samples grown on LSAT.  The contribution from the La $M_4$-edge of the substrate is visible as a small peak around 850 eV, but is strongly suppressed in the more surface sensitive TEY measurement.  The Ni $L$-edge XAS for both superconducting and non-superconducting films on LSAT mirrors the measurements for \NNO/\STO~ presented in Fig. 3(a) of the main text, showing a single sharp peak in the $\varepsilon\parallel a$~ channel and a strong dichroic response.  TEY Measurements at the oxygen $K$-edge for the same samples shown here and in Fig. 3(a) of the main text are presented in Fig. \ref{fig:S:moreXAS}(b), along with a partial fluorescence yield measurement of a bare \STO~ substrate for reference.  All four samples (superconducting and non-superconducting, on LSAT and on \STO) show identical spectra, and no pre-peak feature below 530 eV, consistent with prior measurements.

\section{Lab Based X-ray Diffraction}
\begin{figure*}
  \resizebox{12.0cm}{!}{\includegraphics{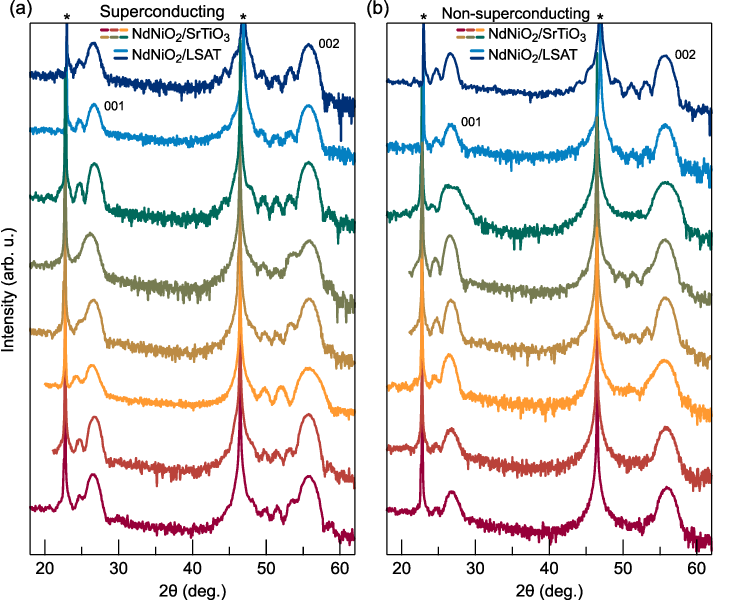}}
  \caption{\label{fig:S:transportXRD} Lab-based x-ray diffraction of NdNiO$_2$ films discussed in the main text. (a) $\theta-2\theta$ scans showing the 001 and 002 peaks of films from Fig. 1(a) of the main text, using the same color key. (b) Same, for the films shown in Fig. 1(b) of the main text. Stars indicate Bragg peaks arising from the \STO~ or LSAT substrates.}
\end{figure*}

In this section we show some additional lab-based x-ray diffraction data for films discussed in this study.  In Figs. \ref{fig:S:transportXRD}(a) and \ref{fig:S:transportXRD}(b) we show x-ray data for those samples whose transport is reported in Figs. 1(a) and 1(b) of the main text, respectively, using the same color key.  We note that all of the samples, save one, exhibit a singular, sharp, 002 peak just above $55^{\circ}$, consistent with fully reduced \NNO.  

\begin{figure*}
  \resizebox{12.0cm}{!}{\includegraphics{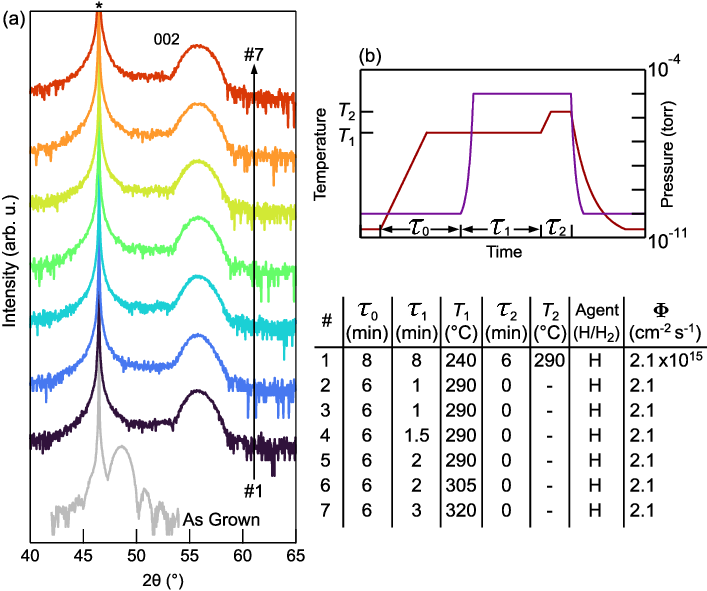}}
  \caption{\label{fig:S:annealing} Additional details of the progressive reduction experiment shown in Fig. 6 of the main text. (a) Lab based x-ray diffraction of the as-grown film and after the subsequent reduction steps. Star indicates the 002 Bragg peak of the \STO~ substrate. (b) Reduction parameters, using the same definitions as in Ref. \cite{Parzyck2024b}, for each of the reduction steps showing the time, temperature, and nominal atomic hydrogen flux.}
\end{figure*}

In Fig. \ref{fig:S:annealing}(a) we show x-ray diffraction measurements of the sequentially annealed sample described in Fig. 6(a) of the main text, performed at each reduction step.  Figure \ref{fig:S:annealing}(b) illustrates the reduction parameters (time, temperature, and flux) that were used at each subsequent step.  As shown in the main text, the superconducting downturn is first enhanced, then suppressed by subsequent reduction steps, coincident with a small change in the residual resistivity but little overall change to the sample resistance.  XRD data, shown here, show essentially no change over this reduction sequence, indicating no gross degradation of the film structure occurs in this sequence. One potentially interesting note which can be made about the subsequent reductions is that their effect, per unit time of exposure, appears substantially less then that of the initial reduction.  As discussed in Ref. \cite{Parzyck2024b}, the atomic hydrogen reduction is sensitive to changes in the reduction time down to the few-minute timescale, with "overreduction" of samples by 5 minutes or more causing substantial degradation.  Following the initial reduction, however, the subsequent steps seem to have a muted effect, with the initial 14 minute reaction time generating a substantial change and the following 10.5 minutes producing no noticeable change in the diffraction patterns and only minimal changes in the normal state resistivity.

\clearpage
\bibliography{Nickelates}